\documentclass[twocolumn]{aastex63}

\usepackage{amsmath}
\usepackage{multirow}
\usepackage{tabularx}
\usepackage{diagbox}
\usepackage{rotating}

\newcommand{\xs}{$\mathtt{XookSuut}$}

\newcommand{\kms}{km\,s$^{-1}$}

\newcommand{\PAdisk}{$\phi_\mathrm{disk}^\prime$}
\newcommand{\PAbar}{$\phi_\mathrm{bar}$}

\newcommand{\PAbarkin}{$\phi_\mathrm{bar,kin}^\prime$}
\newcommand{\PAbarphot}{$\phi_\mathrm{bar,phot}^\prime$}

\newcommand{\SN}{S$/$N}
\newcommand{\hi}{\ion{H}{1}}

\newcommand{\ha}{H$\alpha$}

\shorttitle{AMUSING++ non--circular motions}
\shortauthors{Lopez-Coba et al.}

\begin{document}

\title{Exploring stellar and ionized gas non--circular motions in barred galaxies with MUSE}

\correspondingauthor{C. Lopez-Coba}
\email{calopez@asiaa.sinica.edu.tw}

\author[0000-0003-1045-0702]{Carlos L\'opez-Cob\'a}
\affiliation{Institute of Astronomy and Astrophysics, Academia Sinica, No. 1, Section 4, Roosevelt Road, Taipei 10617, Taiwan}

\author[0000-0001-6444-9307]{Sebasti\'an~F.~S\'anchez}
\affiliation{Instituto de Astronom\'ia, Universidad Nacional Autonoma de M\'exico, Circuito Exterior, Ciudad Universitaria, Ciudad de M\'exico 04510, Mexico}

\author{Lihwai Lin}
\affiliation{Institute of Astronomy and Astrophysics, Academia Sinica, No. 1, Section 4, Roosevelt Road, Taipei 10617, Taiwan}

\author[0000-0003-0227-3451]{Joseph~P.~Anderson}
\affiliation{European Southern Observatory, Alonso de C\'ordova 3107, Vitacura, Casilla 190001, Santiago, Chile}

\author{Kai-Yang Lin}
\affiliation{Institute of Astronomy and Astrophysics, Academia Sinica, No. 1, Section 4, Roosevelt Road, Taipei 10617, Taiwan}

\author[0000-0002-2653-1120]{Irene~Cruz-Gonz\'alez}
\affiliation{Instituto de Astronom\'ia, Universidad Nacional Autonoma de M\'exico, Circuito Exterior, Ciudad Universitaria, Ciudad de M\'exico 04510, Mexico}

\author[0000-0002-1296-6887]{L. Galbany}
\affiliation{Institute of Space Sciences (ICE, CSIC), Campus UAB, Carrer de Can Magrans, s/n, E-08193 Barcelona, Spain.}
\affiliation{Institut d’Estudis Espacials de Catalunya (IEEC), E-08034 Barcelona, Spain.}

\author{Jorge~K.~Barrera-Ballesteros}
\affiliation{Instituto de Astronom\'ia, Universidad Nacional Autonoma de M\'exico, Circuito Exterior, Ciudad Universitaria, Ciudad de M\'exico 04510, Mexico}



\begin{abstract}

We present MUSE integral field stellar and ionized velocity maps for a sample of 14 barred galaxies. Most of these objects exhibit ``S''-shape iso-velocities in the bar region indicative of the presence of streaming motions in the velocity fields.
By applying circular rotation models we observe that bars leave symmetric structures in the residual maps of the stellar velocity.
%
We built non--circular rotation models using the \xs~tool to characterize the observed velocity fields. In particular we adopt bisymmetric models and a harmonic decomposition  for a bar potential for describing the non-axisymmetric velocities. We find that both models reproduce the observed kinematic features. 
The position angle of the oval distortion estimated from the bisymmetric model correlates with the photometric bar position angle $(\rho_{pearson} = 0.95)$, which suggest that non--circular velocities are caused by the bar. However because of the  low amplitudes of the $s_3$ harmonic we can not rule out radial flows as possible source.
Because of the weak detection of \ha~in our objects we are not able to compare gas to stellar non--circular motions in our sample, although we show that when galaxies are gas rich the oval distortion is also observed but with larger amplitudes.
Finally, we do not find evidence that the amplitude of the non--circular motions is dependent on the bar size, stellar mass or the global SFR.

\end{abstract}



\section{Introduction} \label{sec:intro}

Stellar bars are among the multiple non-axisymmetric components observed in galaxies \citep{RC3}, and it is estimated that nearly $\frac{1}{3}$ of disk galaxies exhibit a stellar bar \citep[e.g.,][]{Sellwood1993}, with the number increasing when observing in infrared bands \citep[e.g.,][]{Knapen2000}. 
Commonly, bars are thought to play multiple roles in the evolution of galaxies including  radial migration of stars \citep[e.g.,][]{Minchev2010}, triggering the star--formation (SF) and nuclear activity \citep[e.g.,][]{Combes2001} and for shaping the metallicity gradients in galaxies \citep[e.g.,][]{patri11}, among others.   
All these processes are intrinsically related with the dynamics of the bar. Therefore,
understand their kinematic properties is crucial to unreveal their real influence in galaxy evolution. 
While bars are relatively easy to recognize in continuum images \citep{Wozniak1991}, their kinematic counterpart is not evident until detailed examination of their velocity fields.
Even so, non-axisymmetric motions induced by bars have been identified in a wide variety of objects. These bar--driven motions when projected into the line of sight manifest in the form of oval distortions which have been observed in the velocity field of molecular \citep[e.g.,][]{Weliachew1988}, neutral \cite[e.g.,][]{Bosma1977b,Peterson1978} and ionized gas \citep[e.g.,][]{Fathi2005, Holmes2015}, as well as in stellar velocity maps \citep[e.g.,][]{Kormendy1983, Bettoni1988}. 
{ Such distortion recognized for producing a S--shaped pattern in the velocity field is less evident when the bar axis is closely aligned parallel or perpendicular to the line of nodes; these viewing angles disfavor the detection of bar stream motions \cite[e.g.,][]{Albada1981,Pence1984,Athanassoula2002}, making difficult to identify barlike-flows from the line-of-sight velocities. Hence it is common that bar-flow studies are biased toward galaxies with bars lying at intermediate orientations from the disk major/minor axis \citep[e.g.,][]{Pence1984}.}

With the advent of recent observational techniques such as the integral field spectroscopy (IFS) we are able to spatially resolve their ionized and stellar kinematic properties \citep[e.g.,][]{Fathi2005, jkbb14, Holmes2015,Fraser2020,Gadotti2020}.
In the optical, their kinematic counterpart have been identified mostly using \ha. This, however, tends to bias the studies towards  gas-rich systems.
In addition that \ha~traces in most cases the location of young stars. Conversely, stellar bars are dominated by old stellar populations \citep[e.g.,][]{Sanchez-Blazquez2014}, whose ionization is mostly dominated by old stars \citep[e.g.,][]{Stasinska2008, gomes16a, Lacerda2018, SebastianARAA}.

On the other hand, studies of bars using the stellar velocity maps in most IFS-galaxy surveys are limited by inherent spatial resolution effects,  precluding the identification of bar-driven flows \citep[e.g.,][]{jkbb14}.

In the present work, we address the kinematic study of bars by taking full advantage of integral field spectroscopic data from the multi-unit spectroscopic explorer instrument
to detect bar--like flows on the stellar and \ha~velocity maps of 14 galaxies. We built kinematic models to describe the bar flows  and try to relate the amplitude of the bar non--circular motions with some global properties of galaxies. 

The paper is structured as follows. In section 2 we describe the data; in section 3 the analysis of the velocity maps and the kinematic models adopted; in section 4 we describe the results and the conclusions are presented in section 5. Throughout this paper we adopted a $\Lambda$CDM cosmology with $H_0 = 70$~\kms~Mpc$^{-1}$, $\Omega_m = 0.3$ and $\Omega_{\Lambda} = 0.7$.

\section{Data} \label{sec:data}
In this study we use the dataproducts from the {\sc AMUSING++} galaxy compilation \citep[][]{Galbany2016, AMUSING++}. {\sc AMUSING++} is a compilation of $\sim600$ nearby galaxies observed with the multi-unit spectroscopic explorer (MUSE) instrument \citep[e.g.,][]{MUSE}.  With a field--of--view (FoV) of $1\arcmin \times 1\arcmin$, delivering $\sim$90~K spectra per data--cube, MUSE is
the most advanced integral field spectrocopy (IFS) instrument in the optical range covering from $4800$\AA~to $9300$\AA.
It combines both a moderate spectral resolution ($R\sim3000$), and a high spatial resolution limited by the atmospheric seeing.

The {\sc AMUSING++} cubes were analyzed using the {\sc pipe3d} package \citep[e.g.,][]{pipe3d}. This tool has been extensively used on a number of IFS galaxy surveys such  as CALIFA \citep[e.g.,][]{CALIFA1}, MaNGA \citep[e.g.,][]{MaNGA} and SAMI \citep[e.g.,][]{SAMI}.
{\sc pipe3d} models the stellar population using a combination of simple stellar populations (SSP), and derives the properties of the emission lines once subtracted this model to the original spectra. 

On the other hand, the recovery of the stellar population properties, such as the stellar kinematics, is performed over a tesselation map that follows the galaxy light distribution, typically around the V-band.
In general, the size of these tesselations or voxels depend on the target signal-to-noise (\SN) of the continuum, here chosen to reach \SN$\geq$ 30. 
For the particular case of the MUSE data, these correspond to  $\sim1\arcsec~$  for regions with high \SN, and $\sim 2$\arcsec~for regions with low \SN, typically for the outskirt of the disks. We emphasize that this is one of the major advantage of the current data compared to other IFS-surveys with coarser spatial resolutions.
Then the spectra within each voxel are coadded and over this spectrum is performed a simple stellar population analysis (SSP) to recover the velocity, age of the SSPs, metallicity, among other properties. This procedure is performed over each voxel of the tesselation map.

In addition to the above analysis, the ionized gas properties such as the flux, velocity and velocity dispersion, are recovered spaxel--by--spaxel by implementing a moment analysis on a set of $\sim 50$ emission--lines. We refer the reader to the {\sc AMUSING++} presentation paper \citep[e.g.,][]{AMUSING++} for more details of this procedure. The final outputs of the {\sc pipe3d} analysis are two dimensional maps of the main properties of ionized gas and the underlying stellar populations.

For the current analysis, we selected a sample of barred galaxies from the {\sc AMUSING++} dataset. Galaxies were required to exhibit clear stellar bars in the MUSE $gri$ images. Barred galaxies in interaction were excluded since their kinematics is dominated by the interaction process and not by internal ones; in addition, 
we exclude those objects where the higher \SN~voxels have sizes larger than 3 times the seeing of the observing night (based on the ESO DIMM archive). This condition excludes galaxies observed with non-optimal atmospheric conditions. Finally the apparent bar length must be resolved in the stellar velocity map and it must fit in the MUSE FoV. These criteria lead to a sample of 14 galaxies. Since {\sc AMUSING++} is a compilation of objects, our sample of bars is not an statistical representation of the population of barred galaxies in the Local Universe. Therefore, the results of this analysis are not statistical significant.

\section{Analysis}
\subsection{Photometric properties of bars}\label{sec:phot_sig}

\begin{figure}
\centering\includegraphics[width=1\columnwidth]{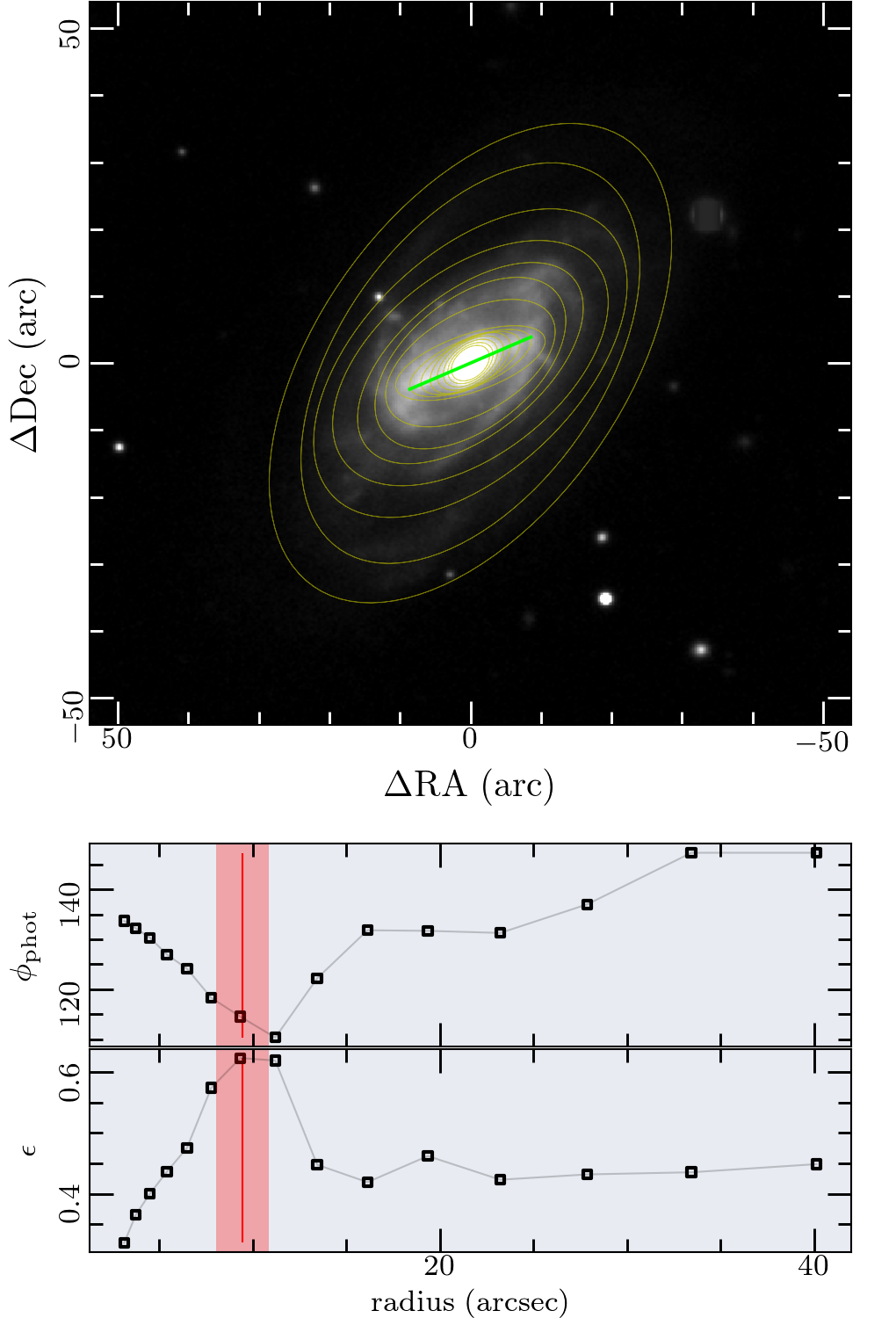}
\caption{Top panel: ESO476-16 DES-z image. Yellow contours represent the isophotes. Bottom panel shows the radial variation of the disk position angle and ellipticity of the isophotes. The vertical red line shows the approximate length and position angle of the bar, estimated in $r'_\mathrm{bar} \sim 9.5\pm1.5'' $ and $\phi = 114\pm3^{\circ}$ respectively. The red shadow region represents the bar length standard deviation. 
The green line ontop of the continuum image represents these values. { The galaxy is oriented North-East with North pointing up and East to the left.}}
\label{fig:isophote}
\end{figure}

Stellar bars are dominated in general by old stellar populations \citep[e.g.,][]{Sanchez-Blazquez2014,Vera2016,Fraser2019,Neumann2020}; 
therefore, it is common to study these structures in the reddest bands of the optical spectrum or in infrared bands \citep[e.g.,][]{Diaz-Garcia2016}, where their morphological properties are enhanced. Among those properties is the bar--length, albeit there is no unambiguous method for estimating it \citep[see][for a thorough description of methods]{Athanassoula2002}.
Parametric methods often decompose the galaxy light distribution into axisymmetric and non-axisymmetric components to recover the disk, bulge and bar properties \cite[e.g.,][]{Laurikainen2018,Jairo2019}.
An easier way consists in analyzing the object light distribution by tracing isophotes. Abrupt changes in the disk position angle ($\phi'_\mathrm{disk}$) and the axial ratio ($\epsilon$) have been commonly used to estimate the bar length, as well as its orientation in the sky \citep[e.g.,][]{Kormendy1983, Pence1984,  BettoniGalletta1988, Wozniak1991,  Perez2009}. Then, the true length of the bar along the major axis can be estimated by pure trigonometric relations as follows: 

\begin{equation}
 \label{Eq:bar_size}
 r_\mathrm{bar}= r'_\mathrm{bar} ( \cos^2\phi + \sin^2\phi/\sin^2 i)^{1/2} 
\end{equation}
where $r'_\mathrm{bar}$ is the apparent length of the bar in the sky, $\phi = \phi'_\mathrm{disk}-$~\PAbarphot, with \PAbarphot~representing the bar position angle, and $i$ is the disk inclination.


Instead of adopting narrow band images from the MUSE cubes, we use $r-$, $i-$ or $z-$ band images from DES \citep[e.g.,][]{DES} or Pan-STARRS \citep[e.g.,][]{Pan-STARRS1} whenever available. These images have a larger FoV and they are deeper than our MUSE data.
For extracting the isophotes we use the {\sc photoutils} Python package \citep{photoutils}, which relies  on the ellipse fitting analysis introduced by \cite{Jedrzejewski1987}.
After background subtraction, we adopt the last faintest isophote to derive the galaxy position angle $\phi'_{\mathrm{disk}}$ and disk inclination. Then we trace their radial variations, and measure the bar position angle and bar length based on the three consecutive isophotes that maximize the difference $\Delta P =  P_{i+1} -  P_{i}$, where $P$ is  $\phi'_{\mathrm{disk}}$ or the ellipticity ($\epsilon$). 
Although this analysis is just a proxy for the bar size and its orientation, it offers a first order estimation of these properties. 
%
As a sanity check, we performed a visual comparison between our estimations of \PAbarphot~and $r'_\mathrm{bar}$ with their apparent projections in the continuum images. In all cases a mutual agreement is reached. 

Figure ~\ref{fig:isophote} shows the implementation of this procedure for one object in our sample, ESO476-16. The estimated photometric bar position angle for this object is \PAbarphot$\sim 114^{\circ}$, which is consistent with the apparent orientation of the bar in the sky. We finally applied this analysis to all our galaxies in our sample.

\subsection{Kinematic signatures of bars}\label{sec:oval}

A common and noticeable effect of bars on the velocity field is the change in orientation of the dynamical major and minor axes \citep[e.g.,][]{jkbb14}, a signature of bar stream flows. 
As mentioned before, oval distortions on the velocity field of barred galaxies are well known from the gas velocity fields (CO, \hi, \ha). These kind of perturbations induce large deviation of the circular motions which can be observed in residual maps of simple axisymmetric models of circular rotation \citep[e.g.,][]{Bettoni1988,BettoniGalletta1988}.
Thus, we adopt kinematic models to identify non--circular motions most likely induced by bars.

\subsubsection{Kinematic models}
Multiple algorithms have been developed for describing the velocity field of galaxies. The vast majority of them are based on the tilted ring model \citep[e.g.,][]{Begeman1989}, in which the velocity field is divided in concentric rings each one rotating with a different velocity around a fixed kinematic centre.
In this work we use \xs~\citep[e.g.,][]{XookSuut}, { which relies on the DiskFit \citep[e.g.,][]{Spekkens2007} and RESWRI \citep[e.g.,][]{Schoenmakers1997} algorithms}. \xs~performs non--parametric circular and non--circular rotation models for describing the line-of-sight velocity (LoS). It adopts Markov Chain Monte Carlo (MCMC) methods for sampling the posterior distribution of the different kinematic components and creates an interpolated model based on the highest probability states. 

\xs~adopts the following $\log$~likelihood with flat priors on the parameters:
\begin{multline}
\label{eq:logpost}
  \log p(V_\mathrm{model}|\vec{\alpha}) =  -\frac{1}{2} \sum_{\substack{xpix \\ ypix}}^N  \Big( \frac{ V_\mathrm{obs} - \sum_{k=1}^K W_{k}  V_{k,\mathrm{model}}  }{\sigma} \Big)^2  \\
          -\log \sigma - \frac{N}{2} \log(2\pi)
\end{multline}
where $\vec{\alpha}$ represents all the parameters describing the considered model; $V_\mathrm{obs}$ is the observed velocity map; { $\sigma$ is the error map of the measured velocities}; $W_{k,n}$ is a series of weights computed at $k$ independent rings and will serve for creating a two-dimensional interpolated model $V_\mathrm{model}$;  $N$ is the total number of pixels included in the model.
{ All \xs~models adopt the thin disk approximation; i.e., it assumes the disk is intrinsically flat, with a constant disk ellipticity, constant position angle and fixed kinematic center. These conditions make it possible to represent kinematic models on two-dimensional maps
(hereafter 2D model).}

The circular model is the simplest model included in \xs, which is described by the following equation:
\begin{equation}
 \label{Eq:circular}
 V_\mathrm{circ,model} = V_\mathrm{sys} + \sin i ~V_t\cos \theta 
\end{equation}
In this expression  $V_t$ is the circular rotation, which is function of the radius. $V_\mathrm{sys}$ is the systemic velocity  assumed constant for all points in the galaxy. The azimuthal angle $\theta$~is measured on the galaxy plane and is related to the sky coordinates through the inclination angle $(i)$, kinematic position angle (\PAdisk) measured from North to East from the galaxy receding side, and the kinematic centre ($x_c,~y_c$).

In order to characterize the non--circular motions in our sample, we adopt the following non--circular models included in \xs.
{ The first one is the bisymmetric model proposed by \cite{Spekkens2007}, and is suitable for describing non--circular motions induced by a bisymmetric distortion to an axisymmetric potential; for instance that caused by a stellar bar.  
 This model fits elliptical stream lines around a fixed angle to the LoS velocities;} the equation describing this model is given by:
\begin{multline}
 \label{Eq:bisymmetric}
 V_\mathrm{bis,model}  = V_\mathrm{sys} + \sin i \Big( V_t\cos \theta - V_{2,t}\cos 2(\theta - \phi_\mathrm{bar}) \cos \theta \\ - V_{2,r} \sin 2(\theta - \phi_\mathrm{bar}) \sin \theta \Big)
\end{multline}
{ In this expression the $V_{2,r}$ and $V_{2,t}$ terms represent the bisymmetric deviations (radial and tangential respectively) from the circular velocity represented by the $V_t$ term.}
The phase $\phi_\mathrm{bar}$, represents the position angle of the oval distortion with respect to the azimuthal angle. Its sky projection is given by the following expression \citep[e.g.,][]{Bettoni1997, Spekkens2007} :
\begin{equation}
\label{Eq:kin_bar_pa}
\phi_\mathrm{bar,kin}^{\prime} =  \phi_{disk}^{\prime} +\arctan(\tan \phi_\mathrm{bar} \cos i)
\end{equation}
where $\phi_{disk}^{\prime}$ is the kinematic position angle major axis of the receding side, and $i$ is the galaxy inclination.

{ The description of the non--circular motions in our second model is based on the epicycle theory, where the radial and tangential components of the non--circular motions are taken into account by inducing small perturbations to circular orbits.}  As shown by \cite{Schoenmakers1997}, the LoS velocity $V_\mathrm{los}$, can be expressed as a Fourier series as follows:
\begin{equation}
 \label{eq1}
 V_\mathrm{los} =  c_0 + \sum_{m^\prime=1}^\infty \sin i \big[ c_{m^\prime} \cos m^\prime\theta + s_{m^\prime} \sin m^\prime\theta \big]
\end{equation}

where $c_0$ is the $0^\mathrm{th}$ harmonic term, assumed to be constant here ($\sim V_\mathrm{sys}$). The harmonic coefficients $c_{m^\prime}$ and $s_{m^\prime}$ describe  different velocity components as a function of the galactocentric radius, and $\theta$ is the azimuthal angle measured in the galaxy plane. 

\cite{Schoenmakers1997} showed that a perturbation to the potential of order $m$ affects only the $m^\prime = m \pm 1$ harmonics in the Fourier expansion (Eq.~\ref{eq1}). For the case of a bar $m=2$ and  the harmonics describing the LoS velocity adopt the following expression \citep[e.g.,][]{Franx1994}:
\begin{equation}
 \label{Eq:harmonic}
 V_\mathrm{los} =  c_0 + \sin i \big[ c_1\cos \theta +  c_3\cos 3\theta  + s_1\sin \theta  + s_3\sin 3\theta   \big]
\end{equation}
%
%
%
Thus, this expression represents the expected behavior of the LoS velocity for an elongated potential. { Note that a bisymmetric perturbation also includes the $m^{\prime }=3$ and $m^{\prime} = 1$ harmonics \citep[e.g.,][]{Spekkens2007, Oman2019}; however, as we will see in further sections the  amplitude of the harmonic coefficients allow us to assess the source of the non-circular motions.  }

\begin{figure}
\centering\includegraphics[width = \columnwidth]{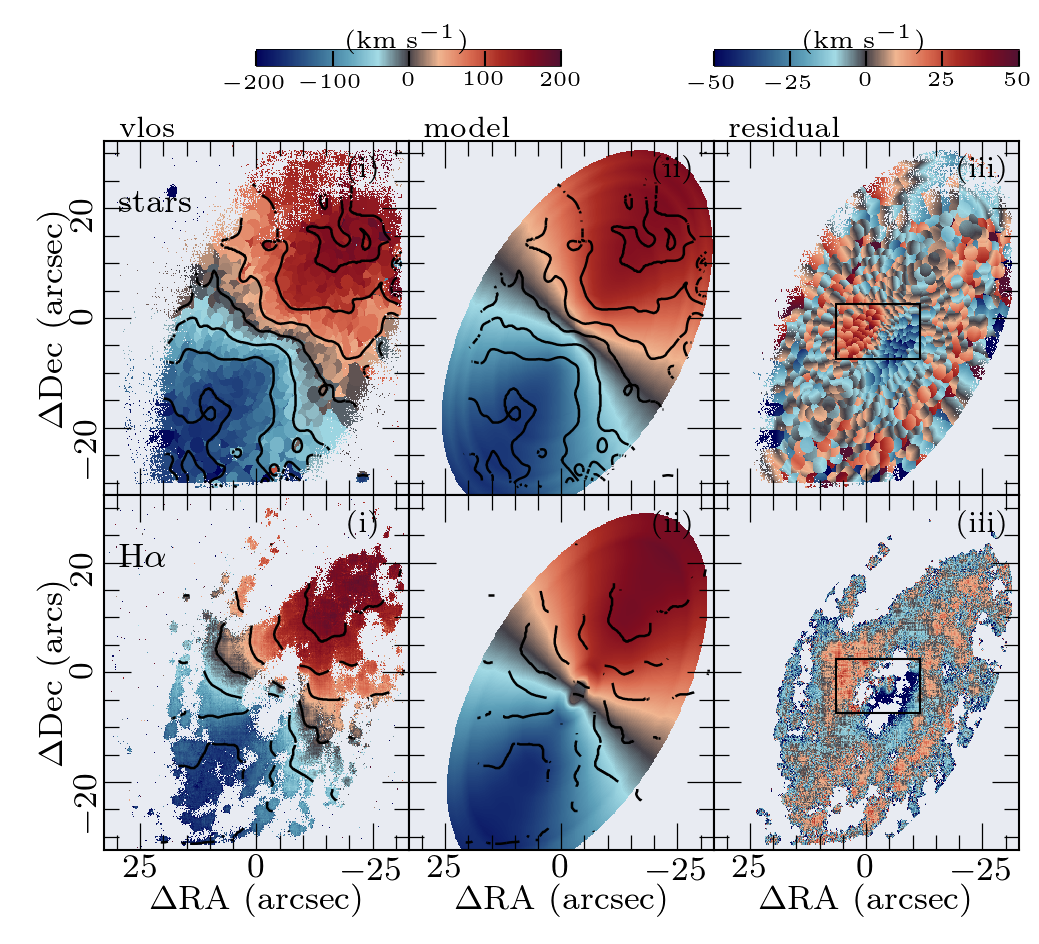}
\caption{Top row figures: (i) stellar velocity map; (ii) best circular rotation model; (iii) residual map of the model. The black rectangle on top of (iii) shows the region with the largest residuals. Iso-velocity contours spaced each 50~\kms~are superimposed on each map. 
Bottom panels have the same meaning but this time for the \ha~velocity map. Note the zero velocity line shows an ``S''--shape distortion.}
\label{fig:circular_model}
\end{figure}

We use \xs~to built circular and non--circular flow models of the stellar and \ha~velocity maps of our sample of galaxies. 
{ Unlike RESWRI \citep[][]{Schoenmakers1997}, \xs~does not allow the kinematic center and projection angles to change across radii. 
In fact, this behavior is not desired in our analysis since we expect that non--circular motions arise due to the bar potential, rather than being induced by the presence of a twisted disk. }
\xs~requires guess values for the disk position angle, inclination and kinematic center. For the first two we use the values estimated from the isophotal analysis described in Sec.~\ref{sec:phot_sig}, while the kinematic center was estimated by eye in each velocity map.
%
%
Since interpolated models are created over rings on the disk plane, we choose the rings to be spaced each $2\arcsec$. This value is grater than the spatial resolution of our objects ($\mathrm{FWHM}_\mathrm{DIMM} \sim 1.5\arcsec$ on average). Finally, we  use the corresponding error maps for discarding  spaxels with low \SN; that is, we remove spaxels with errors larger than 10~\kms.

In the following we use the same galaxy ESO476-16 as a showcase to go through the different kinematic models. 
Figure~\ref{fig:circular_model} shows the circular rotation model on the stellar and \ha~velocity maps. This figure highlights the differences in spatial resolution in both maps. Given that 
the recovery of the gas kinematics does not involve the binning procedure described in Sec.~\ref{sec:data}, the \ha~velocity map shows a better spatial resolution than the stellar velocity.

The stellar velocity reveals a central distortion induced most probably by the presence of the bar, while the iso-velocity contours appear twisted near the minor axis. 
The stellar circular model shows a typical rotating disk with orthogonal major--minor axis. The superimposed iso--velocities show that the outermost regions in the galaxy are compatible with a pure rotating disk.
The residual map in both cases (stars and \ha) show several important features in the inner parts of the disk. A close-up around the minor axis shows symmetric structures with large residuals of the order $\pm40$~\kms. 
On the other side, the \ha~velocity field shows a central ring with no data because of the low \SN~of \ha~in this region. Despite of that, \xs~is able to predict values in these regions by linear interpolation.  
We notice that the previous structure with high residuals is also observed in the gas velocity field, albeit is affected by the low \SN~data. This is more evident in the close-up to this region in the third panel. 
Such symmetric structures with blueshifted and redshifted components have been observed in residual maps of barred galaxies with similar amplitudes \citep[e.g.,][]{Fathi2005,Castillo-Morales2007}.
Thus, we may conclude from figure~\ref{fig:circular_model} that a non-axisymmetric component is present in the residual velocities from both the ionized gas and stars.

\begin{figure}
\centering\includegraphics[width = \columnwidth]{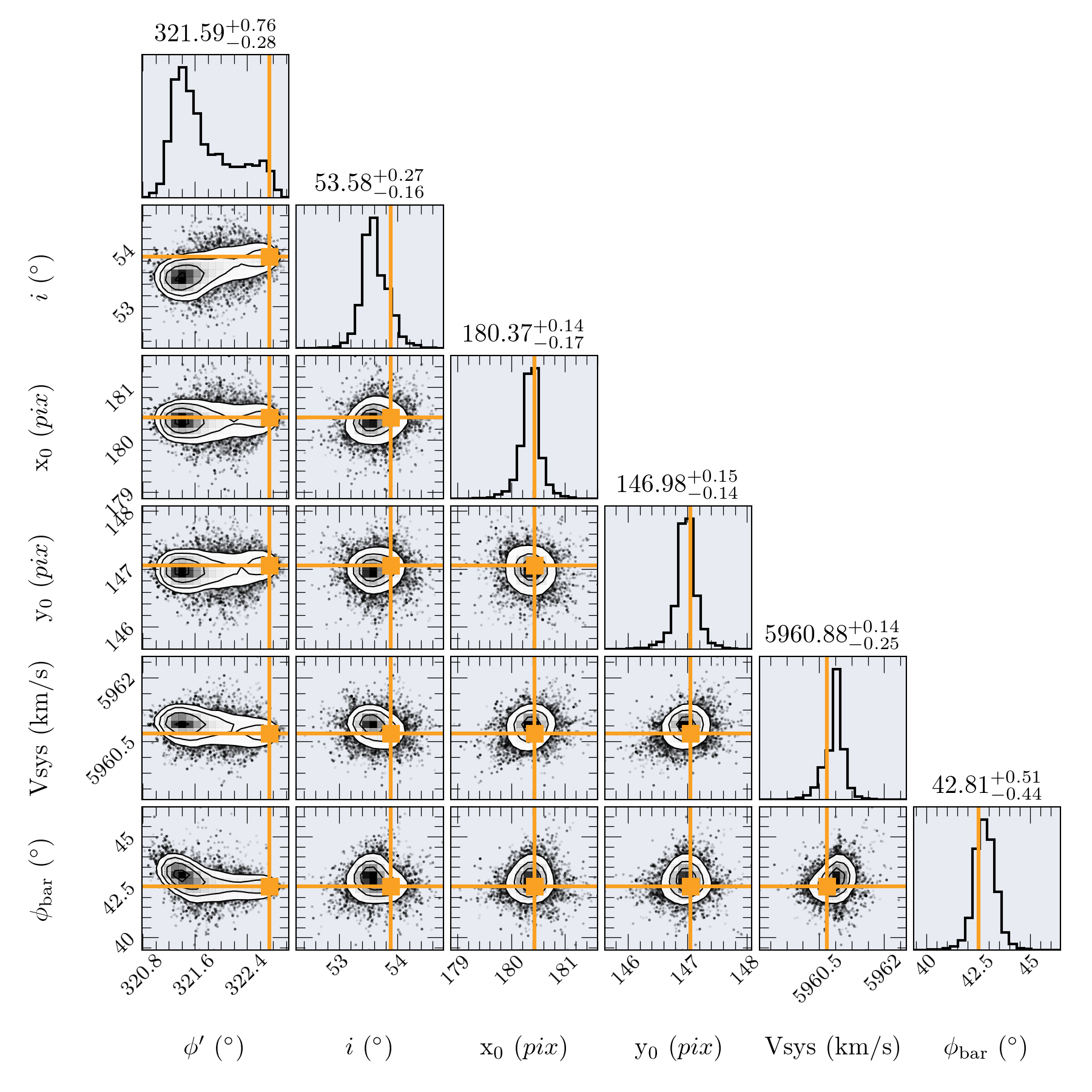}
\caption{Obtaining of the best fit parameters in \xs. Example adopting the bisymmetric model for the stellar velocity map of ESO~476-16. Mean values of the projection angles,  kinematic center, systemic velocity and bar position angle are shown at the top of each histogram.}
\label{fig:corner}
\end{figure}

\begin{figure*}
\centering\includegraphics[width = \textwidth]{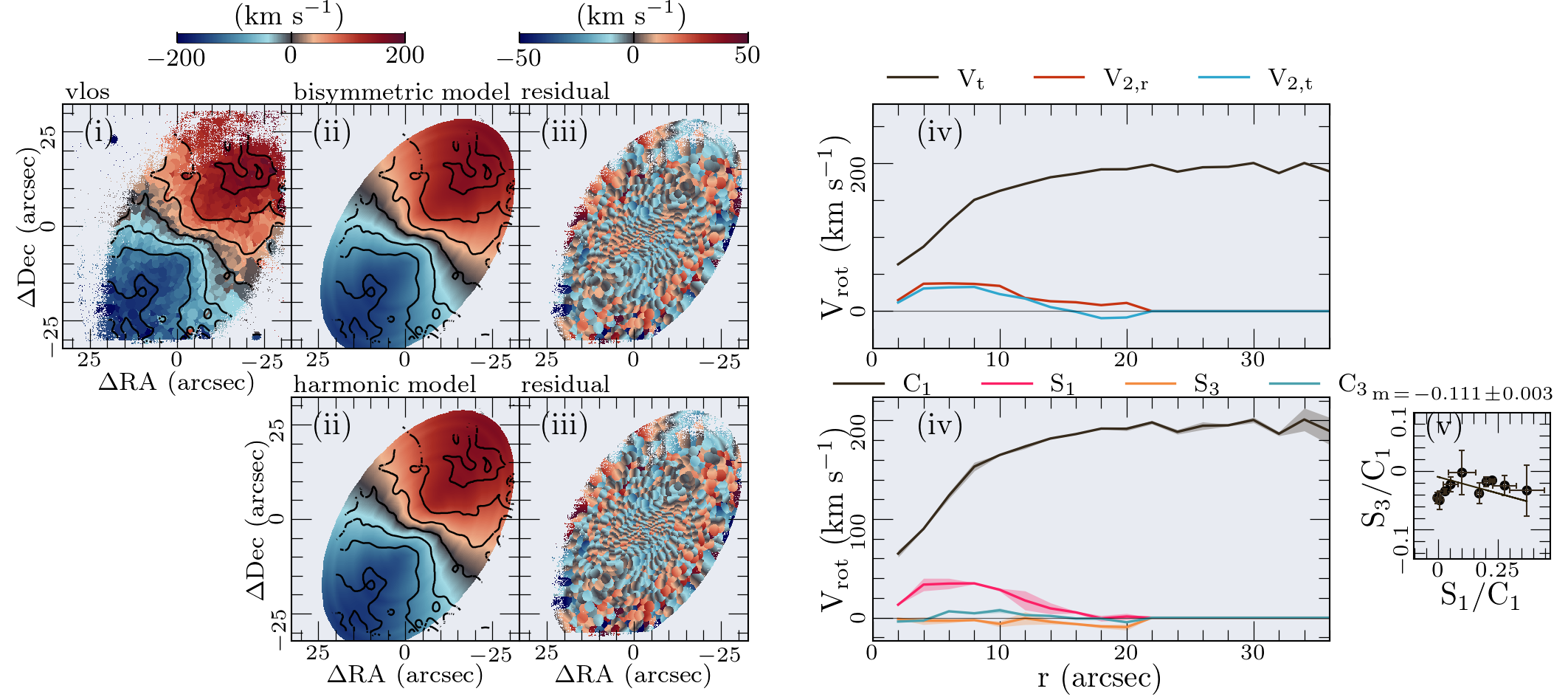}
\caption{Top figures: bisymmetric model applied to the stellar velocity field of ESO~476-16. {\it From left to right:} (i) stellar velocity map: (ii) 2D representation of the best fit bisymmetric model; (iii) residual map defined as observed $-$ model; (iv) radial variation of the tangential velocity (black), $V_{2,r}$ (red) and $V_{2,t}$ velocities as function of the  galactocentric distance. Bottom figures: harmonic decomposition for an $m = 2$ perturbation in the potential. {\it From left to right:} (i) input stellar velocity map; (ii) best 2D kinematic model; (iii) residual map; (iv) radial variation of the inclination corrected $c_1$, $s_1/c_1$ and $s_3/c_3$ harmonics; (v) $s_1/c_1$ vs. $s_s/c_1$ harmonic ratios. The continuous line in this panel represents a linear fit to the data points with slope $(m)$ pointed in the upper--left corner (not to be confused  with the harmonic coefficient).}
\label{fig:harmonic_ASASSN18oa}
\end{figure*}

Whether the velocity field of a barred galaxy is better described by a bisymmetric flow or by a pure axisymmetric radial flow is in general not straightforward to determine \cite[e.g.,][]{Wong2004}. In this work we will not make any {\it a priori} assumption on the kinematic model. Instead, we first adopt a bisymmetric model constraining the radial extension of the non--circular motions up to the the deprojected bar~size plus an additional value ranging from 1\arcsec--2.5\arcsec, which takes into account the spatial resolution of the data. Additionally, we decompose the LoS velocity with a Fourier analysis including only the $m^{\prime} = 1$ and $m^{\prime} = 3$ harmonics (Eq.~\ref{Eq:harmonic}), since 
the ratio of the $s_3$ and $s_1$  coefficients can be used to distinguish between bar--like flows and radial flows. 
\citep[e.g.,][]{Wong2004, Fathi2005, Elson2011}.

The inclusion of complex models such as the bisymmetric model adds extra variables in the fitting procedure that may reduce the residuals, but at the expense of overfiting the data. In order to assess the bisymmetric model we compute the Bayesian information criterion \citep[BIC,][]{BIC}, defined as $\mathrm{BIC}=N\ln (\chi^2/N) +\ln (N)N_{params}$, where $\chi^2$ is the quadratic sum of the residuals and $N_{params}$ the number of parameters to estimate from the model . Unlike $\chi^2$, BIC is more sensitive to the number of parameters to estimate from the model, favoring less complex ones.

We use the outputs of the circular rotation model (namely $x_c$, $y_c$, $i$, \PAdisk, $V_\mathrm{sys}$), as input for the harmonic and bisymmetric models. We note that there is no significant differences in the results if we fix these parameters or we let them vary during fitting. In any case, we let \xs~to derive the best set of parameters.
Results adopting the bisymmetric model on the stellar velocity map of the showcase object is shown in Figure~\ref{fig:corner}. This figure is a corner plot for the parameters describing the disk geometry and the bar orientation. 
The bisymmetric model finds an oval distortion oriented at \PAbar$= 43^{\circ}$, with its corresponding sky projection (Eq.~\ref{Eq:kin_bar_pa}) oriented at \PAbarkin $=289\pm1^\circ$. The position angle of the oval distortion is in agreement with our previous estimation for the photometric bar position angle ($114^\circ\pm3^\circ$)\footnote{ 
Position angles are measured from the North to the East. The photometric P.A. goes from $0^{\circ}-180^{\circ}$, while the kinematic P.A. goes from $0^{\circ}-360^{\circ}$ and is measured from the receding side of the galaxy; Therefore there could be differences of $180^{\circ}$ between both angles.}

The  2D representation of the bisymmetric and harmonic models together with the radial profile of the different velocity components are shown in Figure~\ref{fig:harmonic_ASASSN18oa}.
In the top panels, the bisymmetric model reproduce successfully the ``S''shape distortion observed in the inner regions, furthermore the residual map no longer exhibits the symmetrical patterns observed in Figure~\ref{fig:circular_model}. On the other hand, the non--circular velocities, $V_{2,r}$ and $V_{2,t}$, show maximum amplitudes of the same of order as the residuals in the circular rotation model, i.e., $\sim 40$~\kms, with smooth profiles as those observed in gas velocity fields \citep[e.g.,][]{Sellwood2010,Holmes2015}. 

The bottom panels of figure~\ref{fig:harmonic_ASASSN18oa} show the results of the harmonic decomposition. Again the 2D model seems to be a good representation of the LoS velocities, and no residual structures are observed in the central regions. 
The similar behavior of the $c_1$ harmonic with $V_t$ in the bisymmetric model, suggesting that the disk circular rotation is indistinguishable between both models and that the differences must reside only in the amplitudes of the non--circular components. 
 The inclusion of the $m^\prime = 3$ terms reproduce much better the features observed in the velocity field, however the residual map is, at first sight, indistinguishable from the bisymmetric model indicating that the harmonic decomposition produces results as good as the bisymmetric one. Nevertheless, the interpretation of the harmonic coefficients is not straightforward. Based on the epicyclic theory, \cite{Franx1994} and \cite{Wong2004} computed the expected behavior of the $m^\prime =1 $ and $m^\prime = 3$ harmonics for an elliptical potential (i.e., an $m=2$ perturbation).
 They found that for an elliptical potential, the slope between the $s_3$ and $s_1$ coefficients (hereafter  $ds_3/ds_1$),  is found to be negative, i.e. $ds_3/ds_1<0$, meanwhile for an axisymmetric radial flow $|ds_3/ds_1 | < 0.1$, which means that $s_3\ll s_1$. 
 The bottom rightmost panel of figure~\ref{fig:harmonic_ASASSN18oa} shows the $s_1/c_1$ vs. $s_3/c_1$ ratio together with the best fit line to the points. For the considered galaxy, the slope of the non--circular velocities is found in $-0.11$; therefore, the  $s_1$ and $s_3$ harmonics are compatible with being produced by a bar potential.
 
\section{Results}\label{res}

So far we have described our methodology for a single object. In the following we describe our results when applied to our sample of galaxies. Prior the kinematic modeling we removed spaxels with errors larger than $10$~\kms~to exclude low signal-to-noise data that could affect the final models.
For all objects we first create  circular models with initial values taken from the isophotal analysis. The 2D models for each individual object are shown in Appendix~\ref{sec:appendix_circ}.
Then we use the best fit values of the geometric parameters (namely $\phi^{\prime}$, $i$, $x_c$, $y_c$ and also $V_\mathrm{sys}$) as inputs for the bisymmetric and harmonic decomposition models.

It is expected that bar-like flows are not present across the entire disk, but only in the region influenced by the bar as observed from the residuals of Figure~\ref{Eq:circular}. Therefore, we fit non--circular motions up to the size of the photometric bar, plus a constant value that goes from $1\arcsec-2.5\arcsec$ to compensate for the binning on the stellar velocity maps.
The two-dimensional non--circular models for the stellar and gas velocity maps are shown in Figs.~\ref{fig:appendix1} and \ref{fig:appendix2} from  Appendix~\ref{sec:appendix_noncirc}. 

Table~\ref{Tab:results_amusing} shows \xs~results for the constant parameters namely \PAdisk, $i$ and $V_\mathrm{sys}$. The phometric estimations of the bar position angle and true bar length (i.e., Eq.~\ref{Eq:bar_size}) are shown in the third and fourth columns, respectively.
In general, we do not find large differences in the constant parameters, when adopting a bisymmetric or an harmonic decomposition model. { This means that the flat disk assumption is adequate for our objects, besides that both fits are numerically stable despite the underlying assumptions}. Exception to this might be IC~0004 where a constant ellipticity by \xs~does not seem to reproduce the kinematic orientation of the disk.
Table~\ref{Tab:results_amusing} also shows the BIC value of the bisymmetric model, but normalized to the BIC circular model. Thus, models with BIC $>1$ would not favor the bisymmetric model.  We note that only in one object (PGC\,055442) BIC definitely favors the circular model over the  bisymmetric one.

\subsection{Circular rotation models}

Figure~\ref{fig:appenidixcirc} shows the 2D circular models of the stellar and \ha~velocity maps, as well as the residual maps of the models. We notice that in the vast majority of cases \ha~is not detected in the inner disks; however, a careful analysis of the residual velocities in both stellar and \ha, large scale residuals are observed around the bar region in most cases, for instance in NGC\,692. Furthermore when \ha~is detected, symmetric residual structures with opposite velocities are observed in the inner regions; for instance IC\,2160, IC\,0004, PGC\,055442 and ESO\,18-18. Other objects although show high residual towards the center, they appear to be affected by other sources of non--circular motions, such as spiral arms as in NGC\,289 and NGC\,3464.  

On the other hand, the residuals of the stellar velocity maps show many examples of galaxies with ``S''-shaped iso-velocities together with symmetric residual structures around the bar. Such residuals are not compatible with an axisymmetric rotating disk and are most probably induced by the presence of the stellar bar. 
The fact we are clearly detecting oval distortions in most of our stellar velocity maps is likely a consequence of the high spatial resolution of our data, since similar IFS studies with coarser resolution do not observe such structures \citep[e.g.,][]{jkbb14}.

\subsection{Stellar non--circular motions}

Figure~\ref{fig:appendix1} shows the bisymmetric and harmonic decomposition models for the stellar velocity maps. Bisymmetric models tend to reproduce much better the observed velocity map than circular rotation models do. For most objects, $V_{2,r}$ and $V_{2,t}$ show smooth behaviors, with velocities decaying at large radius, a possible signal of the kinematic ending of the bar.

NGC\,289 is the only object where the bisymmetric model, for both gas and stars, does not seem to be a reasonable physical model given the erratic behavior of the circular and non--circular velocities.
{ As observed in figure~\ref{fig:appendix1} and table~\ref{Tab:results_amusing}, the bar in NGC\,289 is closely aligned with the major axis, being separated by  less than $ 10^{\circ}$; and it is well known that bar stream lines in the velocity field become less evident when the bar lies close the galaxy major-axis \citep[e.g.,][]{Albada1981}. Moreover, when \PAbar $ = 90^{\circ}$ or $ 0^{\circ}$, $V_{2r}$ and $V_{2t}$ from Eq.~\ref{Eq:bisymmetric} become degenerate with the disk circular rotation, making it impossible to separate the non-circular velocities from the circular rotation. As a result, $V_t$, $V_{2r}$ and $V_{2t}$ can lead to unphysical absurd values as is probably the case in NGC\,289; for instance, note that for $r<40\arcsec$ the three velocities shows the same uncommon radial profiles.
On the other hand the harmonic model does not depend on the bar phase; thus, the disk circular rotation related to $c_1$ seems to be offer a better representation of the rotation curve in this case. }

When comparing the BIC ratio from the bisymmetric to circular models we note that, in the case of the stellar maps,  these values are mostly close to 1, with the aforementioned exception. We expect BIC ratios lower than 1 would favour the bisymmetric model. While this is observed in our sample, in some cases the differences with the circular model are only marginal. It should be noted that because of the binning procedure during the SSP analysis, spaxels are correlated in the stellar maps, affecting the estimation of the BIC value. Therefore, we favour the bisymmetric models in those cases where we detect kinematic oval distortions, regardless of the BIC ratio.

\begin{figure}
\centering\includegraphics[width=1\columnwidth]{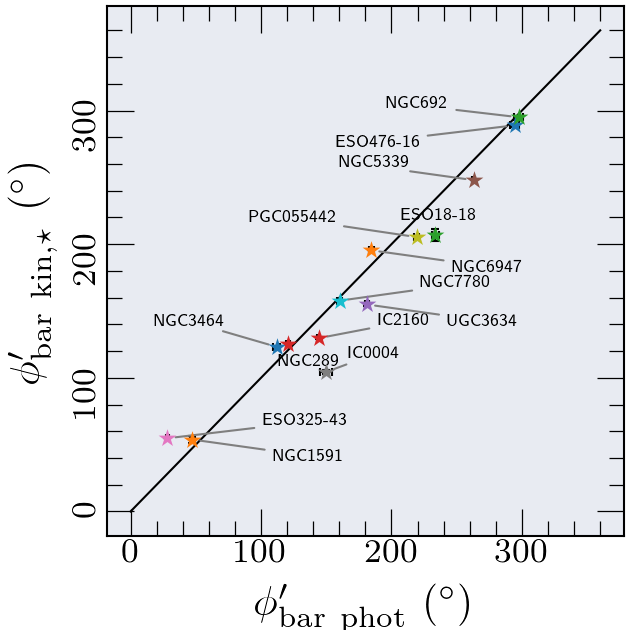}
\caption{Comparison between the sky projected photometric ($x$--axis) and kinematic bar position angle for the stellar velocity maps ($y$--axis). Photometric position angles were shifted $\pm180^{\circ}$ when required to match with the definition of the kinematic position angle. The continuous line represents the 1:1 relation. Each galaxy is observed with a different color. The error bars are of the size of the points.}
\label{fig:pa_kin_phot}
\end{figure}

The bar position angle is the main parameter in the bisymmetric model since it defines the axis where the non--circular motions are expected.
To visualize better this parameter, Figure~\ref{fig:pa_kin_phot} shows the photometric bar position angle vs. the sky projected kinematic bar position angle for the objects in the sample. It can be seen that there is a good agreement between both angles; in deed, the Pearson correlation coefficient for this relation is extremely high, being ${\rho}_{pearson} = 0.95$.

On the other side, the harmonic decomposition models show good representations of the LoS velocities.
Table~\ref{Tab:results_amusing} shows that at least in one velocity map (stellar or ionized), the slopes in the harmonic coefficients $ds_3/ds_1$ are negative, which is in favor of the hypothesis that the non--circular motions are induced by the bar. However there are objects where $|ds_3/ds_1|<0.1$, which means that radial flows could be contributing to the observed non--circular motions.

\subsection{Ionized gas non--circular motions}
As bar-like flows are expected along the bar, the lack of gas in the inner regions affects directly the estimation of \PAbar, and hence the amplitude of the non--circular motions. Because of the weak emission of \ha~in the inner disks in our objects, the bar region provides little or null information of the behavior of the non--circular motions. 
Only in 6 objects is observed plenty of gas in the inner disk to be able to perform reliable non-circular flow models. These objects are shown in Fig.~\ref{fig:appendix2}. 
However, the bars in NGC\,289 and NGC\,3464 are closely aligned parallel to the disk position angle, preventing the bar streams from being separated from the disk circular rotation.
{ Even so, the large residuals observed in fig.~\ref{fig:appenidixcirc} indicate the presence of strong non--circular motions, although a large fraction of them could be attributed to the prominent spiral arms in NGC\,289 \citep[e.g.,][]{Pence1984}.}

We notice however that oval distortions appear in the \ha~velocity maps when galaxies are gas rich; and such distortions are pronounced when the bar is elongated at an intermediate position angle from \PAdisk.
A remarkable example is observed in IC~2160 where the bar is oriented $\sim 40^{\circ}$ away the kinematic position angle. As consequence,  a strong twist in the kinematic major and minor axes is observed.

\begin{figure}
\centering\includegraphics[width = \columnwidth]{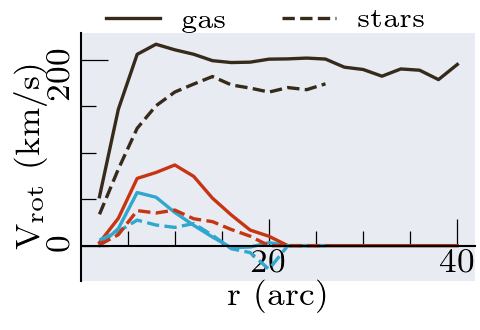}
\caption{Comparison between ionized gas and stellar bisymmetric models for IC~2160. Black, blue and red curves represent the $V_t$, $V_{2,r}$ and $V_{2,t}$ components for the \ha~ (continuous line) and stellar (dashed line) velocity maps.}
\label{fig:SN2009J}
\end{figure}

\subsection{Stellar and ionized gas bar-like flows}

As to whether the ionized gas and stars have similar amplitudes of the non--circular motions, 
the lack of ionized gas in the bar regions  precludes a direct comparison between them.
Figure~\ref{fig:SN2009J} shows bisymmetric models for  IC~2160, where oval distortions are observed simultaneously in the \ha~and stellar velocity maps. We notice similar behaviors between the circular and non--circular motions of gas and stars, albeit ionized gas have much larger amplitudes.
Such differences in amplitudes could be related to the dynamical origin of the tracers. 
Since stars are dynamically hot, they are more affected by the asymmetric drift (AD) than the gas. 
Although the AD is more important in pressure support systems, like dwarf galaxies \citep[e.g.,][]{Se-Heon2011}, in spiral galaxies the bulge and stellar bars, composed largely by old stellar populations, dominate the AD contribution \citep[e.g.,][]{Shetty2020}. 
Thus, the lower amplitudes observed in the stellar non--circular motions could be explained by the AD.

\begin{longrotatetable}
\begin{deluxetable*}{lccccccccccc}
\tablecaption{Main properties \label{Tab:results_amusing}}
\tablewidth{0pt}
\tablehead{
\colhead{AMUSING++} & \colhead{$\log~\mathrm{ M_\star/M_{\odot} }$} & \colhead{$\phi_\mathrm{bar,phot}^{\prime}$}  & \colhead{$r_\mathrm{bar}$} & \colhead{model} & \colhead{\PAdisk} & \colhead{$i$} & \colhead{$V_{sys}$} & \colhead{$\phi_\mathrm{bar}$} & \colhead{$\phi_\mathrm{bar,kin}^{\prime}$}   & \colhead{BIC} & \colhead{slope}\\
\colhead{} & \colhead{} & \colhead{$(^{\circ})$}& \colhead{(kpc)}  & \colhead{} & \colhead{($^{\circ}$)} & \colhead{($^{\circ}$)} & \colhead{(\kms)} & \colhead{($^{\circ}$)} & \colhead{($^{\circ}$)} & \colhead{}  & \colhead{$(ds_3/ds_1)$} 
}
\startdata
ESO476-16 stars & 10.8 & 114.4 $\pm$ 3.2 & 4.0 $\pm$ 0.6 & bis & 321.6 $\pm$ 0.8 & 53.6 $\pm$ 0.3 & 5960.9 $\pm$ 0.3 & 42.8 $\pm$ 0.5 & 288.9 $\pm$ 0.9 & 0.89 &  \nodata \\
ESO476-16 stars & \nodata & \nodata & \nodata & hrm & 321.0 $\pm$ 0.4 & 54.0 $\pm$ 0.7 & 5960.9 $\pm$ 0.6 &  \nodata &  \nodata &  \nodata & -0.04 \\
\hline
NGC6947 stars & 10.9 & 4.3 $\pm$ 1.9 & 6.6 $\pm$ 1.0 & bis & 238.8 $\pm$ 1 & 47.8 $\pm$ 1 & 5602.3 $\pm$ 0.2 & 35.4 $\pm$ 0.7 & 195.4 $\pm$ 1.2 & 0.93 &  \nodata \\
NGC6947 stars & \nodata & \nodata & \nodata & hrm & 235.9 $\pm$ 0.4 & 48.1 $\pm$ 0.7 & 5602.3 $\pm$ 0.6 &  \nodata &  \nodata &  \nodata & -0.25 \\
\hline
NGC692 stars & 11.2 & 117.6 $\pm$ 3.3 & 7.3 $\pm$ 1.1 & bis & 259.4 $\pm$ 0.2 & 27.0 $\pm$ 0.3 & 6343.4 $\pm$ 0.7 & 38.8 $\pm$ 0.3 & 295.0 $\pm$ 0.4 & 1.01 &  \nodata \\
NGC692 stars & \nodata & \nodata & \nodata & hrm & 259.0 $\pm$ 0.7 & 28.0 $\pm$ 0.5 & 6343.4 $\pm$ 0.7 &  \nodata &  \nodata &  \nodata & 0.13 \\
\hline
IC2160 stars & 10.6 & 144.2 $\pm$ 1.1 & 5.8 $\pm$ 0.5 & bis & 106.0 $\pm$ 0.0 & 47.6 $\pm$ 0.2 & 4729.6 $\pm$ 0.5 & 33.1 $\pm$ 1.1 & 129.7 $\pm$ 0.9 & 0.96 &  \nodata \\
IC2160 gas & \nodata & \nodata & \nodata & bis & 110.4 $\pm$ 1.2 & 47.2 $\pm$ 0.0 & 4730.5 $\pm$ 10.0 & 38.9 $\pm$ 0.7 & 139.2 $\pm$ 1.3 & 0.97 &  \nodata \\
IC2160 stars & \nodata & \nodata & \nodata & hrm & 108.9 $\pm$ 0.5 & 44.9 $\pm$ 0.7 & 4728.5 $\pm$ 0.6 &  \nodata &  \nodata &  \nodata & -0.07 \\
IC2160 gas & \nodata & \nodata & \nodata & hrm & 113.3 $\pm$ 0.4 & 43.4 $\pm$ 0.7 & 4727.9 $\pm$ 0.6 &  \nodata &  \nodata &  \nodata & 0.02 \\
\hline
UGC3634 stars & 11.0 & 1.1 $\pm$ 1.5 & 6.0 $\pm$ 0.9 & bis & 120.5 $\pm$ 0.2 & 39.1 $\pm$ 1.1 & 7876.0 $\pm$ 5.0 & 41.4 $\pm$ 0.9 & 154.9 $\pm$ 1.0 & 1.0 &  \nodata \\
UGC3634 stars & \nodata & \nodata & \nodata & hrm & 120.5 $\pm$ 0.4 & 48.9 $\pm$ 2.7 & 7876.5 $\pm$ 1.0 &  \nodata &  \nodata &  \nodata & -0.24 \\
\hline
NGC5339 stars & 10.4 & 83.1 $\pm$ 1.0 & 5.7 $\pm$ 0.8 & bis & 208.9 $\pm$ 0.1 & 35.5 $\pm$ 0.3 & 2736.2 $\pm$ 0.1 & 44.9 $\pm$ 0.1 & 247.9 $\pm$ 0.2 & 1.03 &  \nodata \\
NGC5339 stars & \nodata & \nodata & \nodata & hrm & 208.5 $\pm$ 1.2 & 33.6 $\pm$ 0.8 & 2736.2 $\pm$ 0.6 &  \nodata &  \nodata &  \nodata & -0.23 \\
\hline
ESO325-43 stars & 11.1 & 28.2 $\pm$ 1.3 & 4.3 $\pm$ 0.5 & bis & 109.2 $\pm$ 1 & 36.6 $\pm$ 1 & 10460.2 $\pm$ 0.9 & 29.9 $\pm$ 0.7 & 54.8 $\pm$ 1.3 & 0.99 &  \nodata \\
ESO325-43 stars & \nodata & \nodata & \nodata & hrm & 109.2 $\pm$ 0.4 & 36.8 $\pm$ 1.0 & 10460.4 $\pm$ 0.6 &  \nodata &  \nodata &  \nodata & -0.21 \\
\hline
IC0004 stars & 10.5 & 150.0 $\pm$ 4.5 & 4.7 $\pm$ 0.5 & bis & 178.3 $\pm$ 0.7 & 52.5 $\pm$ 1.4 & 4991.2 $\pm$ 1.1 & 9.9 $\pm$ 0.9 & 104.4 $\pm$ 0.9 & 1.0 &  \nodata \\
IC0004 gas & \nodata & \nodata & \nodata & bis & 178.4 $\pm$ 1 & 45.8 $\pm$ 1 & 5002.3 $\pm$ 16.8 & 61.3 $\pm$ 1.7 & 157.5 $\pm$ 2.3 & 0.97 &  \nodata \\
IC0004 stars & \nodata & \nodata & \nodata & hrm & 177.6 $\pm$ 0.6 & 45.8 $\pm$ 1.3 & 4996.1 $\pm$ 0.7 &  \nodata &  \nodata &  \nodata & 0.12 \\
IC0004 gas & \nodata & \nodata & \nodata & hrm & 176.3 $\pm$ 1.3 & 49.7 $\pm$ 1.3 & 4996.9 $\pm$ 0.8 &  \nodata &  \nodata &  \nodata & -0.37 \\
\hline
PGC055442 stars & 10.9 & 39.3 $\pm$ 1.9 & 2.5 $\pm$ 0.4 & bis & 193.8 $\pm$ 0.1 & 37.7 $\pm$ 0.6 & 7036.3 $\pm$ 0.9 & 104.4 $\pm$ 1.1 & 205.3 $\pm$ 1.4 & 1.06 &  \nodata \\
PGC055442 gas & \nodata & \nodata & \nodata & bis & 194.7 $\pm$ 1 & 46.7 $\pm$ 1 & 7031.8 $\pm$ 21.7 & 87.2 $\pm$ 3.1 & 280.7 $\pm$ 4.6 & 1.16 &  \nodata \\
PGC055442 stars & \nodata & \nodata & \nodata & hrm & 193.9 $\pm$ 0.5 & 38.9 $\pm$ 1.6 & 7036.5 $\pm$ 0.8 &  \nodata &  \nodata &  \nodata & 0.05 \\
PGC055442 gas & \nodata & \nodata & \nodata & hrm & 194.1 $\pm$ 1.1 & 38.3 $\pm$ 1.9 & 7028.8 $\pm$ 1.0 &  \nodata &  \nodata &  \nodata & -0.49 \\
\hline
NGC7780 stars & 10.5 & 160.2 $\pm$ 2.2 & 3.9 $\pm$ 0.6 & bis & 188.2 $\pm$ 1 & 63.3 $\pm$ 1 & 5185.7 $\pm$ 0.9 & 37.4 $\pm$ 0.0 & 157.7 $\pm$ 1.0 & 1.0 &  \nodata \\
NGC7780 stars & \nodata & \nodata & \nodata & hrm & 189.1 $\pm$ 0.5 & 61.2 $\pm$ 0.5 & 5185.0 $\pm$ 0.7 &  \nodata &  \nodata &  \nodata & -0.14 \\
\hline
NGC3464 stars & 10.6 & 112.0 $\pm$ 2.5 & 3.9 $\pm$ 0.6 & bis & 109.4 $\pm$ 1 & 45.6 $\pm$ 1 & 3740.0 $\pm$ 1.1 & 19.2 $\pm$ 0.7 & 123.2 $\pm$ 1.1 & 1.0 &  \nodata \\
NGC3464 gas & \nodata & \nodata & \nodata & bis & 109.4 $\pm$ 1 & 45.6 $\pm$ 1 & 3740.0 $\pm$ 1.1 & 19.2 $\pm$ 0.7 & 123.2 $\pm$ 1.1 & 1.03 &  \nodata \\
NGC3464 stars & \nodata & \nodata & \nodata & hrm & 109.6 $\pm$ 0.5 & 47.6 $\pm$ 0.6 & 3740.3 $\pm$ 0.6 &  \nodata &  \nodata &  \nodata & -0.26 \\
NGC3464 gas & \nodata & \nodata & \nodata & hrm & 109.2 $\pm$ 0.4 & 54.8 $\pm$ 0.7 & 3730.6 $\pm$ 0.5 &  \nodata &  \nodata &  \nodata & -0.49 \\
\hline
NGC1591 stars & 10.4 & 47.2 $\pm$ 2.0 & 2.6 $\pm$ 0.4 & bis & 24.0 $\pm$ 1.2 & 56.8 $\pm$ 0.4 & 4120.7 $\pm$ 1.9 & 46.3 $\pm$ 3.0 & 53.8 $\pm$ 2.9 & 0.68 &  \nodata \\
NGC1591 gas & \nodata & \nodata & \nodata & bis & 25.4 $\pm$ 0.0 & 50.8 $\pm$ 0.2 & 4129.6 $\pm$ 5.6 & 26.8 $\pm$ 1.0 & 43.1 $\pm$ 0.8 & 0.97 &  \nodata \\
NGC1591 stars & \nodata & \nodata & \nodata & hrm & 29.2 $\pm$ 0.6 & 60.5 $\pm$ 1.0 & 4117.1 $\pm$ 0.6 &  \nodata &  \nodata &  \nodata & -0.04 \\
NGC1591 gas & \nodata & \nodata & \nodata & hrm & 25.8 $\pm$ 0.5 & 47.4 $\pm$ 2.8 & 4128.0 $\pm$ 0.7 &  \nodata &  \nodata &  \nodata & 0.06 \\
\hline
ESO18-18 stars & 10.5 & 53.6 $\pm$ 2.5 & 5.7 $\pm$ 0.6 & bis & 281.4 $\pm$ 0.0 & 38.9 $\pm$ 0.9 & 5117.2 $\pm$ 4.8 & 12.0 $\pm$ 3.5 & 206.7 $\pm$ 2.8 & 0.53 &  \nodata \\
ESO18-18 stars & \nodata & \nodata & \nodata & hrm & 277.3 $\pm$ 0.6 & 18.1 $\pm$ 0.4 & 5113.2 $\pm$ 0.7 &  \nodata &  \nodata &  \nodata & 0.01 \\
\hline
NGC289 stars & 10.5 & 120.7 $\pm$ 1.7 & 1.9 $\pm$ 0.1 & bis & 128.3 $\pm$ 1 & 45.1 $\pm$ 1 & 1623.8 $\pm$ 0.0 & 85.9 $\pm$ 0.3 & 125.4 $\pm$ 1.1 & 0.93 &  \nodata \\
NGC289 gas & \nodata & \nodata & \nodata & bis & 125.0 $\pm$ 0.1 & 43.6 $\pm$ 0.1 & 1629.3 $\pm$ 14.3 & 61.9 $\pm$ 6.1 & 178.7 $\pm$ 7.0 & 0.98 &  \nodata \\
NGC289 stars & \nodata & \nodata & \nodata & hrm & 130.0 $\pm$ 1.0 & 37.7 $\pm$ 0.7 & 1624.0 $\pm$ 0.6 &  \nodata &  \nodata &  \nodata & -0.24 \\
NGC289 gas & \nodata & \nodata & \nodata & hrm & 126.2 $\pm$ 0.9 & 43.8 $\pm$ 0.6 & 1624.7 $\pm$ 0.7 &  \nodata &  \nodata &  \nodata & 0.1 \\
\hline
\hline
\enddata
\tablecomments{In some cases $V_{2t}$ and $V_{2r}$ have both negative values. When this occurs, $\phi_\mathrm{bar}$ no longer represents the bar major axis, but the bar minor axis in the galaxy plane. This is taken into account when we compute position angle of the bar in the sky plane by shifting  $\phi_\mathrm{bar}$ by $90^{\circ}$. This guarantees that $\phi_\mathrm{bar,kin}^{\prime}$ is measuring the bar major axis in column 10. The reported photometric bar position angle takes into account the possible shift in $180^{\circ}$ due to the different definitions between kinematic and photometric position angle. The BIC value is normalized to the BIC value from the circular rotation model. }
\end{deluxetable*}
\end{longrotatetable}

\section{Discussion}

\subsection{Bar-like or radial flows?}

\cite{Wong2004} categorized different source of non--circular motions based on the slope of the $s_1$ and $s_3$ coefficients. For radial flow dominated systems, they found that $|ds_3/ds_1|<0.1$. Table~\ref{Tab:results_amusing} show objects that fall in this category; however, for describing barred galaxies the bisymmetric model is usually preferred over radial flows. In this model the perturbation is driven along a fixed axis (\PAbar) and the non--circular velocities correspond to those of elliptical orbits.
In the epicyclic theory, radial stellar motions are not expected to contribute significantly to the non--circular motions \citep[e.g.,][]{Sellwood2002}. { In addition, the large speeds on the observed non--circular velocities could not persist for a very long time without rearranging the mass distribution of the disk \citep[e.g.,][]{Wong2004, Spekkens2007}; therefore, invoking pure axisymmetric radial flows (inflows/outflows) brings considerable consequences in the stability of stellar disks. }

In addition, numerical simulations also show that inflows in bars are expected to be $<5$~\kms \citep[e.g.,][]{atha92}, but not of tens of kilometers per second as observed in most objects.
%
Recent studies however, observe radial motions in spirals of the order of 10-30 \kms \cite[e.g.,][]{DiTeodoro2021}. Such large speeds of gas inflow/outflow would invoke necessary efficient mechanisms of gas depletion through star formation processes or galactic scale winds to remove gas out of the galaxy. The former scenario, although it has been widely observed \citep[e.g.,][]{clc2018,AMUSING++}, the link of axisymmetric radial flows with out of plane outflows is a subject that has not yet been addressed. For all the above reasons pure axisymmetric radial flows are infrequently considered for describing the non--circular flows in barred galaxies.

\subsection{Deficit of ionized gas in bars}

The absence of ionized gas in the central regions observed in halve of our sample, suggests it could be related with the presence of the bar itself. In fact, in most barred galaxies there is observed a lack of gas in the inner disk \citep[e.g.,][]{Erroz-Ferrer2015, Fraser2020}.
However there is no clear explanations in the literature for why bars tend to not exhibit ionized gas. Two scenarios are related with the absence of molecular gas. If there is absence of molecular gas along the bars, no new stars are formed. Closely related might be the presence of, indeed, radial flows that could have transported cold gas towards the center. In such case one should expect a higher star formation rate (SFR) towards the center, which is not yet observed \citep[e.g.,][]{Erroz-Ferrer2015}. It is therefore important to trace the cold gas abundance in bars to help to distinguish between both scenarios.

\subsection{Connection between non--circular motions and galaxy properties}

\begin{figure}
\centering\includegraphics[width=1\columnwidth]{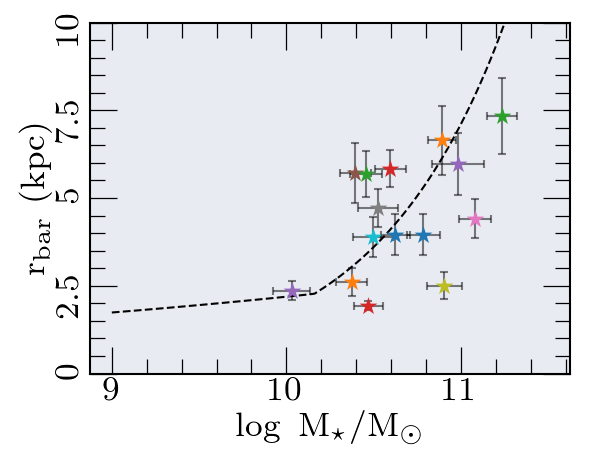}
\caption{Stellar mass versus the deprojected bar semi--major axis for our sample. The overplotted discontinuous line shows the relation observed by \cite{Erwin2019} with more than $1000$ barred galaxies from the $\mathrm{S^4G}$ survey \citep[e.g.,][]{Diaz-Garcia2016}.}
\label{fig:M_vs_bar}
\end{figure}

As noted in Figure~\ref{fig:appendix1} and \ref{fig:appendix2}, the amplitudes of the non--circular velocities $V_{2,r}$ and $V_{2,t}$ vary widely from galaxy to galaxy. Thus, we ask 
whether the amplitude of the non--circular motions depend on some galaxy properties or they are local processes. For instance recent studies show a two-slope relation between the stellar mass and the bar length \cite{Erwin2019}. Since bars are composed largely of old stellar populations, the bar must contribute to a large fraction of the total stellar mass, as observed in some barred galaxies \citep[e.g,][]{patri11}. Thus the relation between $r_{bar}$ and the host galaxy stellar mass, should be expected as well with the bar-mass.

\begin{figure*}
\centering\includegraphics[]{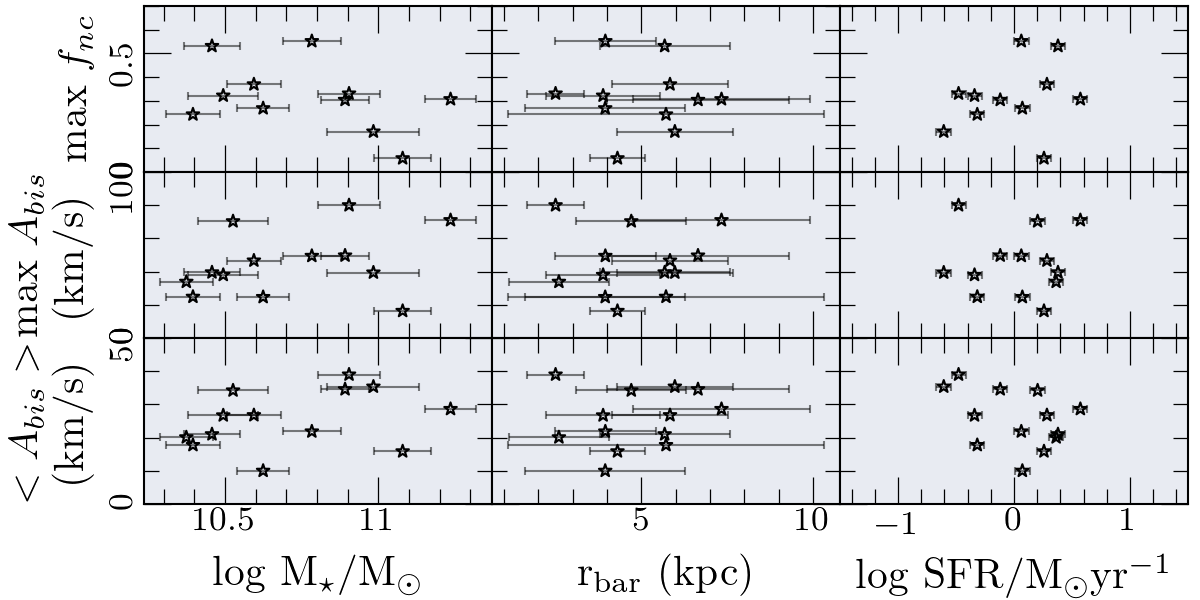}
\caption{Stellar mass, bar length and global SFR versus some descriptions of the non--circular motions. The top set of figures show the maximum value of the non--circular to circular motions ratio (Eq.~\ref{fbis}). Middle and bottom figures show  maximum and average value of the amplitude of the non--circular motions (Eq.~\ref{Abis})}
\label{fig:correlation}
\end{figure*}
{
\centering
\begin{table*}
\begin{tabular}{|p{1.5cm}|p{1.6cm}|p{1.6cm}|p{1.6cm}|p{1.6cm}|p{1.6cm}|p{1.6cm}|}\hline
& \multicolumn{3}{c|}{$\rho_\mathrm{pearson}$} & \multicolumn{3}{c|}{$\rho_\mathrm{spearman}$}\\
& $\log~\mathrm{ M_{\star}/M_{\odot} }$  & $r_\mathrm{bar}$ & $\log$ SFR & $\log~\mathrm{ M_{\star}/M_{\odot} }$  & $r_\mathrm{bar}$ & $\log$ SFR  \\\hline
$f_\mathrm{nc}$             & -0.42 (0.13) & -0.25 (0.39) & 0.24 (0.44)  & -0.58 (0.03) & -0.31 (0.28) & 0.36 (0.22) \\
$\max~A_\mathrm{bis}$       & 0.37 (0.19)  & 0.11 (0.71)  & 0.01 (0.98)  & 0.39 (0.17) & 0.3 (0.3) & 0.01 (0.99)\\
$<A_\mathrm{bis}>$          & 0.37 (0.19)  & 0.37 (0.19)  & -0.39 (0.18) & 0.40 (0.16)& 0.45 (0.11) & -0.38 (0.2)\\
\hline
\end{tabular}
\caption{Pearson and Spearem correlation coefficients between non--circular motions and galaxy global properties. The first column from top to bottom correspond to the fraction of non--circular over the local circular rotation, the maximum amplitude of the non--circular motions and the average velocity of non--circular motions. These parameters are compared with the galaxy stellar mass, the size of the bar and the SFR integrated across the MUSE FoV. For each set of parameters we have computed the Pearson correlation coefficient and the Spearman rank correlation with p--values shown in parenthesis.  }
\label{tab:correlations}
\end{table*}
}
Figure~\ref{fig:M_vs_bar} shows the deprojected stellar bar length (Eq.~\ref{Eq:bar_size}) versus  the  stellar mass obtained from the SSP analysis.
We notice that our objects fall around the steeper slope of this relation where galaxies with larger bars are located. Indeed most of our objects have deprojected bar sizes larger than $2.5$~kpc. 

We investigate if the stellar mass, size of the bar and the star formation rate (SFR) are coupled with the amplitude of the non--circular motions.
%
%
Instead of adopting the average value of the circular rotation residuals, we characterize the non--circular motions with the amplitude of the bisymmetric components $V_{2,r}$ and $V_{2,t}$. As these terms  are function of radius we define their amplitud, $A_\mathrm{bis}$, as the quadratic sum of both components:
\begin{equation}
 \label{Abis}
 A_{\mathrm{bis}}(r) = \sqrt{V_{2,t}^2(r) + V_{2,r}^2(r)}
\end{equation}
and the fraction of non--circular over circular motions as:
\begin{equation}
 \label{fbis}
 f_{\mathrm{nc}}(r) = A_{\mathrm{bis}}(r)/V_t(r)~~\mathrm{for}~r \leq r_{bar} 
\end{equation}
where $V_{t}$ is the tangential velocity component of the bisymmetric model, which gives a better description of the circular rotation than if we consider the pure circular model. Note again that these two parameters depend on the galactocentric distance. Thus in the second expression we are comparing at each distance the non--circular motions with the local circular velocity.  

Figure~\ref{fig:correlation} shows $A_{bis}$ and $f_{nc}$ computed for the stellar velocity maps against the stellar mass and the bar lengths of our objects. In this figure we also include the integrated SFR derived with \ha, with the dust attenuation correction using the \cite{Cardelli1989} extinction law and case B of recombination \citep[e.g.,][]{Osterbrock}, and use the distances reported in the AMUSING++ paper \citep[e.g.,][]{AMUSING++}.  
The first thing to notice is the strength of non--circular motions ranges between 10--50\% of the local circular rotation on average, which is similar to what other studies have found  \citep[e.g.,][]{Bettoni1997}. 

In figure~\ref{fig:correlation} we tried to find possible correlations between kinematic properties and global properties, while Table~\ref{tab:correlations} shows the Pearson and Spearman coefficients for those relations. We find however only weak correlations between these parameters. Even more, the large p-values show that such correlations are not statistically significant. These results are in line with recent studies \citep[e.g.,][]{Erroz-Ferrer2015}, where no obvious relation is observed with the residuals of circular rotation and morphological properties of bars.
However, we are aware of the low statistic provided by this sample. 

\section{Conclusions}

In this study we analyzed the incidence of non--circular flows in the stellar and \ha~ velocity field of a sample of non-interacting barred galaxies observed with the MUSE spectrograph. The exquisite resolution of the data allowed us to detect oval distortions in the stellar velocity maps most probably associated to the presence of the bar. 
Such perturbations are recognized in the residual maps of pure circular rotation models as symmetric structures with blueshifted and redshifted velocities, clearly evidencing the presence of a non-axisymmetric potential affecting the expected disk rotation.
Despite not observing ionized gas in the inner disk in most of the objects, the bar leave imprints in the circular residual velocities in regions close to the bar. When ionized gas is observed along the bar, the oval distortion is also revealed in the \ha~velocity field. This evidence that both stars and gas are affected by the bar potential.

We characterize the kinematics of these galaxies with models that include non--circular motions, in particular the bisymmetric model and a harmonic decomposition for a $m=2$ potential perturbation.
We use the slope of the harmonic coefficients, the $ds_3/ds_1$ ratio, to determine the source of the observed non--circular motions.

Based on this parameter we find that the harmonic decomposition on the LoS velocity is compatible in most cases with being produced by an elongated potential. However, some radial flows could be present in some objects since $|ds_3/ds_1| < 0.1$ is observed in 4/14 cases. In these cases, we attribute to the dominance of other sources of non--circular motions such as spiral arms.

We find that the bisymmetric model produces successful fittings in the velocity maps within the bar region. We also find that the position angle of the oval distortion  correlates with the photometric position angle of the bar. This supports the scenario of a stellar bar potential inducing deviations of the circular rotation in the observed velocity fields.
We find that the lack of ionized gas along the bars restricts the robustness of a modeling of the non--circular motions, in which case the stellar velocity could be considered for studying bar-like flows when no gas is detected.
However we notice that not only the stellar circular rotation is affected by the asymmetric drift, but also the non--circular motions since gas velocities appear rotate faster.

We find the average amplitude of the non--circular motions in our sample is $\sim 30$~\kms~ for stars and ionized gas, while the strength of the non--circular motions reaches values of up to 50\% of the local circular velocity, although this fraction varies among galaxies. 

When trying to relate the non--circular motions with galaxy properties we do not find any clear correlation with the stellar mass, the bar size or the global SFR. These results point that bar-flows are rather local process; however, we stress that larger statistical samples of barred galaxies with high spatial resolution are required to reveal possible correlations between kinematic properties of bars and global properties of galaxies.

We would like to thank the anonymous referee for their comments and suggestions that contribute to improve the quality of this paper.

L.~L and C.~L.~C. thank the supports by the Academia Sinica under Career Development Award CDA-107-M03 and the Ministry of Science \& Technology of Taiwan under the grant MOST 108-2628-M-001-001-MY3.
ICG acknowledges support from DGAPA-UNAM grant IN113320.
L.G. acknowledges financial support from the Spanish Ministry of Science, Innovation and Universities (MICIU) under the 2019 Ram\'on y Cajal program RYC2019-027683 and from the Spanish MICIU project HOSTFLOWS PID2020-115253GA-I00.

\appendix
\section{Circular rotation models}
\label{sec:appendix_circ}

\begin{figure}
\figurenum{9}
\plotone{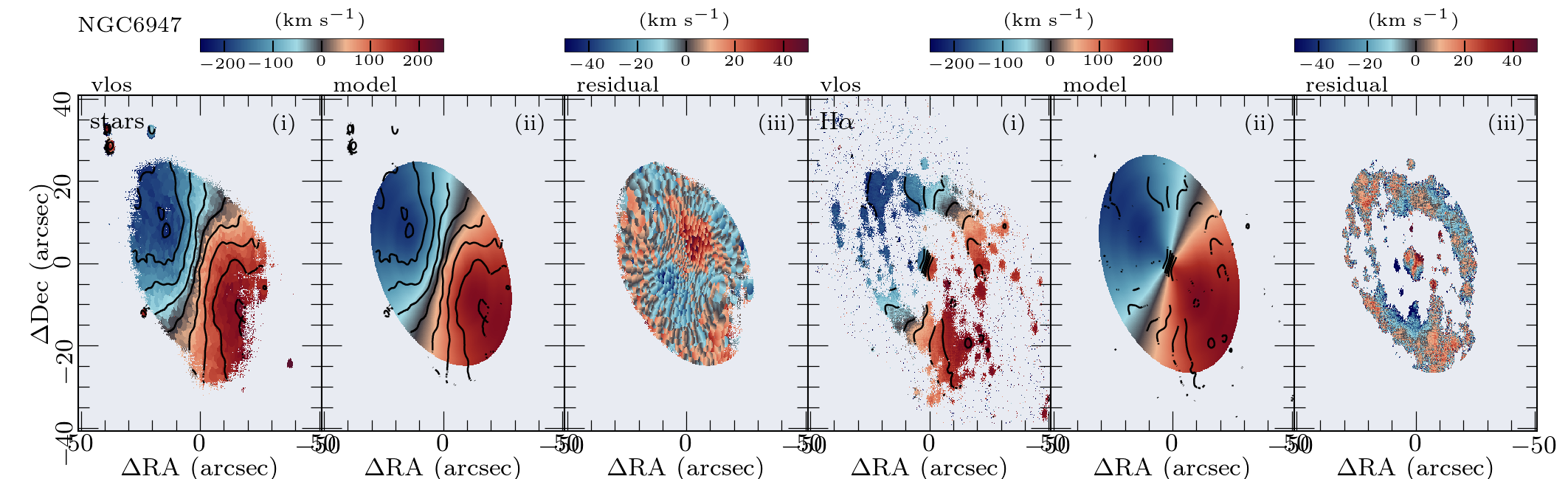}
\caption{NGC\,6947 galaxy. Each block of figures refers to the results of the kinematic modelling in each galaxy. The three leftmost (rightmost) figures show the observed velocity field; the best circular rotation model obtained by \xs; and the residual map for the stellar (\ha) velocity map. Overlayed contours are spaced each $\pm50$~\kms. }
\label{fig:appenidixcirc}
\end{figure}

Following are shown the maps of the circular and non--circular kinematic models for the sample of galaxies shown in Table~\ref{Tab:results_amusing}. 
In each figure, only one object is shown as an example, the remaining figures are available in the online journal. Figure~\ref{fig:appenidixcirc} shows the circular rotation models derived by \xs~for the ionized-gas and stellar velocity maps of each object. The three figures on the left (right) correspond to the observed velocity map, best kinematic model and the residual map for the ionized (stellar) velocity maps. 

%
%

\figsetstart
\figsetnum{9}
\figsettitle{Circular rotation models for the \ha~and stellar velocity maps.}

\figsetgrpstart
\figsetgrpnum{9.1}
\figsetgrptitle{ESO476-16}
\figsetplot{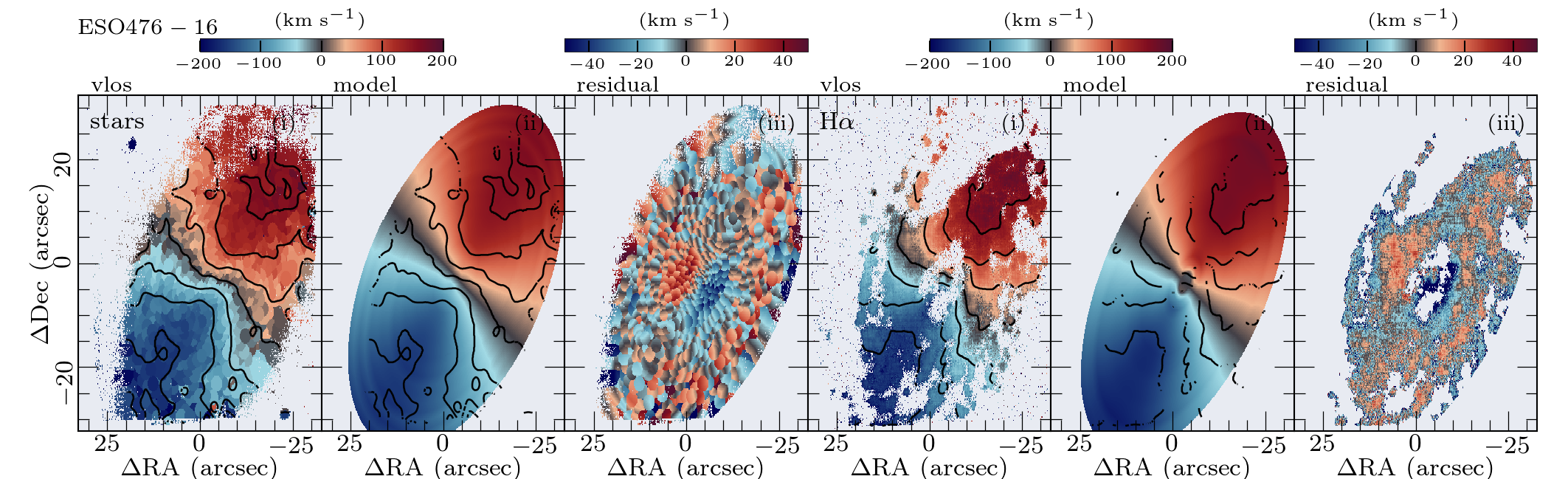}
\figsetgrpnote{Each block of figures refers to the results of the kinematic modelling in each galaxy. The three leftmost (rightmost) figures show the observed velocity field; the best circular rotation model obtained by \xs; and the residual map for the stellar (\ha) velocity map. Overlayed contours are spaced each $\pm50$~\kms. }
\figsetgrpend

\figsetgrpstart
\figsetgrpnum{9.2}
\figsetgrptitle{NGC6947}
\figsetplot{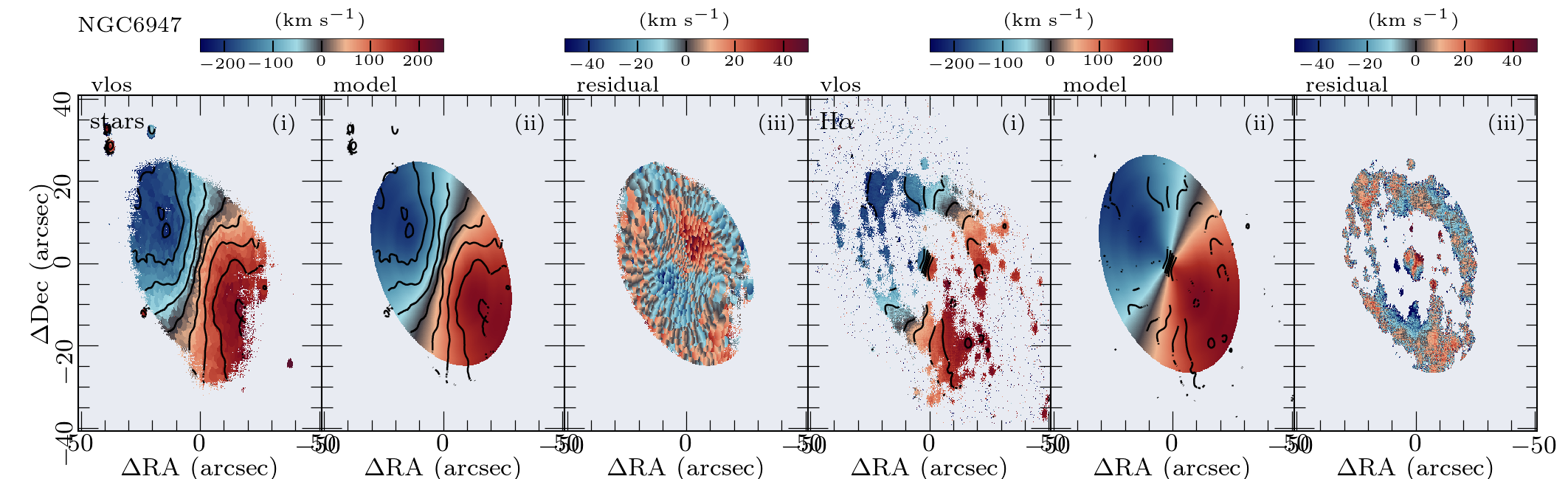}
\figsetgrpnote{Each block of figures refers to the results of the kinematic modelling in each galaxy. The three leftmost (rightmost) figures show the observed velocity field; the best circular rotation model obtained by \xs; and the residual map for the stellar (\ha) velocity map. Overlayed contours are spaced each $\pm50$~\kms. }
\figsetgrpend

\figsetgrpstart
\figsetgrpnum{9.3}
\figsetgrptitle{NGC692}
\figsetplot{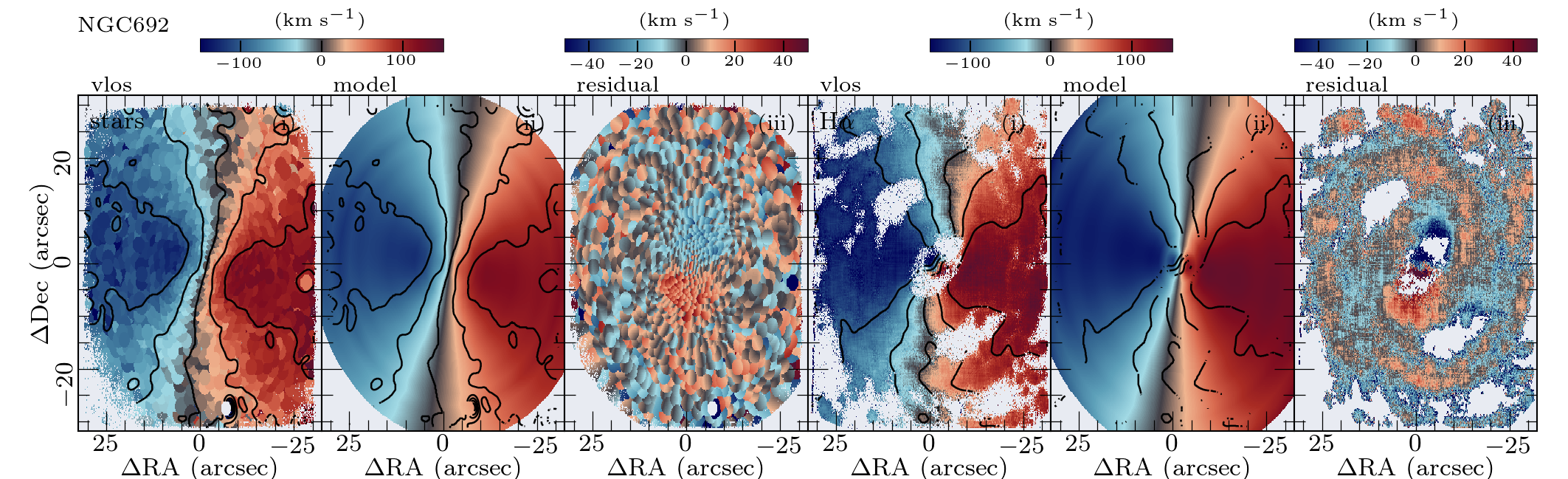}
\figsetgrpnote{Each block of figures refers to the results of the kinematic modelling in each galaxy. The three leftmost (rightmost) figures show the observed velocity field; the best circular rotation model obtained by \xs; and the residual map for the stellar (\ha) velocity map. Overlayed contours are spaced each $\pm50$~\kms. }
\figsetgrpend

\figsetgrpstart
\figsetgrpnum{9.4}
\figsetgrptitle{IC2160}
\figsetplot{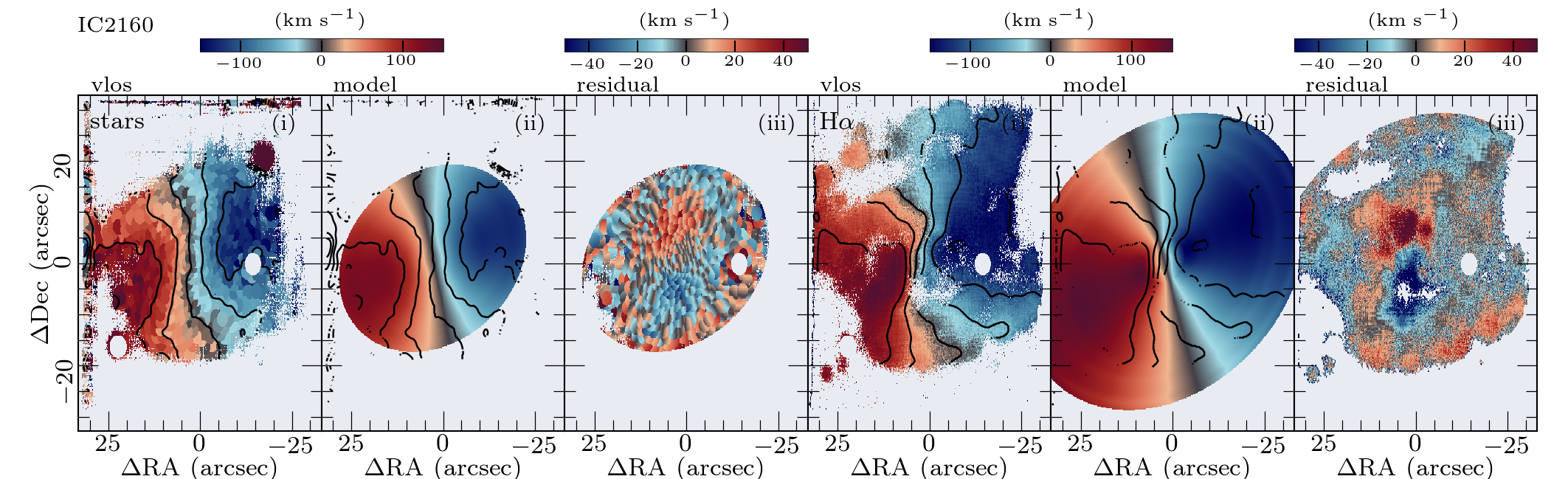}
\figsetgrpnote{Each block of figures refers to the results of the kinematic modelling in each galaxy. The three leftmost (rightmost) figures show the observed velocity field; the best circular rotation model obtained by \xs; and the residual map for the stellar (\ha) velocity map. Overlayed contours are spaced each $\pm50$~\kms. }
\figsetgrpend

\figsetgrpstart
\figsetgrpnum{9.5}
\figsetgrptitle{UGC3634}
\figsetplot{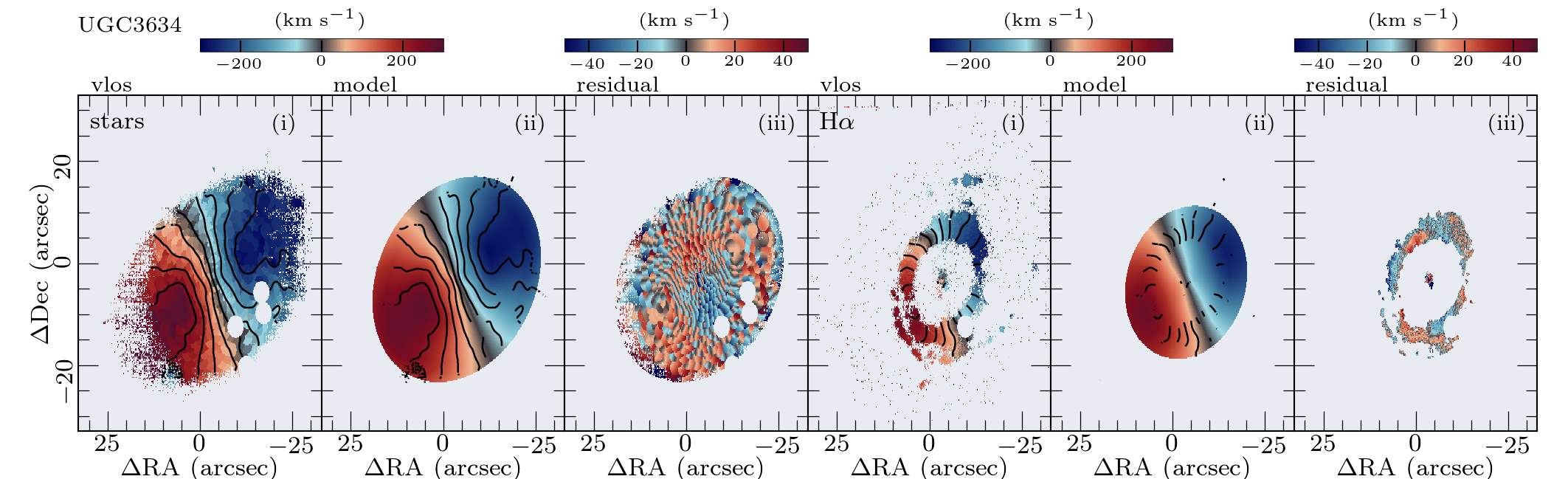}
\figsetgrpnote{Each block of figures refers to the results of the kinematic modelling in each galaxy. The three leftmost (rightmost) figures show the observed velocity field; the best circular rotation model obtained by \xs; and the residual map for the stellar (\ha) velocity map. Overlayed contours are spaced each $\pm50$~\kms. }
\figsetgrpend

\figsetgrpstart
\figsetgrpnum{9.6}
\figsetgrptitle{NGC5339}
\figsetplot{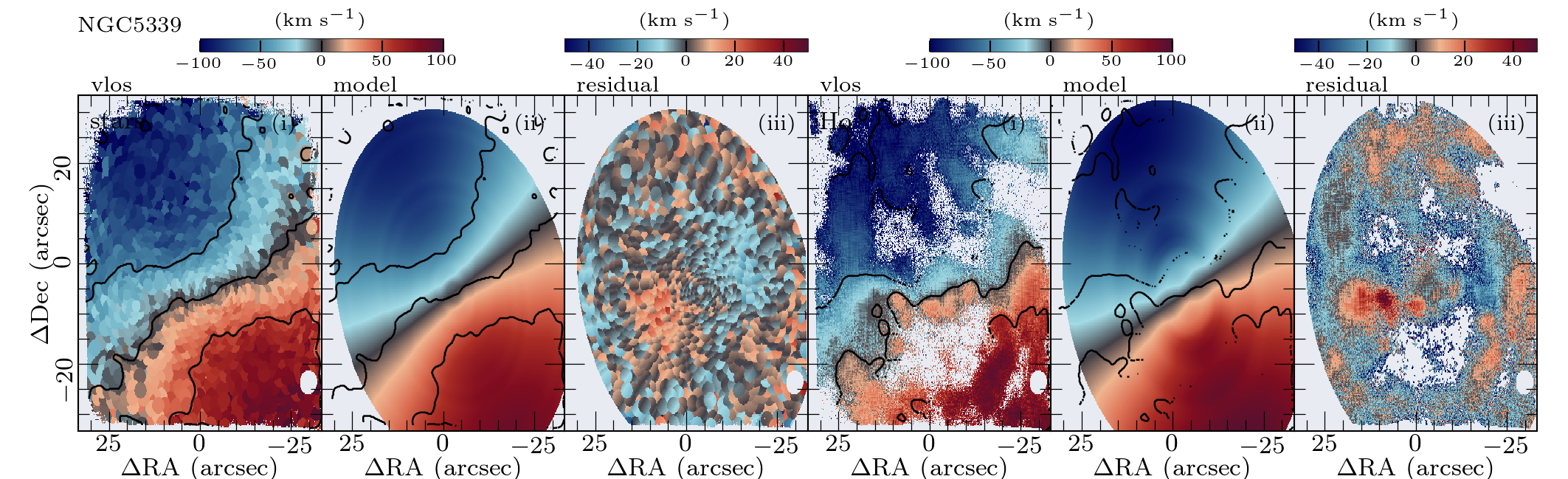}
\figsetgrpnote{Each block of figures refers to the results of the kinematic modelling in each galaxy. The three leftmost (rightmost) figures show the observed velocity field; the best circular rotation model obtained by \xs; and the residual map for the stellar (\ha) velocity map. Overlayed contours are spaced each $\pm50$~\kms. }
\figsetgrpend

\figsetgrpstart
\figsetgrpnum{9.7}
\figsetgrptitle{ESO325-43}
\figsetplot{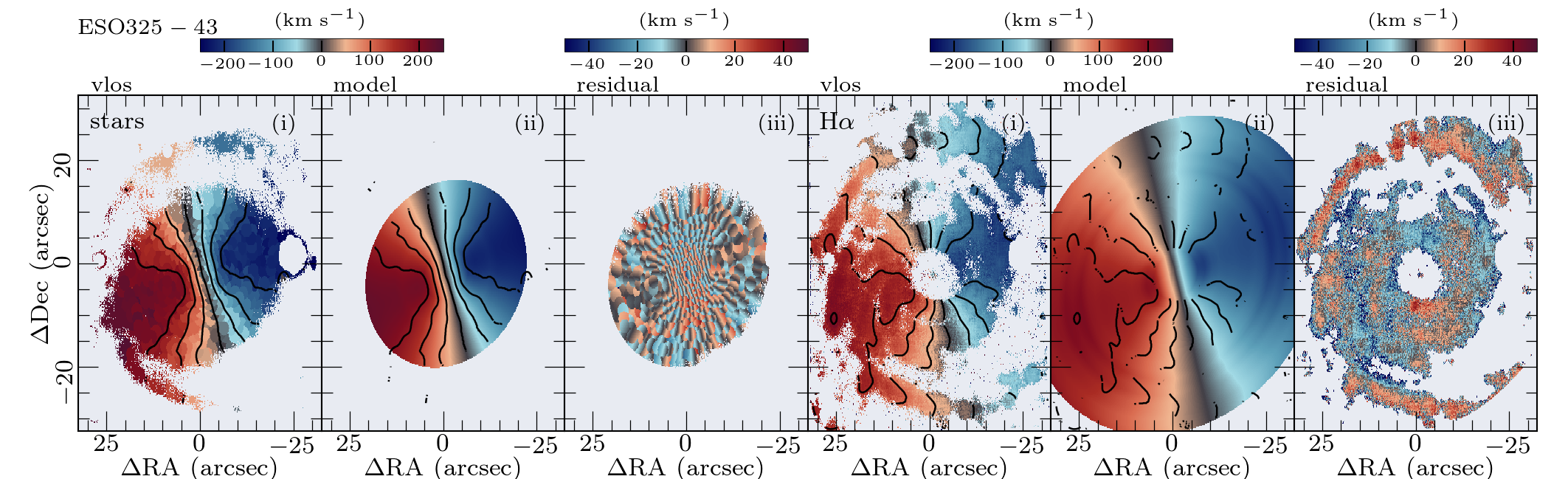}
\figsetgrpnote{Each block of figures refers to the results of the kinematic modelling in each galaxy. The three leftmost (rightmost) figures show the observed velocity field; the best circular rotation model obtained by \xs; and the residual map for the stellar (\ha) velocity map. Overlayed contours are spaced each $\pm50$~\kms. }
\figsetgrpend

\figsetgrpstart
\figsetgrpnum{9.8}
\figsetgrptitle{IC0004}
\figsetplot{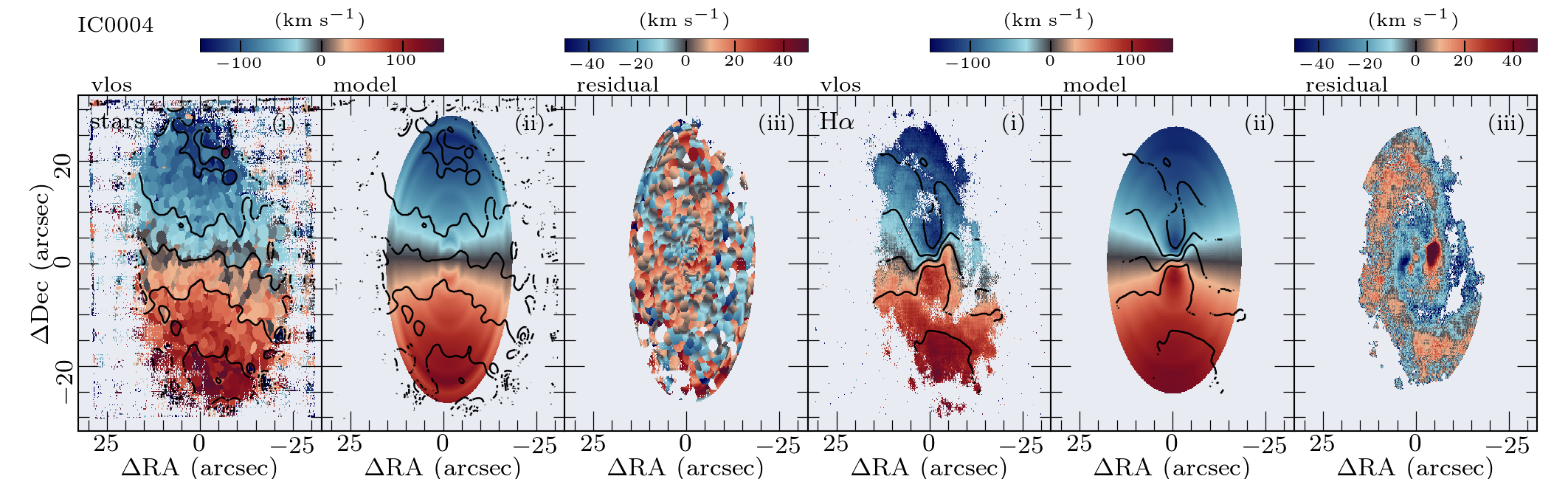}
\figsetgrpnote{Each block of figures refers to the results of the kinematic modelling in each galaxy. The three leftmost (rightmost) figures show the observed velocity field; the best circular rotation model obtained by \xs; and the residual map for the stellar (\ha) velocity map. Overlayed contours are spaced each $\pm50$~\kms. }
\figsetgrpend

\figsetgrpstart
\figsetgrpnum{9.9}
\figsetgrptitle{PGC055442}
\figsetplot{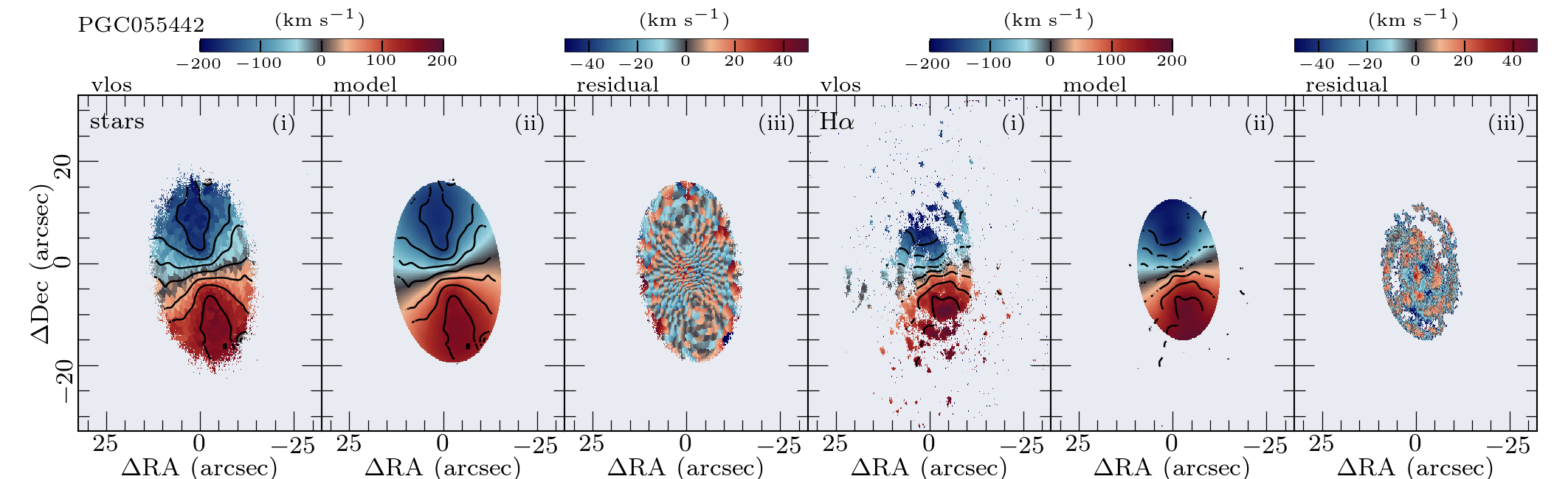}
\figsetgrpnote{Each block of figures refers to the results of the kinematic modelling in each galaxy. The three leftmost (rightmost) figures show the observed velocity field; the best circular rotation model obtained by \xs; and the residual map for the stellar (\ha) velocity map. Overlayed contours are spaced each $\pm50$~\kms. }
\figsetgrpend

\figsetgrpstart
\figsetgrpnum{9.10}
\figsetgrptitle{NGC7780}
\figsetplot{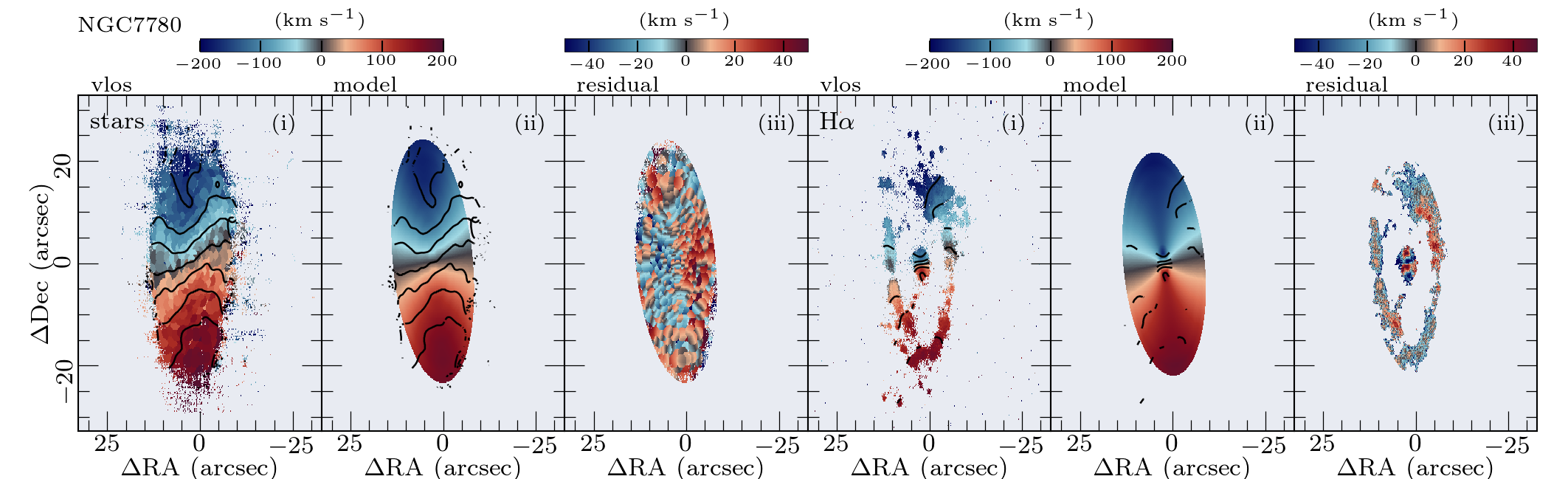}
\figsetgrpnote{Each block of figures refers to the results of the kinematic modelling in each galaxy. The three leftmost (rightmost) figures show the observed velocity field; the best circular rotation model obtained by \xs; and the residual map for the stellar (\ha) velocity map. Overlayed contours are spaced each $\pm50$~\kms. }
\figsetgrpend

\figsetgrpstart
\figsetgrpnum{9.11}
\figsetgrptitle{NGC3464}
\figsetplot{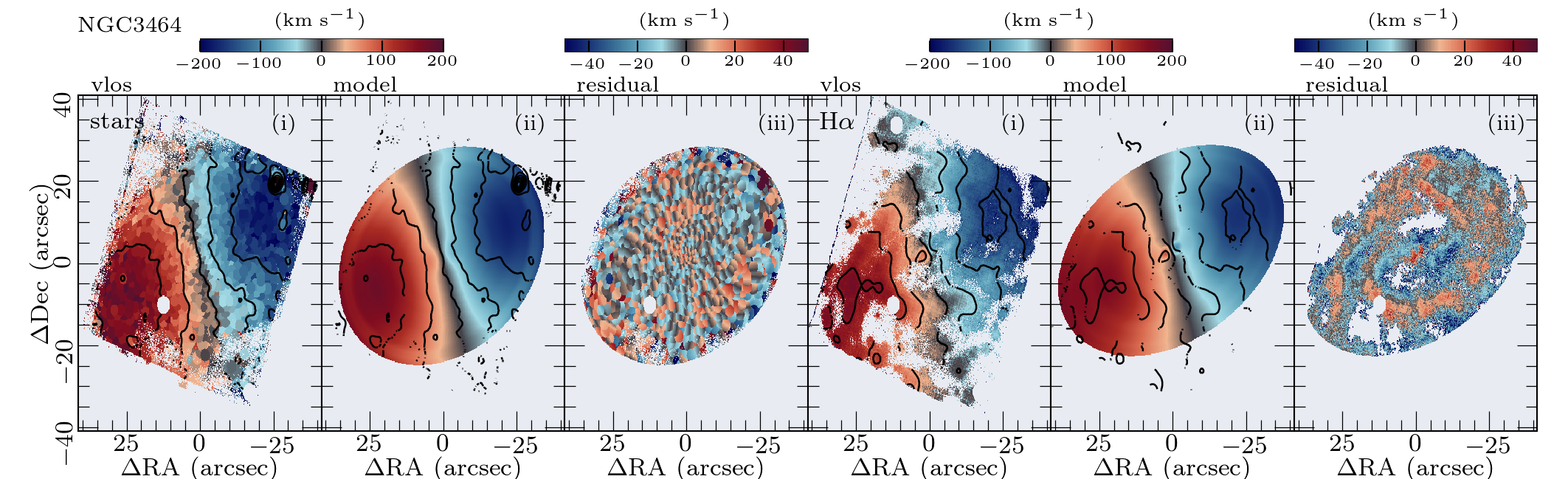}
\figsetgrpnote{Each block of figures refers to the results of the kinematic modelling in each galaxy. The three leftmost (rightmost) figures show the observed velocity field; the best circular rotation model obtained by \xs; and the residual map for the stellar (\ha) velocity map. Overlayed contours are spaced each $\pm50$~\kms. }
\figsetgrpend

\figsetgrpstart
\figsetgrpnum{9.12}
\figsetgrptitle{NGC1591}
\figsetplot{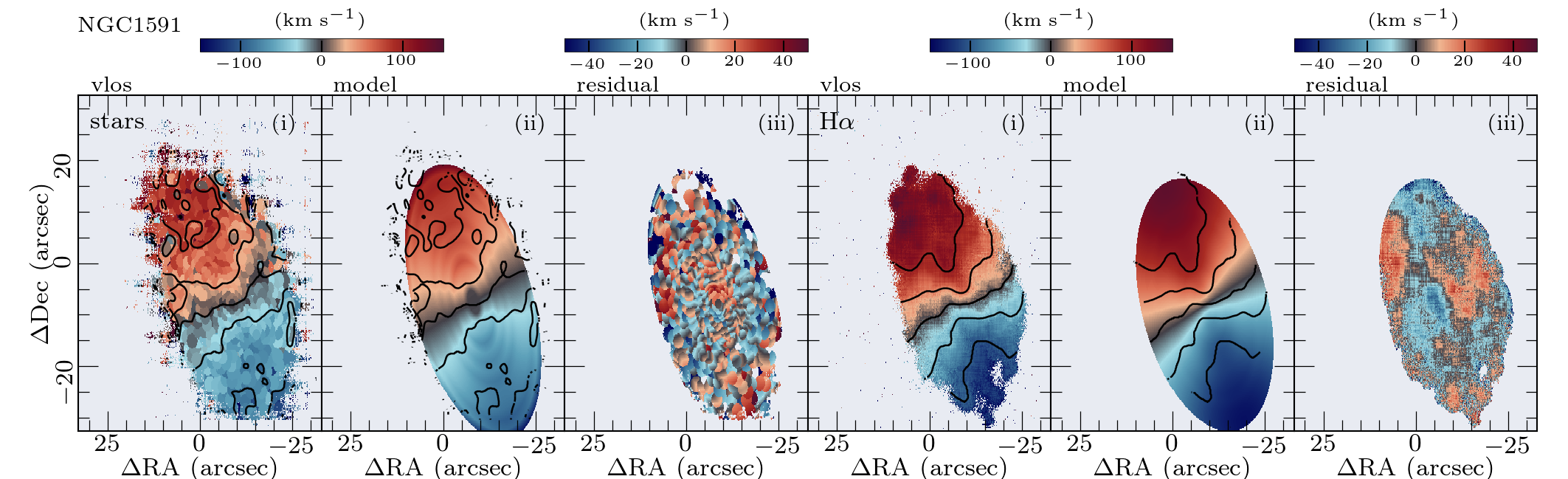}
\figsetgrpnote{Each block of figures refers to the results of the kinematic modelling in each galaxy. The three leftmost (rightmost) figures show the observed velocity field; the best circular rotation model obtained by \xs; and the residual map for the stellar (\ha) velocity map. Overlayed contours are spaced each $\pm50$~\kms. }
\figsetgrpend

\figsetgrpstart
\figsetgrpnum{9.13}
\figsetgrptitle{ESO18-18}
\figsetplot{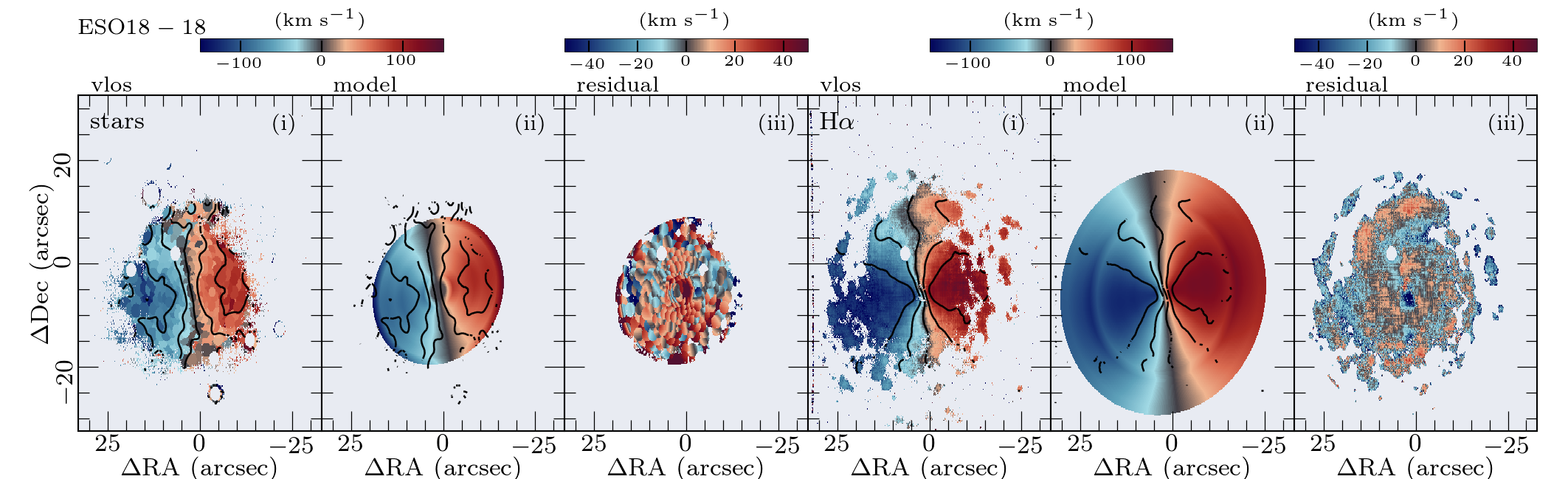}
\figsetgrpnote{Each block of figures refers to the results of the kinematic modelling in each galaxy. The three leftmost (rightmost) figures show the observed velocity field; the best circular rotation model obtained by \xs; and the residual map for the stellar (\ha) velocity map. Overlayed contours are spaced each $\pm50$~\kms. }
\figsetgrpend

\figsetgrpstart
\figsetgrpnum{9.14}
\figsetgrptitle{NGC289}
\figsetplot{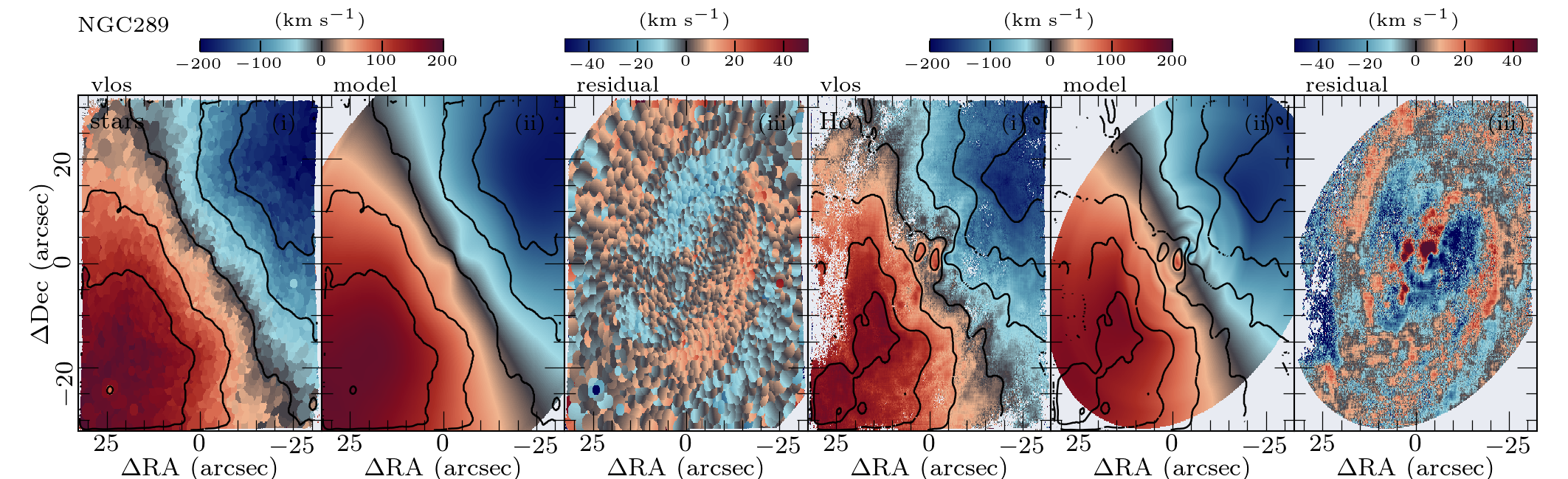}
\figsetgrpnote{Each block of figures refers to the results of the kinematic modelling in each galaxy. The three leftmost (rightmost) figures show the observed velocity field; the best circular rotation model obtained by \xs; and the residual map for the stellar (\ha) velocity map. Overlayed contours are spaced each $\pm50$~\kms. }
\figsetgrpend

\figsetend

\section{Noncircular rotation models}
\label{sec:appendix_noncirc}
\begin{figure}
\figurenum{10}
\plotone{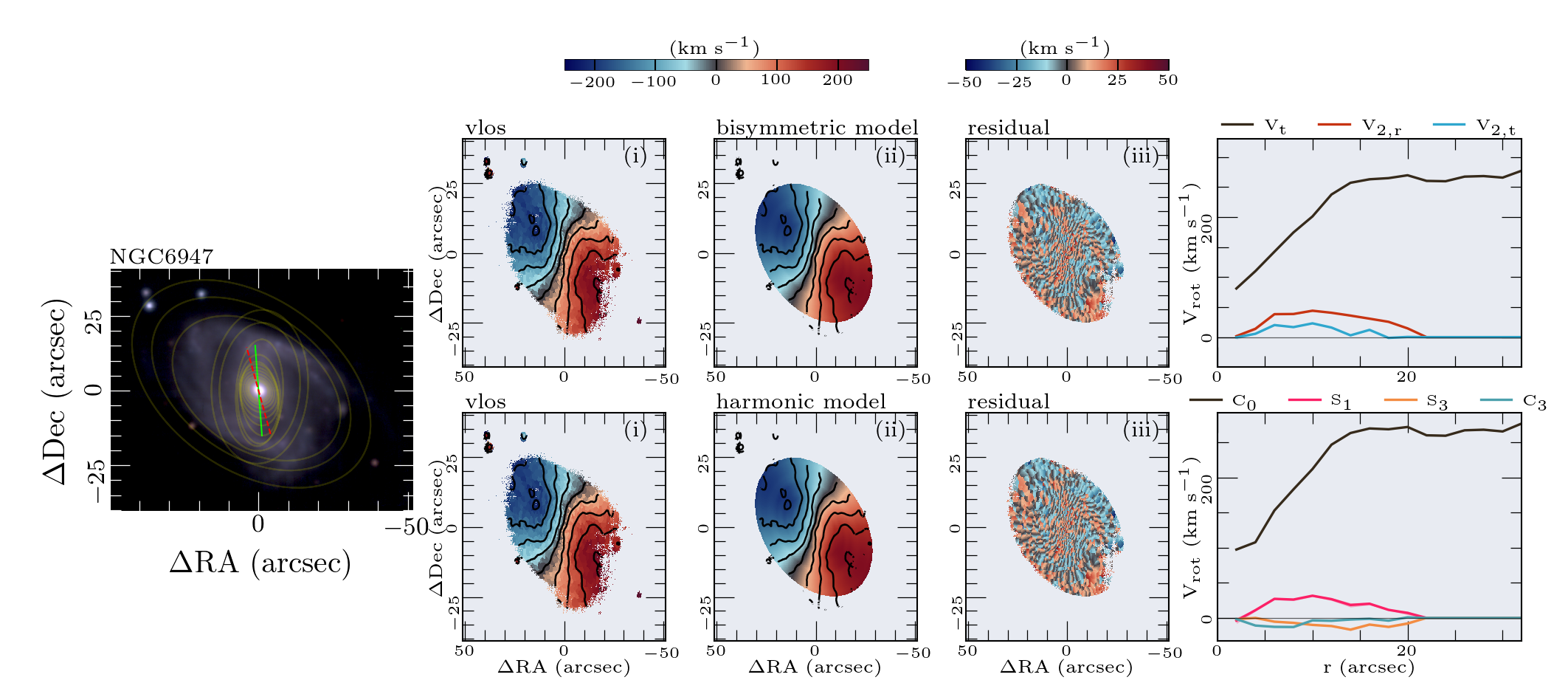}
\caption{The left-most figure shows the reconstructed continuum image from the MUSE datacube using  $gri$ filters. The photometric position angle of the bar is shown with a green line, while the kinematic one derived by \xs~is shown with a red dashed line; the size of these lines represent the aparent size of the bar. Overlayed yellow lines represent isophotes estimated with DES or Pan-STARRS images. The right {\it top (bottom)} figures show the bisymmetric (harmonic) model for the stellar velocity map.  {\it From left to right:} (i) observed stellar velocity field; (ii) best kinematic model; (iii) residual map; (iv) radial distribution of the different kinematic components describing the considered model. Iso-velocity contours are spaced by $\pm 50$ \kms.{ Objects are oriented North-East with North pointing up and East to the left.}}
\label{fig:appendix1}
\end{figure}

Figure~\ref{fig:appendix1} shows the results from the isophotal analysis together with the kinematic modeling of \xs~adopting bisymmetric and harmonic decomposition models. In this set of figures, only one object is shown as an example, the remaining figures are available in the online journal. This figure  shows results only for the stellar velocity maps.

%
%

\figsetstart
\figsetnum{10}
\figsettitle{Non--circular rotation models for the stellar velocity maps.}

\figsetgrpstart
\figsetgrpnum{10.1}
\figsetgrptitle{ESO476-16}
\figsetplot{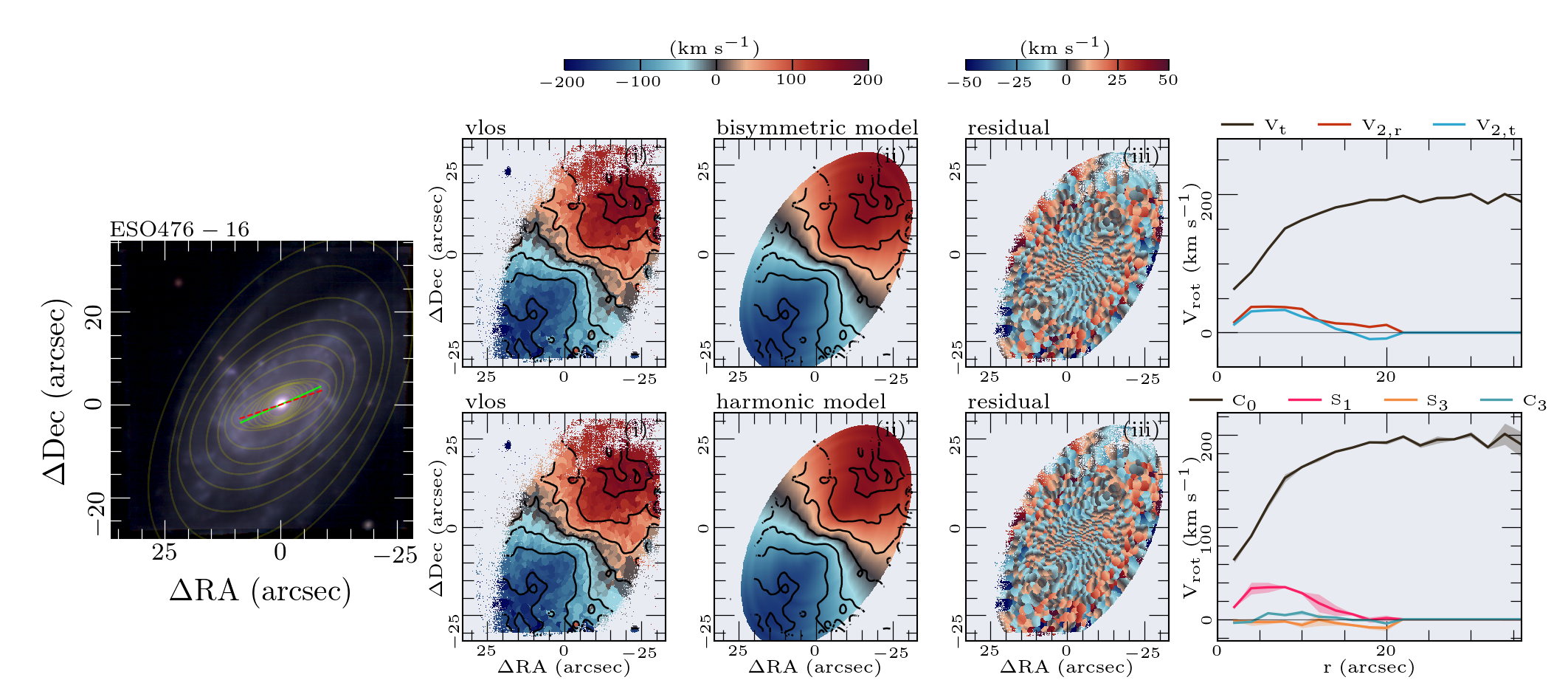}
\figsetgrpnote{The left-most figure shows the reconstructed continuum image from the MUSE datacube with the $gri$ filters. The green continuous line represents the photometric position angle of the bar while the red dashed line is the kinematic position angle of the bar estimated with \xs. The length of these lines represents the projected length of the bar. Overlayed yellow lines represent isophotes estimated with DES or Pan-STARRS images. The right {\it top (bottom)} figures show the bisymmetric (harmonic) model for the stellar velocity map.  {\it From left to right:} the observed velocity field; the best 2D model recovered by \xs; the residual map and the radial variation of the different kinematic components. Superimposed on these maps are iso-velocities spaced by $\pm 50$ \kms. }
\figsetgrpend

\figsetgrpstart
\figsetgrpnum{10.2}
\figsetgrptitle{NGC6947}
\figsetplot{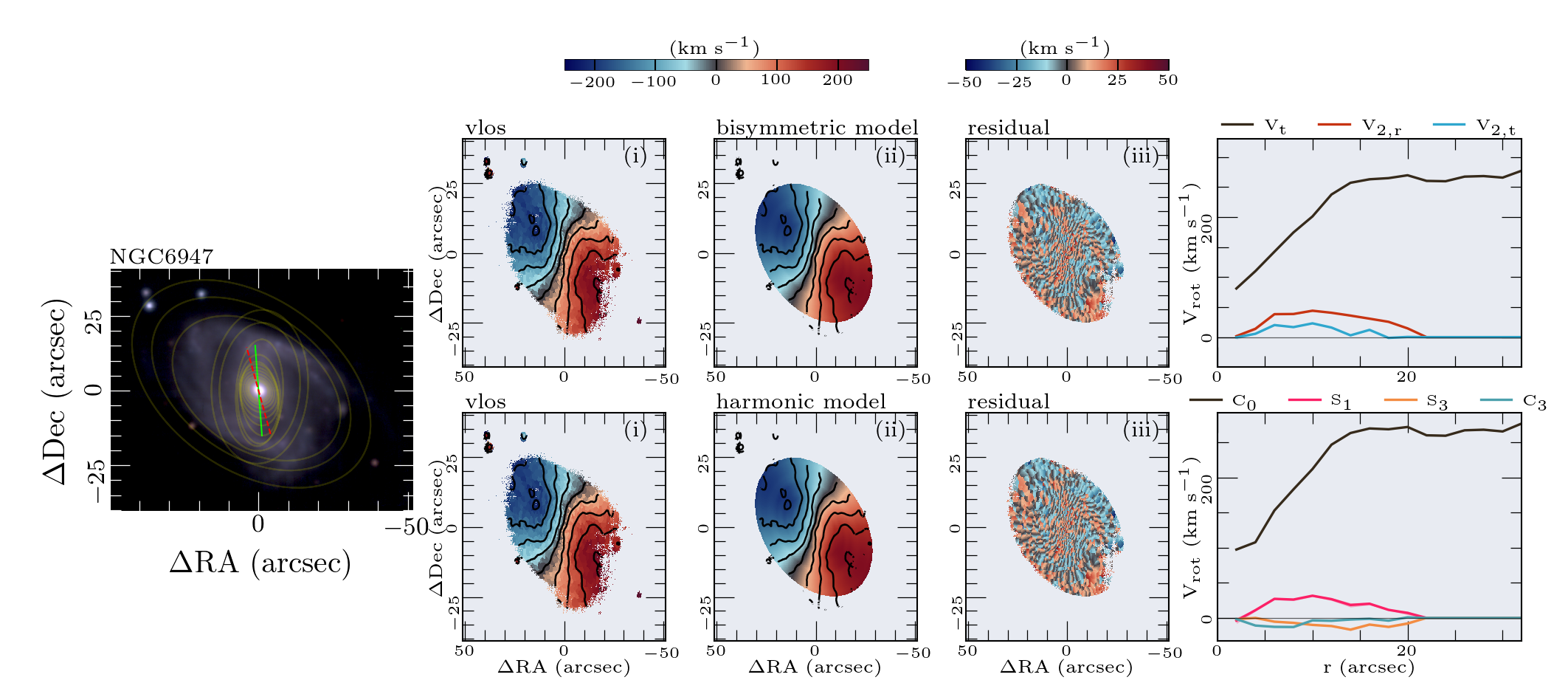}
\figsetgrpnote{The left-most figure shows the reconstructed continuum image from the MUSE datacube with the $gri$ filters. The green continuous line represents the photometric position angle of the bar while the red dashed line is the kinematic position angle of the bar estimated with \xs. The length of these lines represents the projected length of the bar. Overlayed yellow lines represent isophotes estimated with DES or Pan-STARRS images. The right {\it top (bottom)} figures show the bisymmetric (harmonic) model for the stellar velocity map.  {\it From left to right:} the observed velocity field; the best 2D model recovered by \xs; the residual map and the radial variation of the different kinematic components. Superimposed on these maps are iso-velocities spaced by $\pm 50$ \kms. }
\figsetgrpend

\figsetgrpstart
\figsetgrpnum{10.3}
\figsetgrptitle{NGC692}
\figsetplot{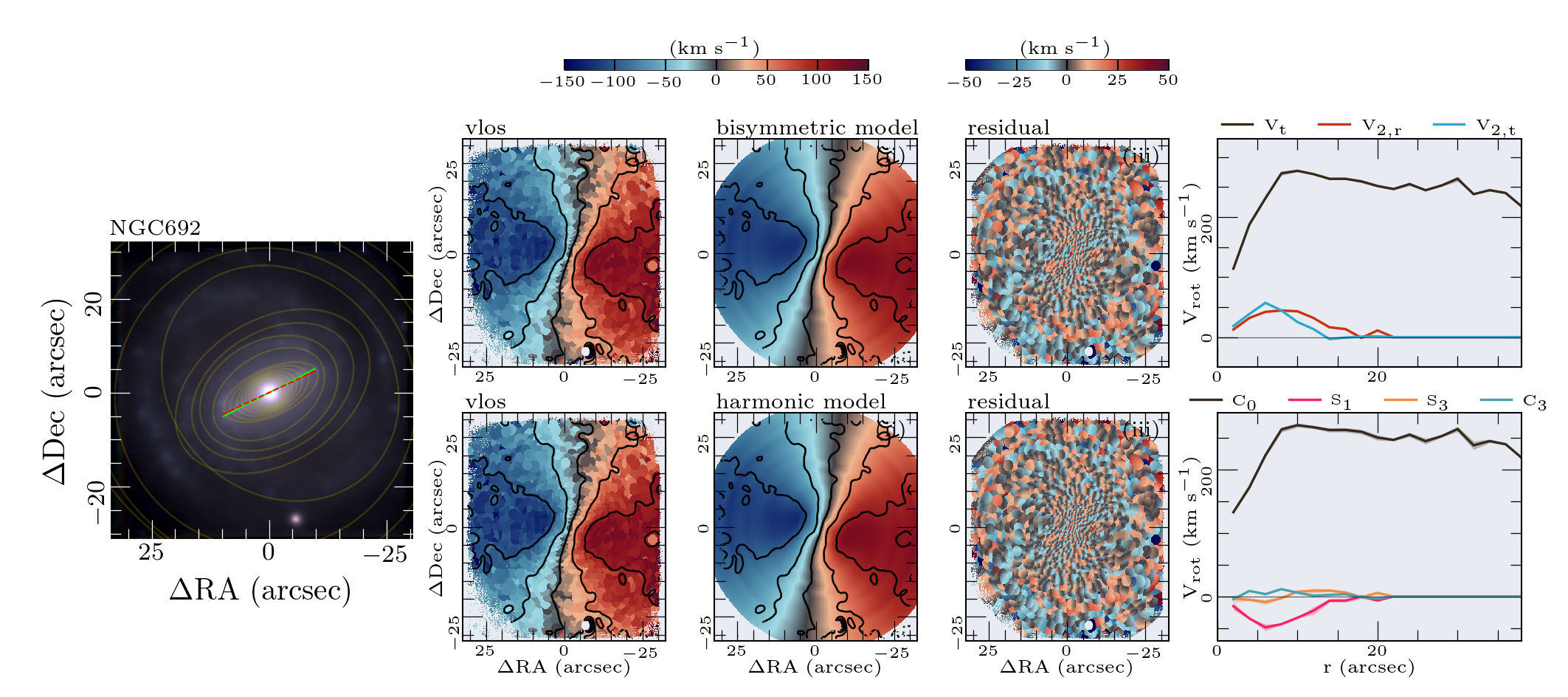}
\figsetgrpnote{The left-most figure shows the reconstructed continuum image from the MUSE datacube with the $gri$ filters. The green continuous line represents the photometric position angle of the bar while the red dashed line is the kinematic position angle of the bar estimated with \xs. The length of these lines represents the projected length of the bar. Overlayed yellow lines represent isophotes estimated with DES or Pan-STARRS images. The right {\it top (bottom)} figures show the bisymmetric (harmonic) model for the stellar velocity map.  {\it From left to right:} the observed velocity field; the best 2D model recovered by \xs; the residual map and the radial variation of the different kinematic components. Superimposed on these maps are iso-velocities spaced by $\pm 50$ \kms. }
\figsetgrpend

\figsetgrpstart
\figsetgrpnum{10.4}
\figsetgrptitle{IC2160}
\figsetplot{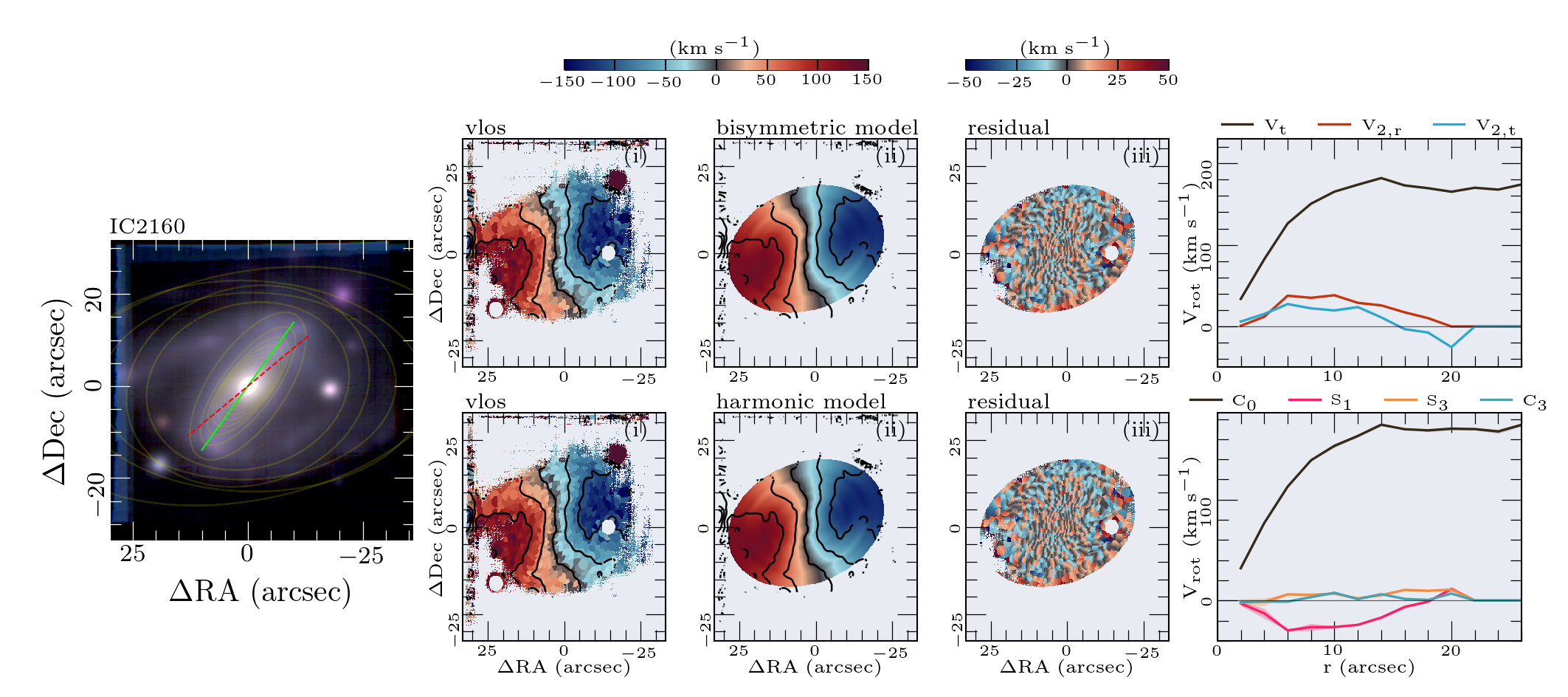}
\figsetgrpnote{The left-most figure shows the reconstructed continuum image from the MUSE datacube with the $gri$ filters. The green continuous line represents the photometric position angle of the bar while the red dashed line is the kinematic position angle of the bar estimated with \xs. The length of these lines represents the projected length of the bar. Overlayed yellow lines represent isophotes estimated with DES or Pan-STARRS images. The right {\it top (bottom)} figures show the bisymmetric (harmonic) model for the stellar velocity map.  {\it From left to right:} the observed velocity field; the best 2D model recovered by \xs; the residual map and the radial variation of the different kinematic components. Superimposed on these maps are iso-velocities spaced by $\pm 50$ \kms. }
\figsetgrpend

\figsetgrpstart
\figsetgrpnum{10.5}
\figsetgrptitle{UGC3634}
\figsetplot{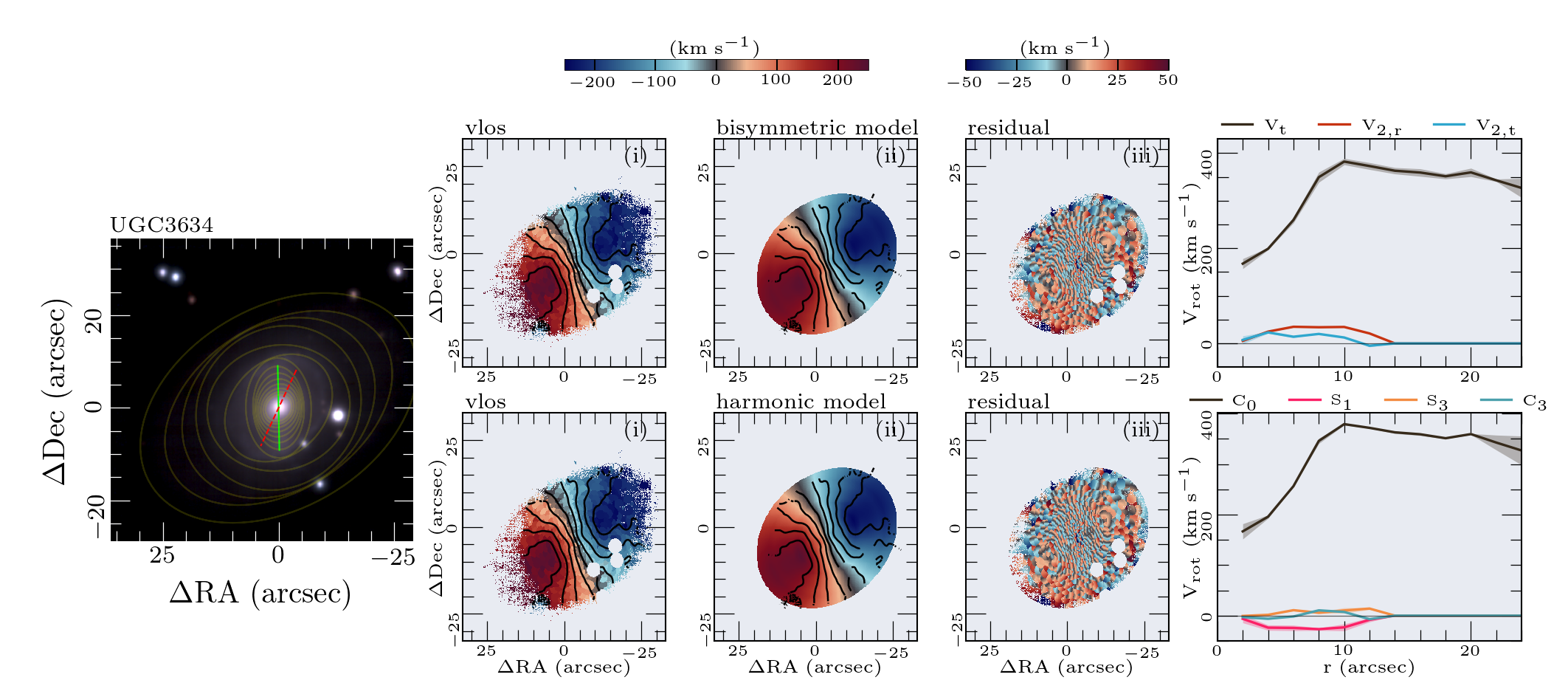}
\figsetgrpnote{The left-most figure shows the reconstructed continuum image from the MUSE datacube with the $gri$ filters. The green continuous line represents the photometric position angle of the bar while the red dashed line is the kinematic position angle of the bar estimated with \xs. The length of these lines represents the projected length of the bar. Overlayed yellow lines represent isophotes estimated with DES or Pan-STARRS images. The right {\it top (bottom)} figures show the bisymmetric (harmonic) model for the stellar velocity map.  {\it From left to right:} the observed velocity field; the best 2D model recovered by \xs; the residual map and the radial variation of the different kinematic components. Superimposed on these maps are iso-velocities spaced by $\pm 50$ \kms. }
\figsetgrpend

\figsetgrpstart
\figsetgrpnum{10.6}
\figsetgrptitle{NGC5339}
\figsetplot{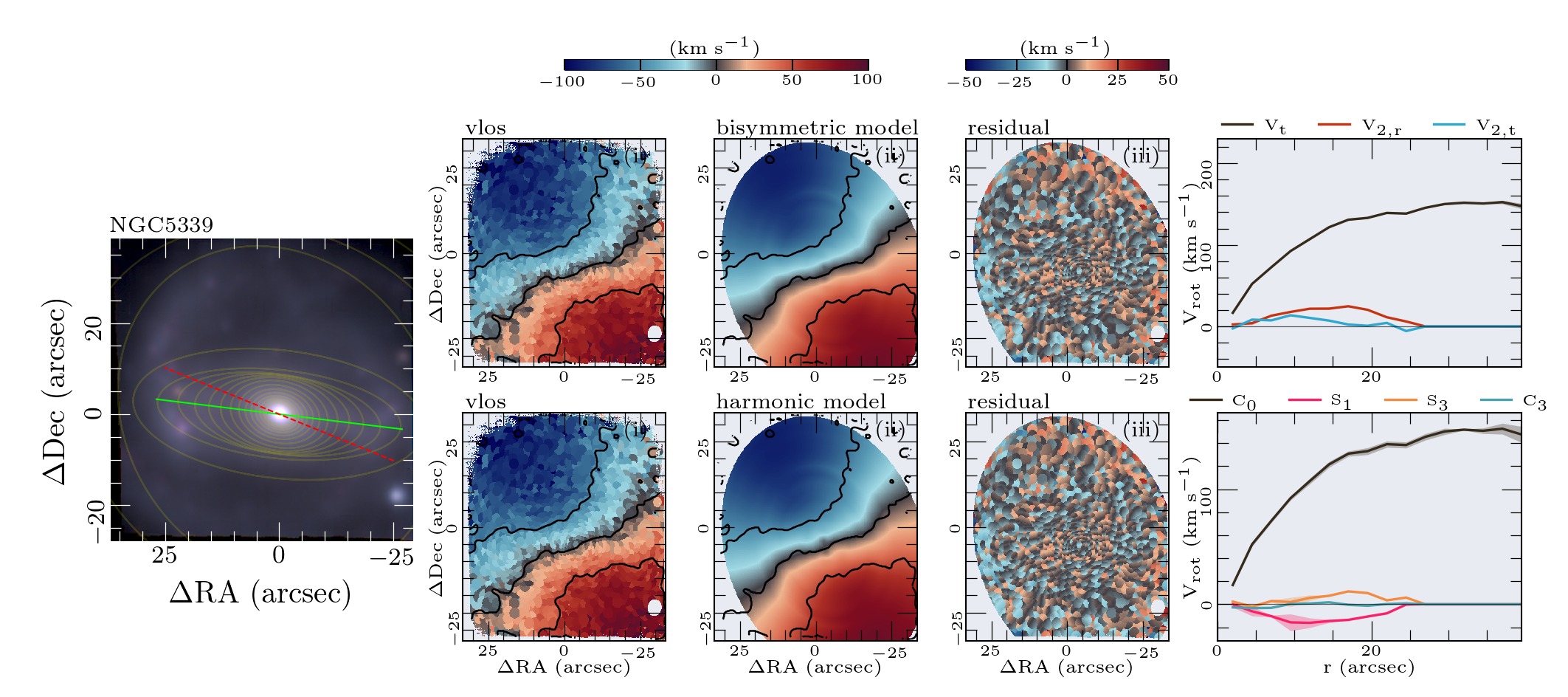}
\figsetgrpnote{The left-most figure shows the reconstructed continuum image from the MUSE datacube with the $gri$ filters. The green continuous line represents the photometric position angle of the bar while the red dashed line is the kinematic position angle of the bar estimated with \xs. The length of these lines represents the projected length of the bar. Overlayed yellow lines represent isophotes estimated with DES or Pan-STARRS images. The right {\it top (bottom)} figures show the bisymmetric (harmonic) model for the stellar velocity map.  {\it From left to right:} the observed velocity field; the best 2D model recovered by \xs; the residual map and the radial variation of the different kinematic components. Superimposed on these maps are iso-velocities spaced by $\pm 50$ \kms. }
\figsetgrpend

\figsetgrpstart
\figsetgrpnum{10.7}
\figsetgrptitle{ESO325-43}
\figsetplot{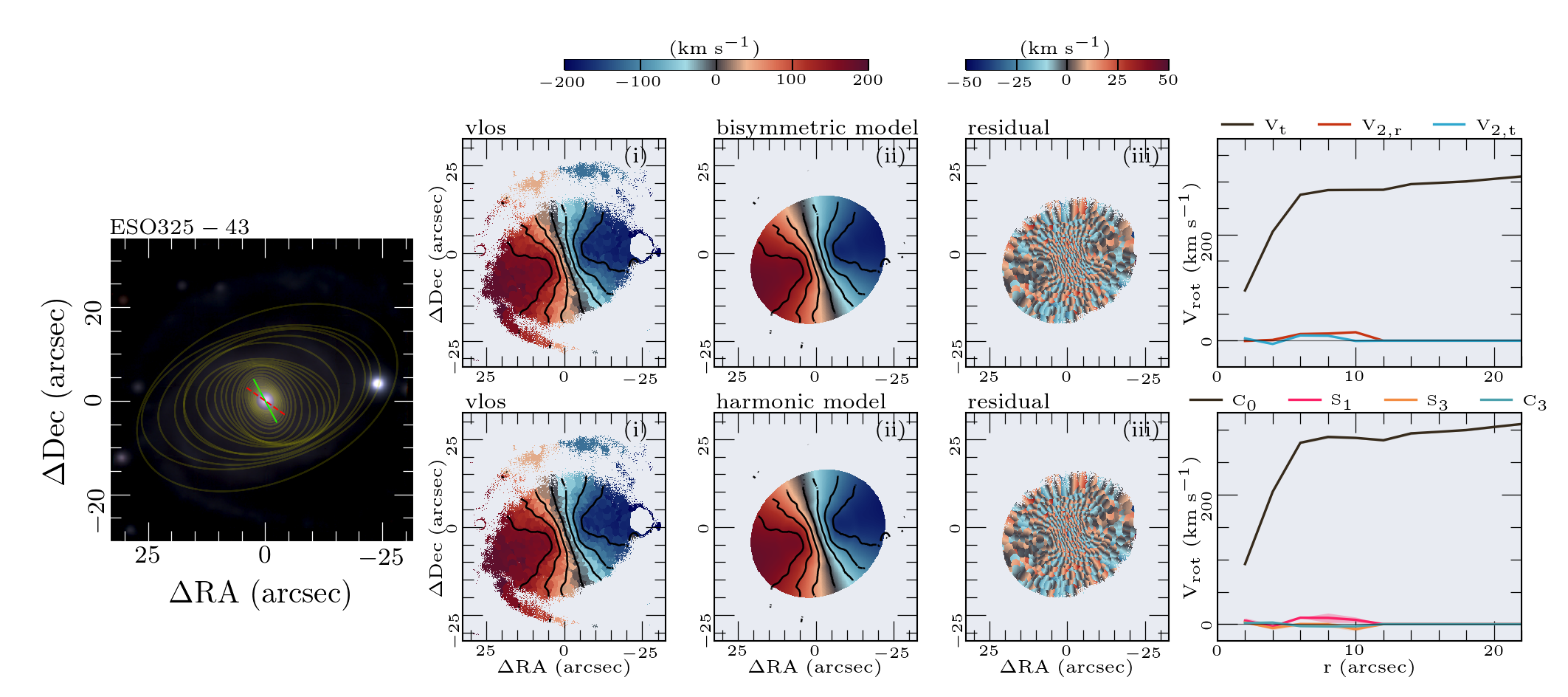}
\figsetgrpnote{The left-most figure shows the reconstructed continuum image from the MUSE datacube with the $gri$ filters. The green continuous line represents the photometric position angle of the bar while the red dashed line is the kinematic position angle of the bar estimated with \xs. The length of these lines represents the projected length of the bar. Overlayed yellow lines represent isophotes estimated with DES or Pan-STARRS images. The right {\it top (bottom)} figures show the bisymmetric (harmonic) model for the stellar velocity map.  {\it From left to right:} the observed velocity field; the best 2D model recovered by \xs; the residual map and the radial variation of the different kinematic components. Superimposed on these maps are iso-velocities spaced by $\pm 50$ \kms. }
\figsetgrpend

\figsetgrpstart
\figsetgrpnum{10.8}
\figsetgrptitle{IC0004}
\figsetplot{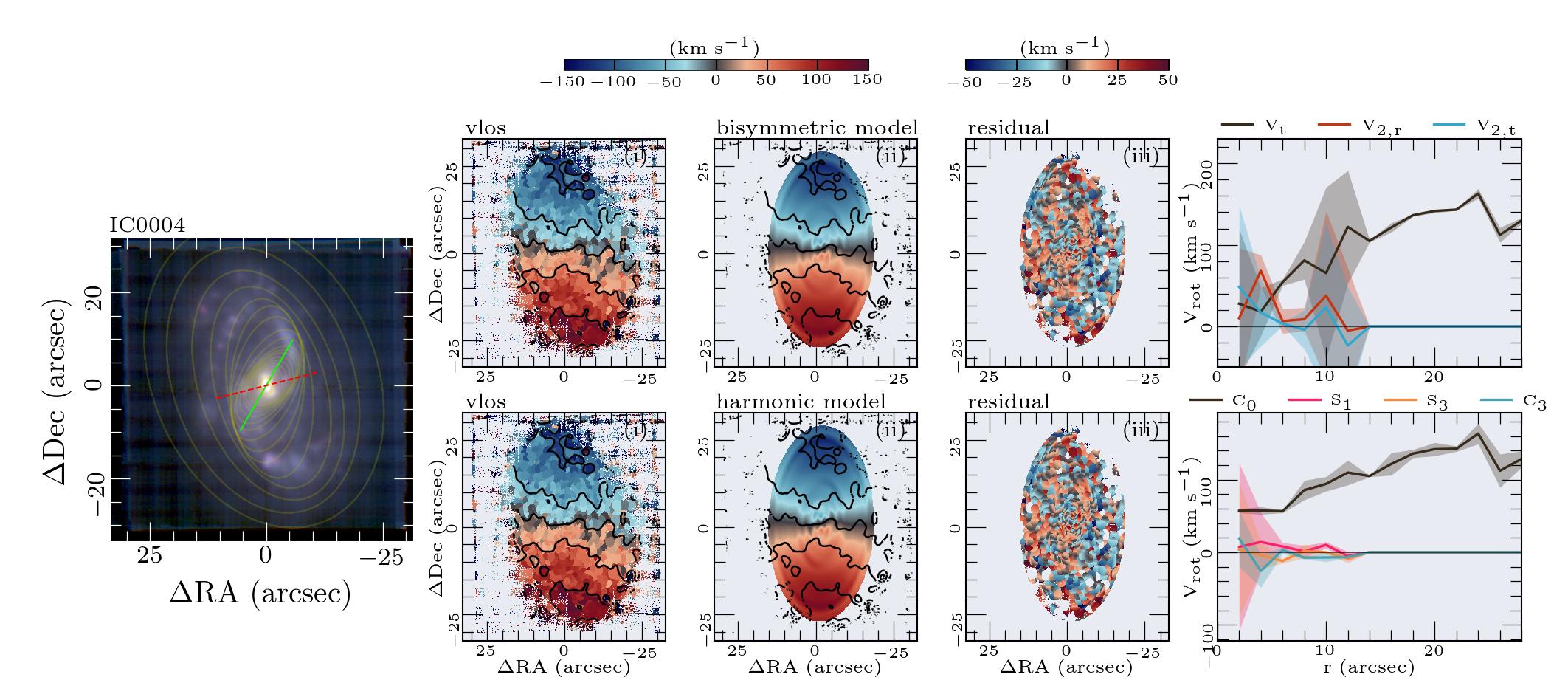}
\figsetgrpnote{The left-most figure shows the reconstructed continuum image from the MUSE datacube with the $gri$ filters. The green continuous line represents the photometric position angle of the bar while the red dashed line is the kinematic position angle of the bar estimated with \xs. The length of these lines represents the projected length of the bar. Overlayed yellow lines represent isophotes estimated with DES or Pan-STARRS images. The right {\it top (bottom)} figures show the bisymmetric (harmonic) model for the stellar velocity map.  {\it From left to right:} the observed velocity field; the best 2D model recovered by \xs; the residual map and the radial variation of the different kinematic components. Superimposed on these maps are iso-velocities spaced by $\pm 50$ \kms. }
\figsetgrpend

\figsetgrpstart
\figsetgrpnum{10.9}
\figsetgrptitle{PGC055442}
\figsetplot{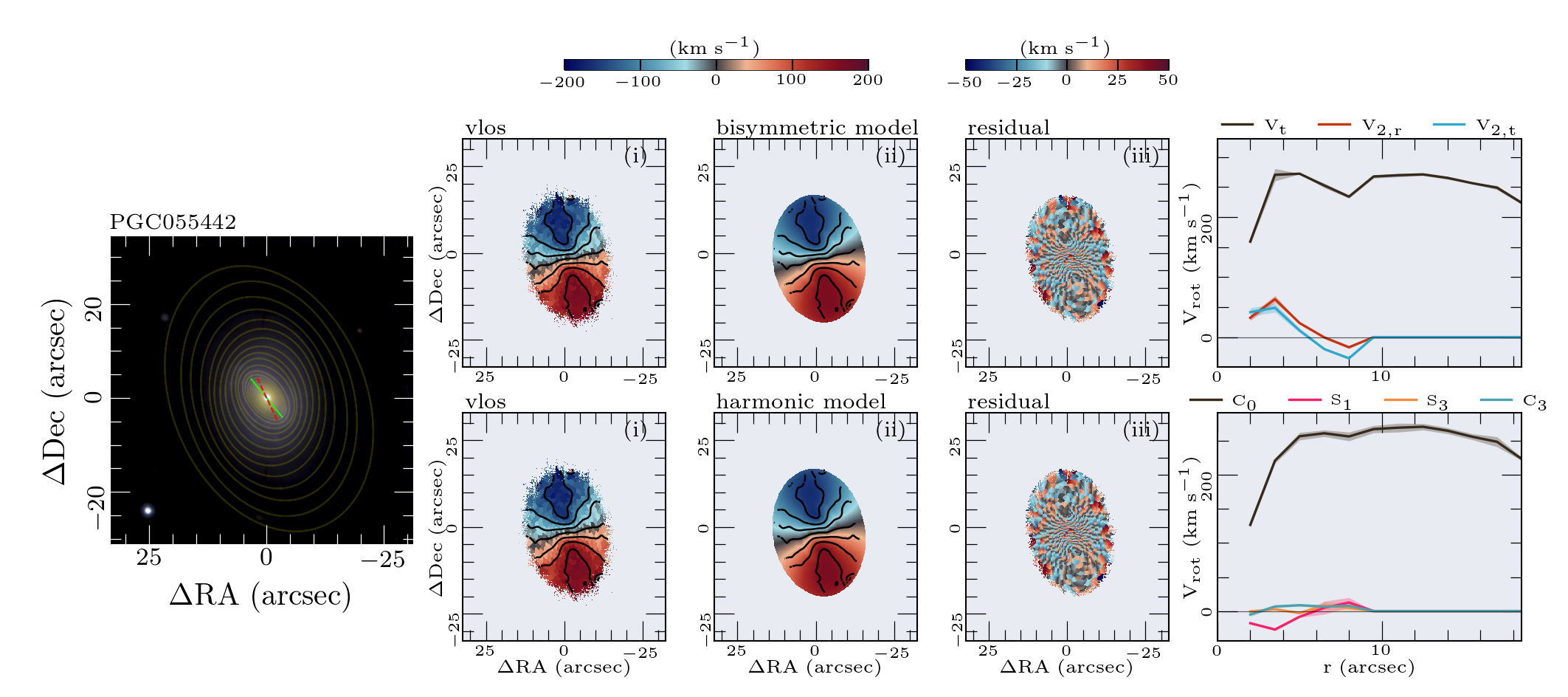}
\figsetgrpnote{The left-most figure shows the reconstructed continuum image from the MUSE datacube with the $gri$ filters. The green continuous line represents the photometric position angle of the bar while the red dashed line is the kinematic position angle of the bar estimated with \xs. The length of these lines represents the projected length of the bar. Overlayed yellow lines represent isophotes estimated with DES or Pan-STARRS images. The right {\it top (bottom)} figures show the bisymmetric (harmonic) model for the stellar velocity map.  {\it From left to right:} the observed velocity field; the best 2D model recovered by \xs; the residual map and the radial variation of the different kinematic components. Superimposed on these maps are iso-velocities spaced by $\pm 50$ \kms. }
\figsetgrpend

\figsetgrpstart
\figsetgrpnum{10.10}
\figsetgrptitle{NGC7780}
\figsetplot{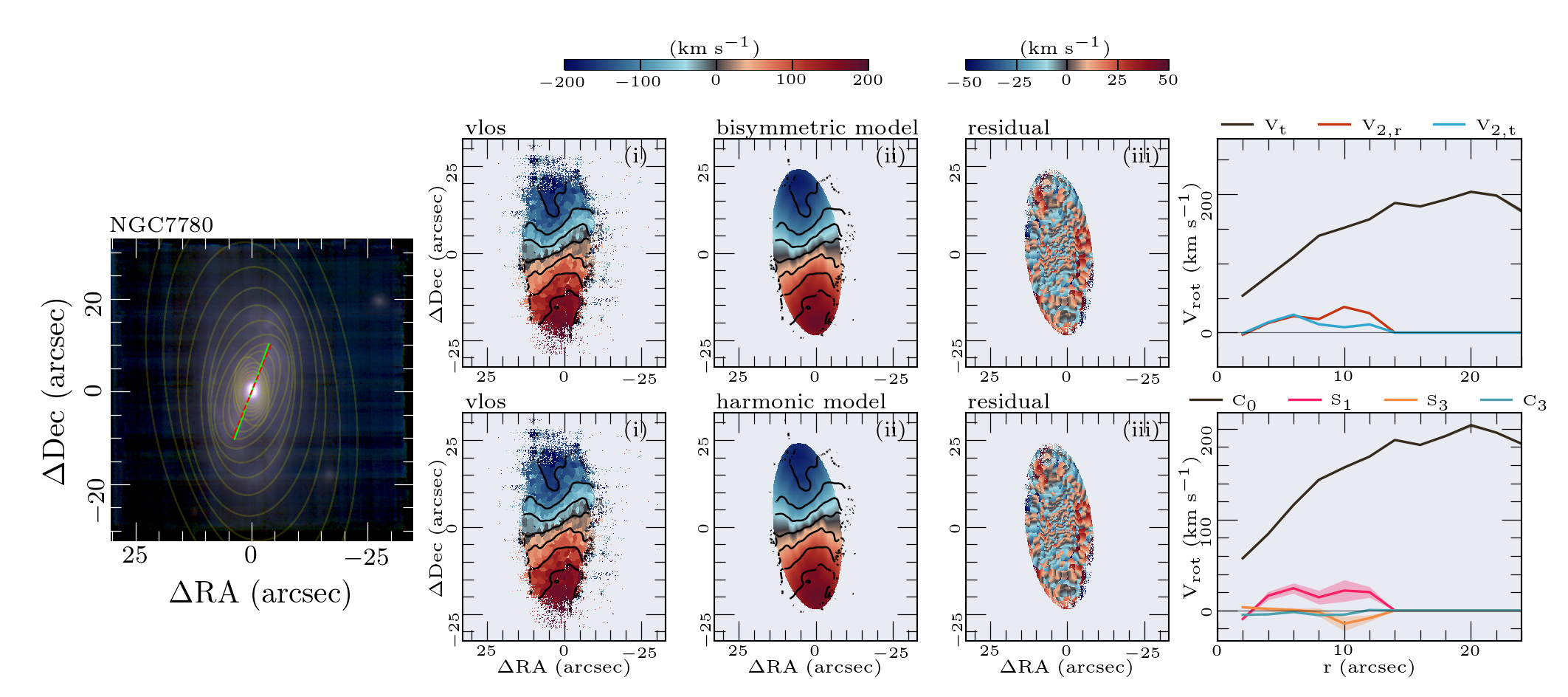}
\figsetgrpnote{The left-most figure shows the reconstructed continuum image from the MUSE datacube with the $gri$ filters. The green continuous line represents the photometric position angle of the bar while the red dashed line is the kinematic position angle of the bar estimated with \xs. The length of these lines represents the projected length of the bar. Overlayed yellow lines represent isophotes estimated with DES or Pan-STARRS images. The right {\it top (bottom)} figures show the bisymmetric (harmonic) model for the stellar velocity map.  {\it From left to right:} the observed velocity field; the best 2D model recovered by \xs; the residual map and the radial variation of the different kinematic components. Superimposed on these maps are iso-velocities spaced by $\pm 50$ \kms. }
\figsetgrpend

\figsetgrpstart
\figsetgrpnum{10.11}
\figsetgrptitle{NGC3464}
\figsetplot{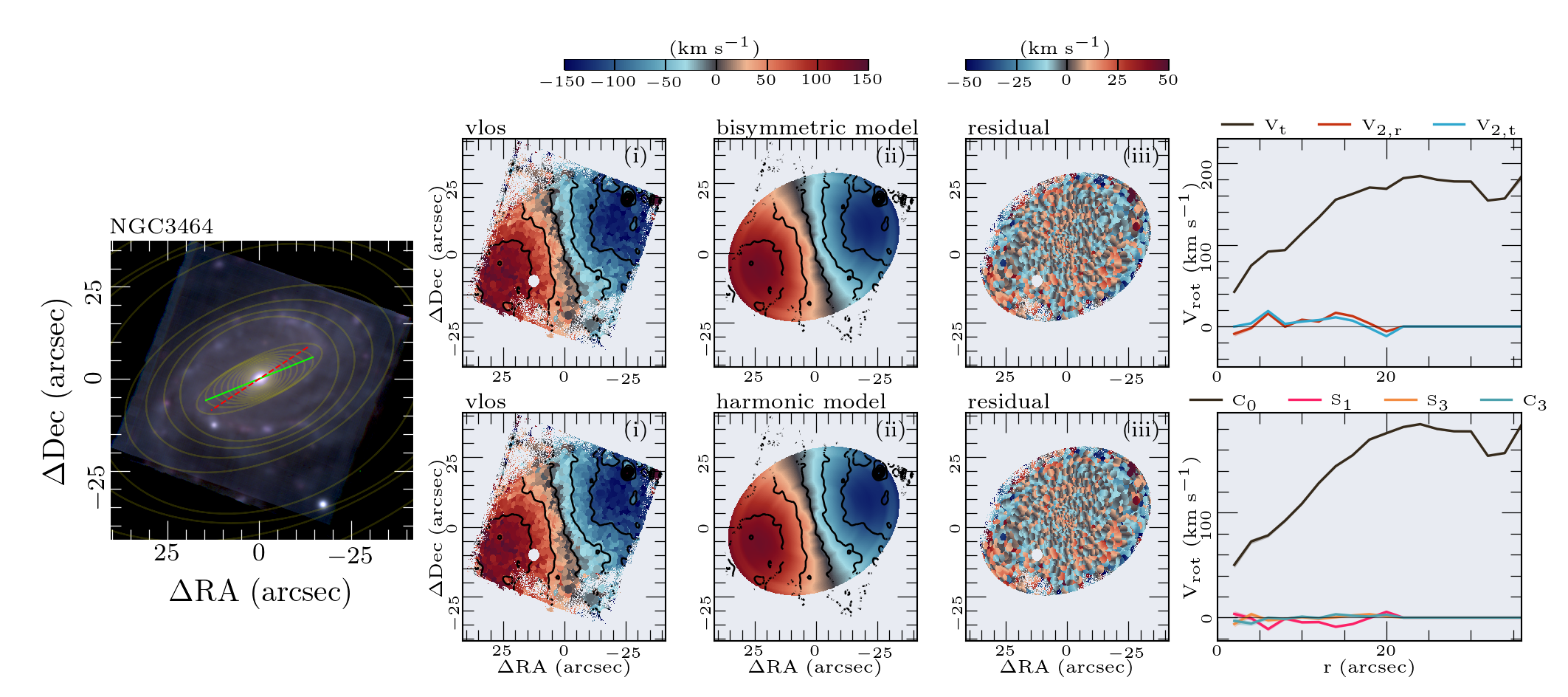}
\figsetgrpnote{The left-most figure shows the reconstructed continuum image from the MUSE datacube with the $gri$ filters. The green continuous line represents the photometric position angle of the bar while the red dashed line is the kinematic position angle of the bar estimated with \xs. The length of these lines represents the projected length of the bar. Overlayed yellow lines represent isophotes estimated with DES or Pan-STARRS images. The right {\it top (bottom)} figures show the bisymmetric (harmonic) model for the stellar velocity map.  {\it From left to right:} the observed velocity field; the best 2D model recovered by \xs; the residual map and the radial variation of the different kinematic components. Superimposed on these maps are iso-velocities spaced by $\pm 50$ \kms. }
\figsetgrpend

\figsetgrpstart
\figsetgrpnum{10.12}
\figsetgrptitle{NGC1591}
\figsetplot{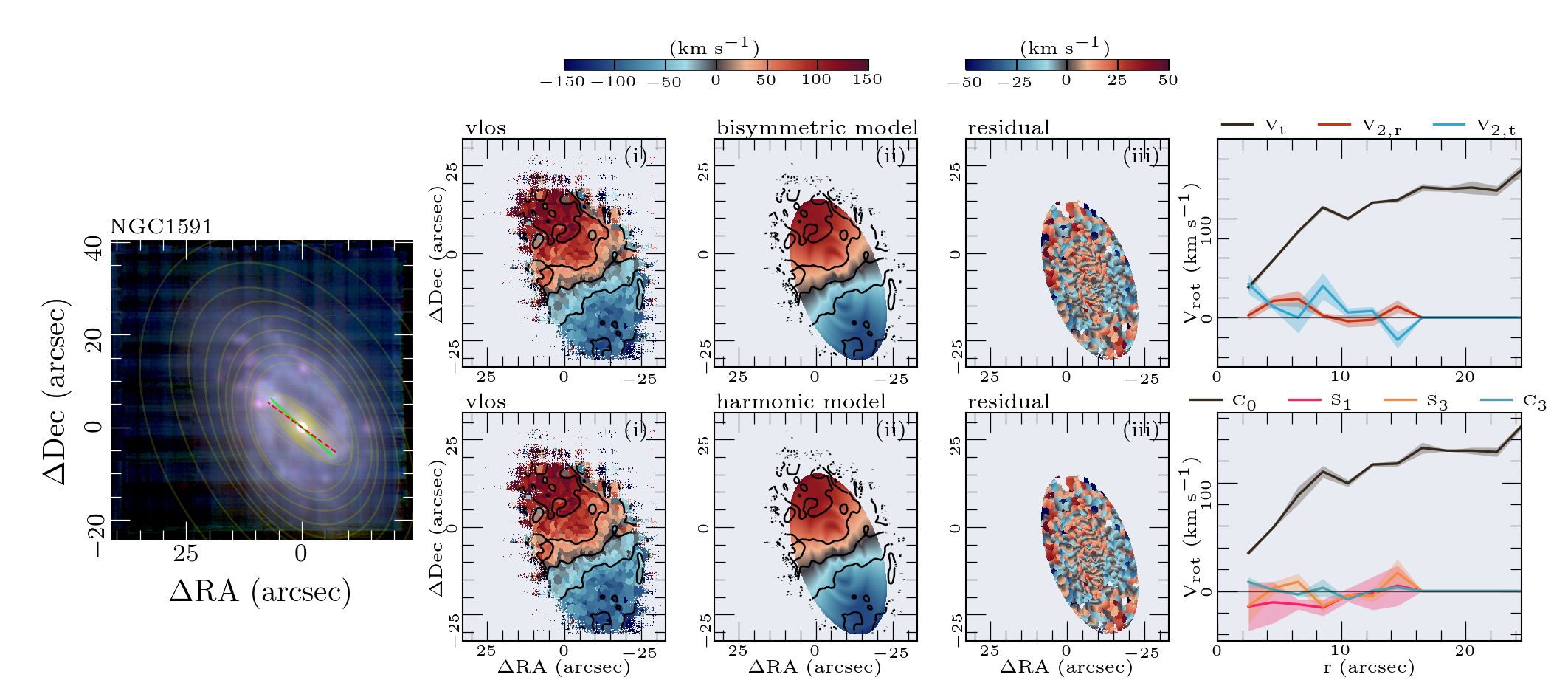}
\figsetgrpnote{The left-most figure shows the reconstructed continuum image from the MUSE datacube with the $gri$ filters. The green continuous line represents the photometric position angle of the bar while the red dashed line is the kinematic position angle of the bar estimated with \xs. The length of these lines represents the projected length of the bar. Overlayed yellow lines represent isophotes estimated with DES or Pan-STARRS images. The right {\it top (bottom)} figures show the bisymmetric (harmonic) model for the stellar velocity map.  {\it From left to right:} the observed velocity field; the best 2D model recovered by \xs; the residual map and the radial variation of the different kinematic components. Superimposed on these maps are iso-velocities spaced by $\pm 50$ \kms. }
\figsetgrpend

\figsetgrpstart
\figsetgrpnum{10.13}
\figsetgrptitle{ESO18-18}
\figsetplot{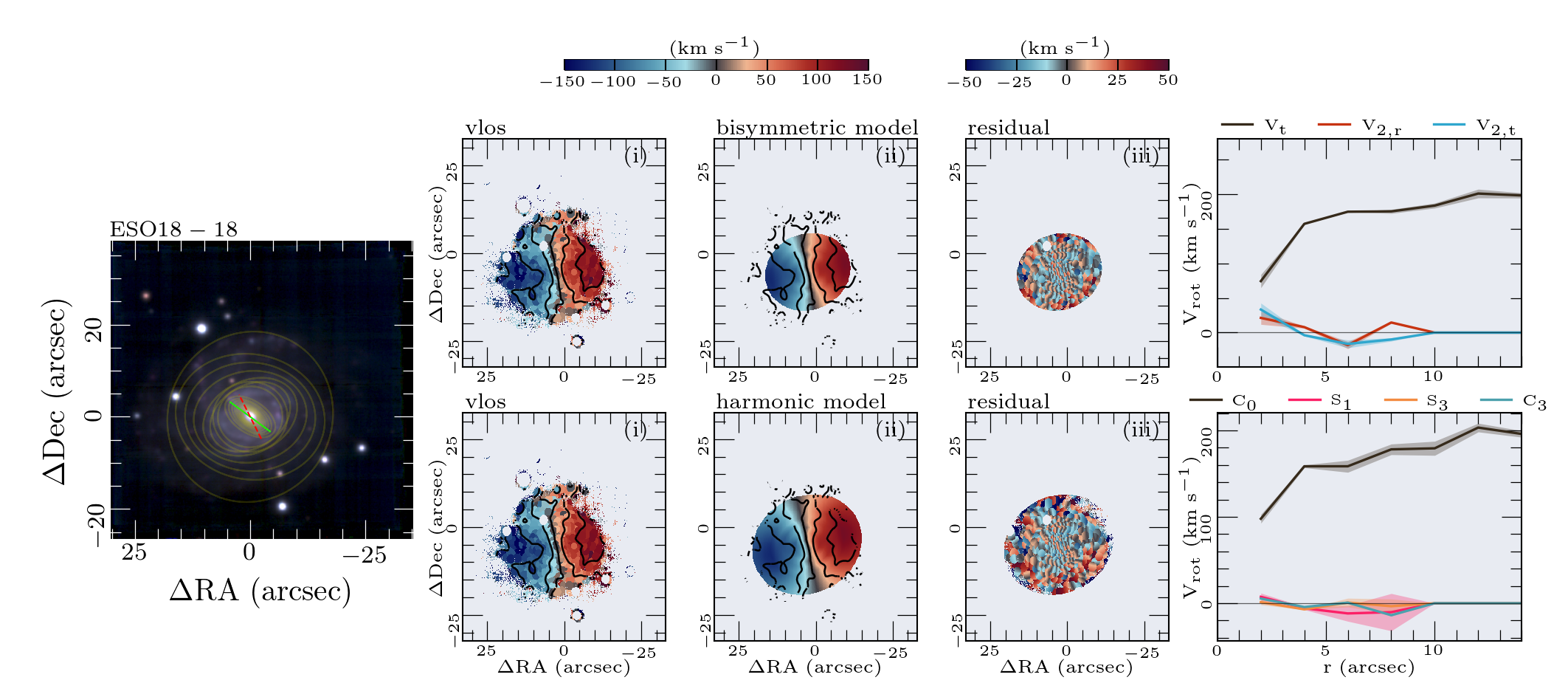}
\figsetgrpnote{The left-most figure shows the reconstructed continuum image from the MUSE datacube with the $gri$ filters. The green continuous line represents the photometric position angle of the bar while the red dashed line is the kinematic position angle of the bar estimated with \xs. The length of these lines represents the projected length of the bar. Overlayed yellow lines represent isophotes estimated with DES or Pan-STARRS images. The right {\it top (bottom)} figures show the bisymmetric (harmonic) model for the stellar velocity map.  {\it From left to right:} the observed velocity field; the best 2D model recovered by \xs; the residual map and the radial variation of the different kinematic components. Superimposed on these maps are iso-velocities spaced by $\pm 50$ \kms. }
\figsetgrpend

\figsetgrpstart
\figsetgrpnum{10.14}
\figsetgrptitle{NGC289}
\figsetplot{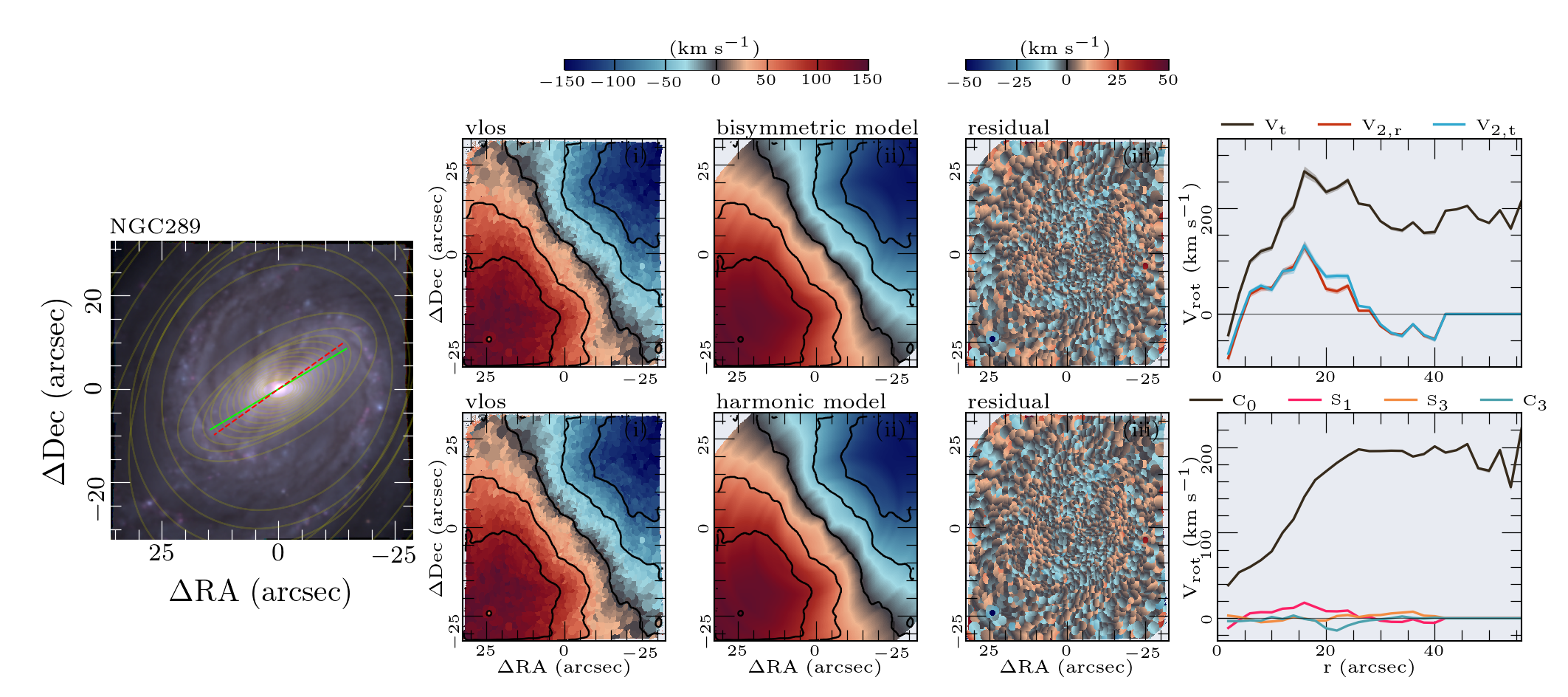}
\figsetgrpnote{The left-most figure shows the reconstructed continuum image from the MUSE datacube with the $gri$ filters. The green continuous line represents the photometric position angle of the bar while the red dashed line is the kinematic position angle of the bar estimated with \xs. The length of these lines represents the projected length of the bar. Overlayed yellow lines represent isophotes estimated with DES or Pan-STARRS images. The right {\it top (bottom)} figures show the bisymmetric (harmonic) model for the stellar velocity map.  {\it From left to right:} the observed velocity field; the best 2D model recovered by \xs; the residual map and the radial variation of the different kinematic components. Superimposed on these maps are iso-velocities spaced by $\pm 50$ \kms. }
\figsetgrpend

\figsetend

\begin{figure}
\figurenum{11}
\plotone{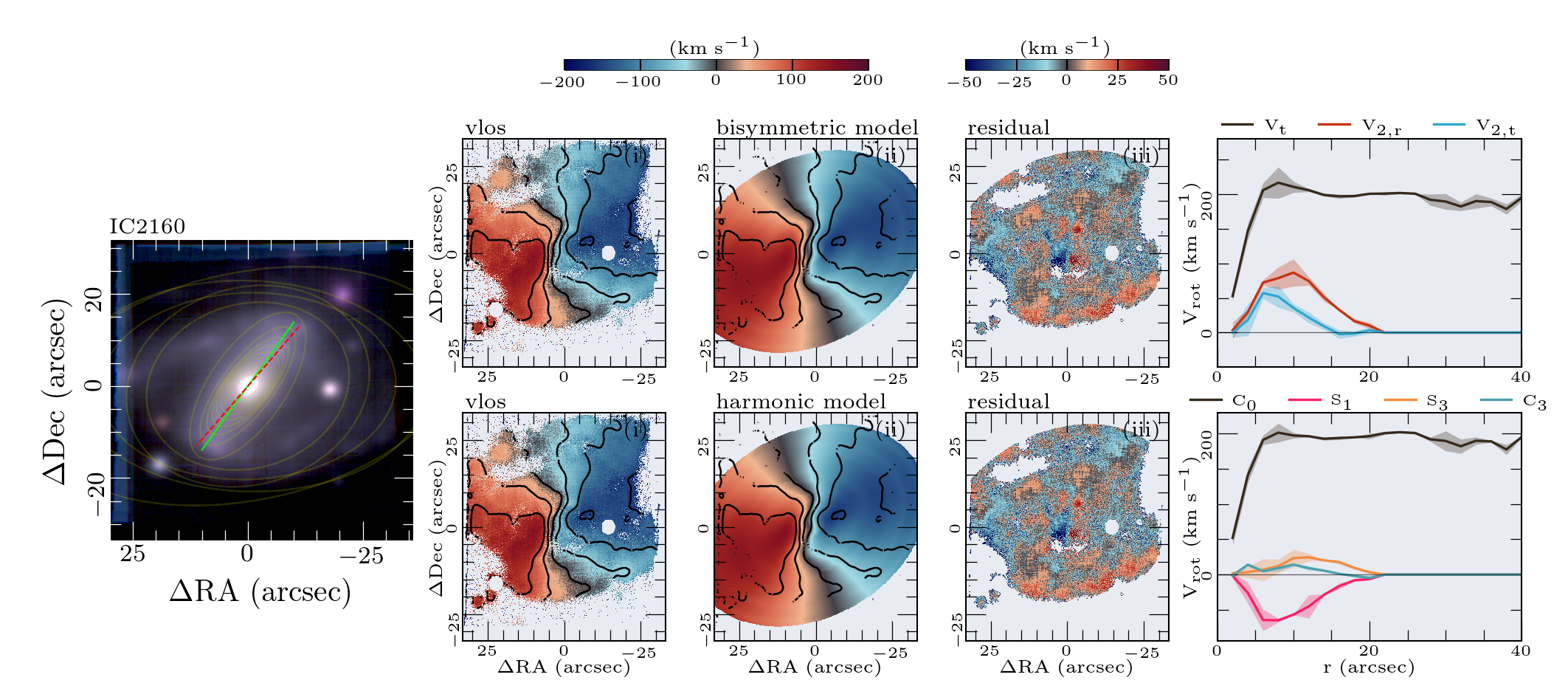}
\caption{Same  as Fig.~\ref{fig:appendix1} but this time for the \ha~velocity maps. We show results only for those galaxies with \ha~emission detected along the bars.}
\label{fig:appendix2}
\end{figure}

%
%
Similarly, Fig~\ref{fig:appendix2} shows results for the \ha~velocity maps, but only for those objects with \ha~emission detected along the bars. The complete figures set is available in the online journal.

\figsetstart
\figsetnum{11}
\figsettitle{Non--circular rotation models for the \ha~velocity maps.}

\figsetgrpstart
\figsetgrpnum{11.1}
\figsetgrptitle{IC2160}
\figsetplot{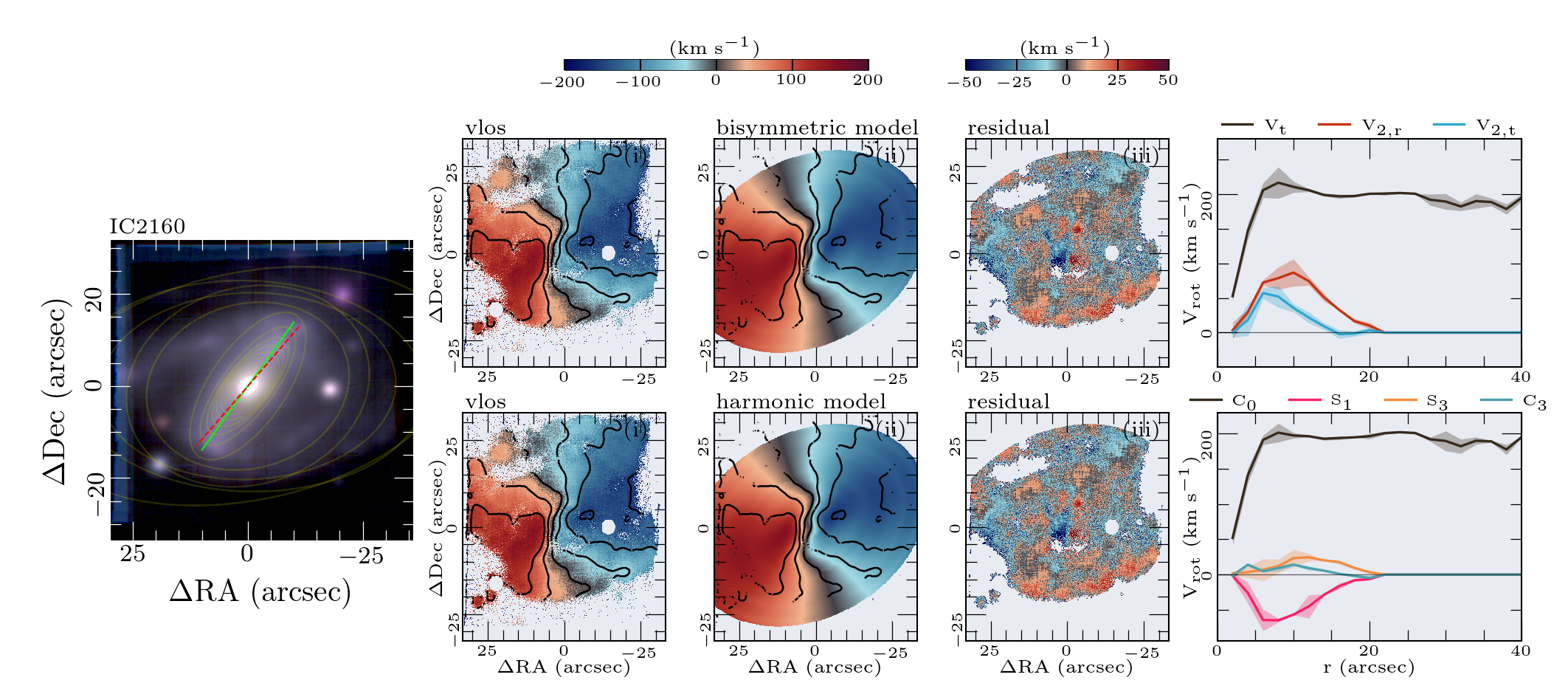}
\figsetgrpnote{Same  as Fig.~\ref{fig:appendix1} but this time for the \ha~velocity maps. We show results only for those galaxies with \ha~emission detected along the bars.}
\figsetgrpend

\figsetgrpstart
\figsetgrpnum{11.2}
\figsetgrptitle{IC0004}
\figsetplot{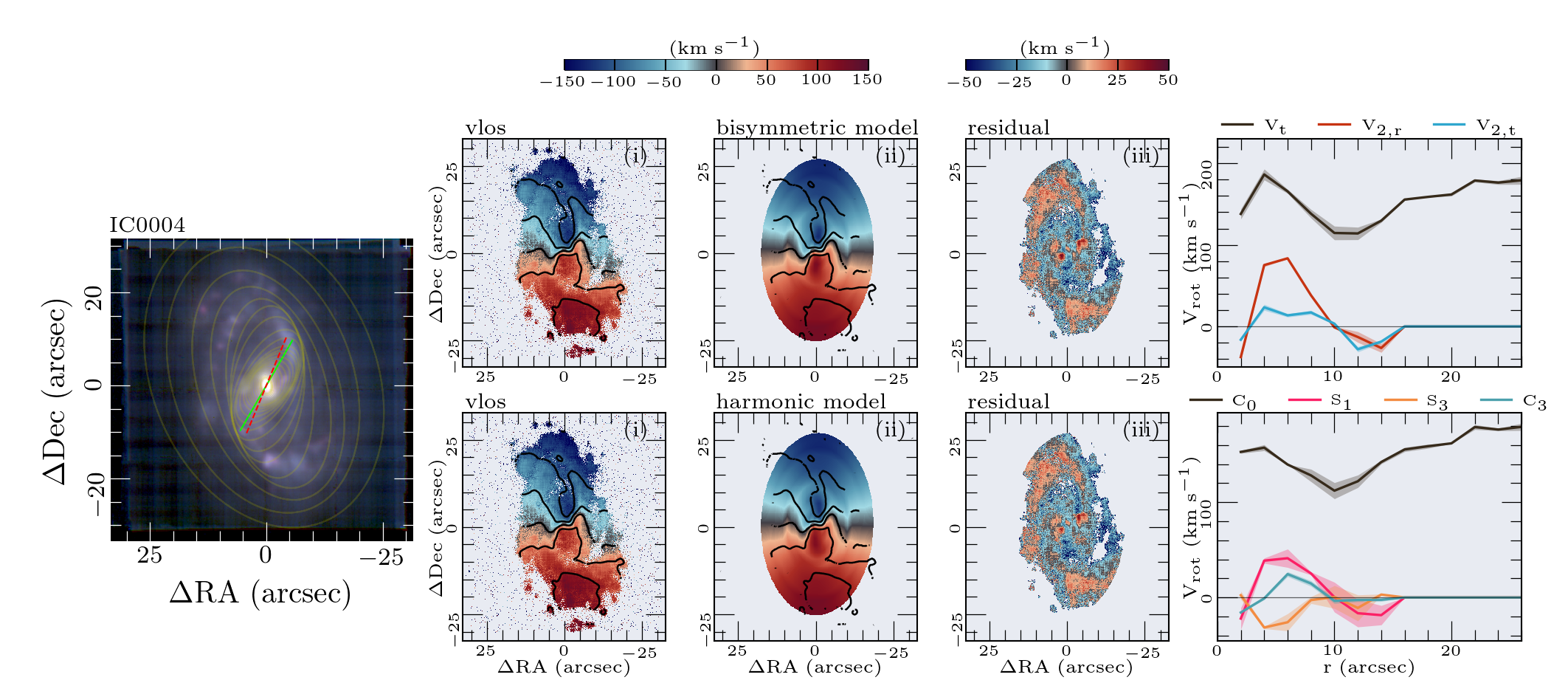}
\figsetgrpnote{Same  as Fig.~\ref{fig:appendix1} but this time for the \ha~velocity maps. We show results only for those galaxies with \ha~emission detected along the bars.}
\figsetgrpend

\figsetgrpstart
\figsetgrpnum{11.3}
\figsetgrptitle{NGC1591}
\figsetplot{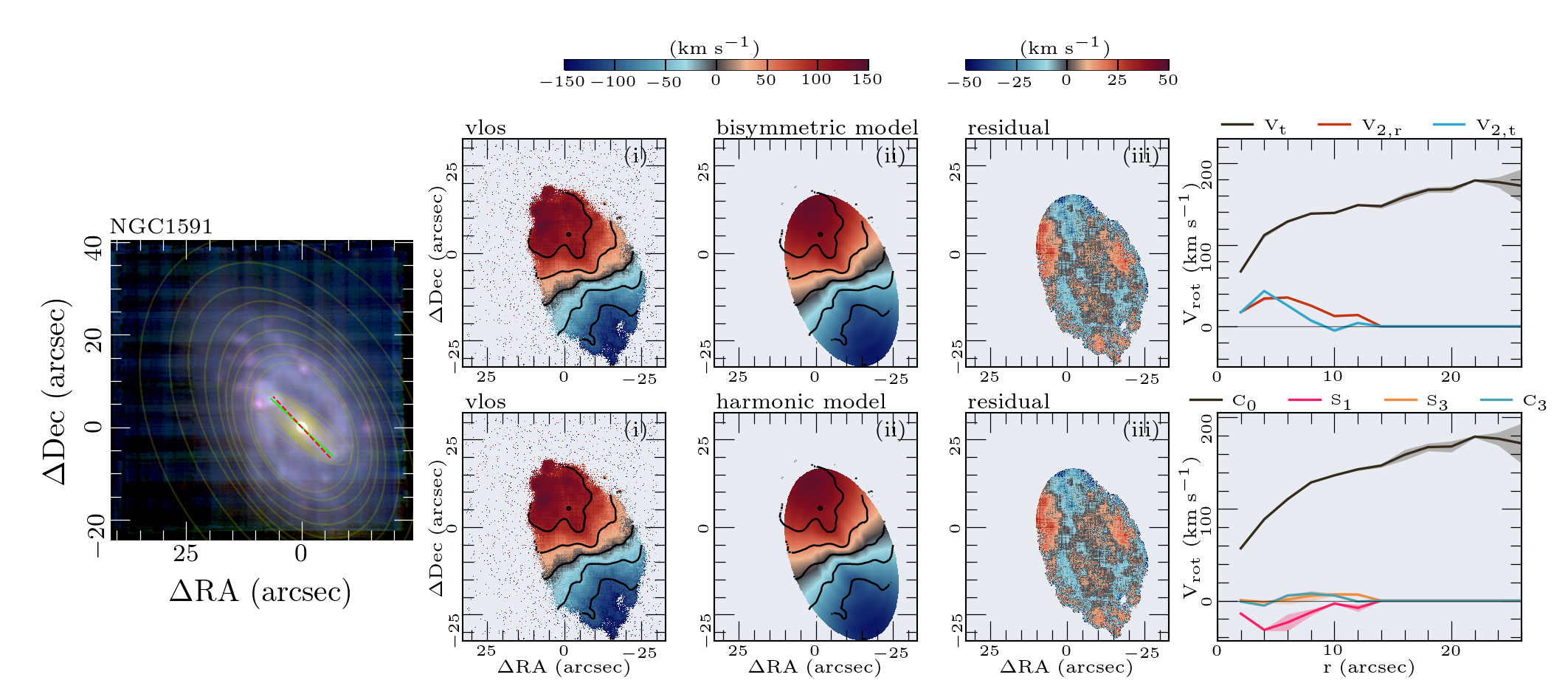}
\figsetgrpnote{Same  as Fig.~\ref{fig:appendix1} but this time for the \ha~velocity maps. We show results only for those galaxies with \ha~emission detected along the bars.}
\figsetgrpend

\figsetgrpstart
\figsetgrpnum{11.4}
\figsetgrptitle{NGC3464}
\figsetplot{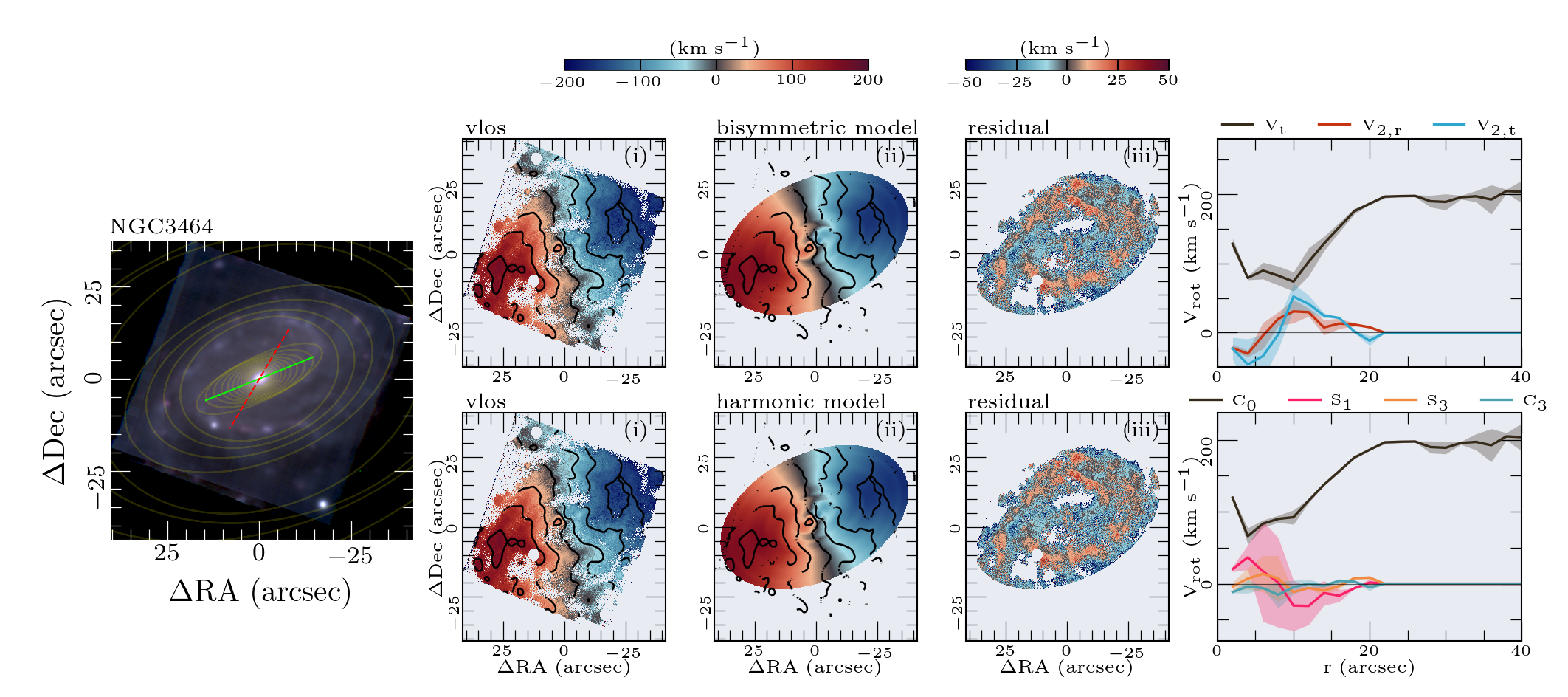}
\figsetgrpnote{Same  as Fig.~\ref{fig:appendix1} but this time for the \ha~velocity maps. We show results only for those galaxies with \ha~emission detected along the bars.}
\figsetgrpend

\figsetgrpstart
\figsetgrpnum{11.5}
\figsetgrptitle{PGC055442}
\figsetplot{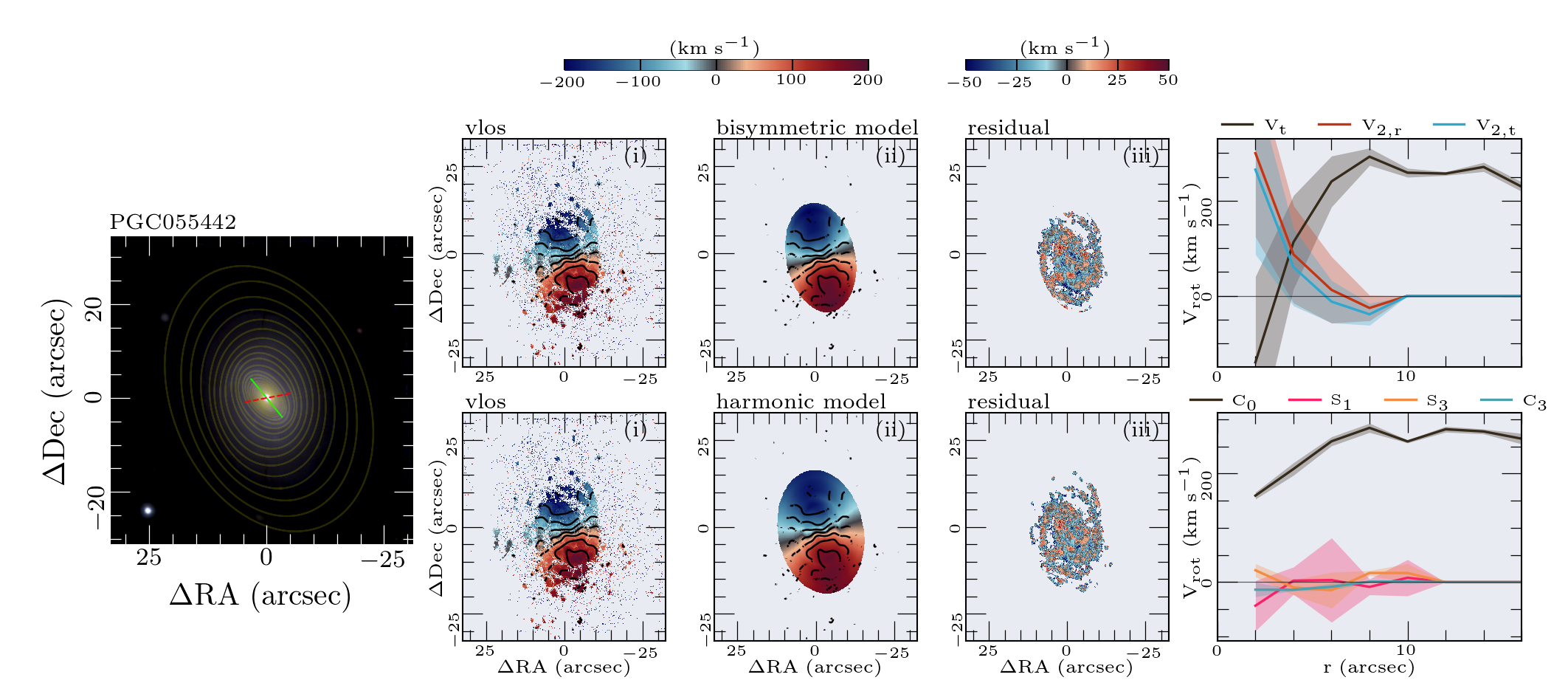}
\figsetgrpnote{Same  as Fig.~\ref{fig:appendix1} but this time for the \ha~velocity maps. We show results only for those galaxies with \ha~emission detected along the bars.}
\figsetgrpend

\figsetgrpstart
\figsetgrpnum{11.6}
\figsetgrptitle{NGC289}
\figsetplot{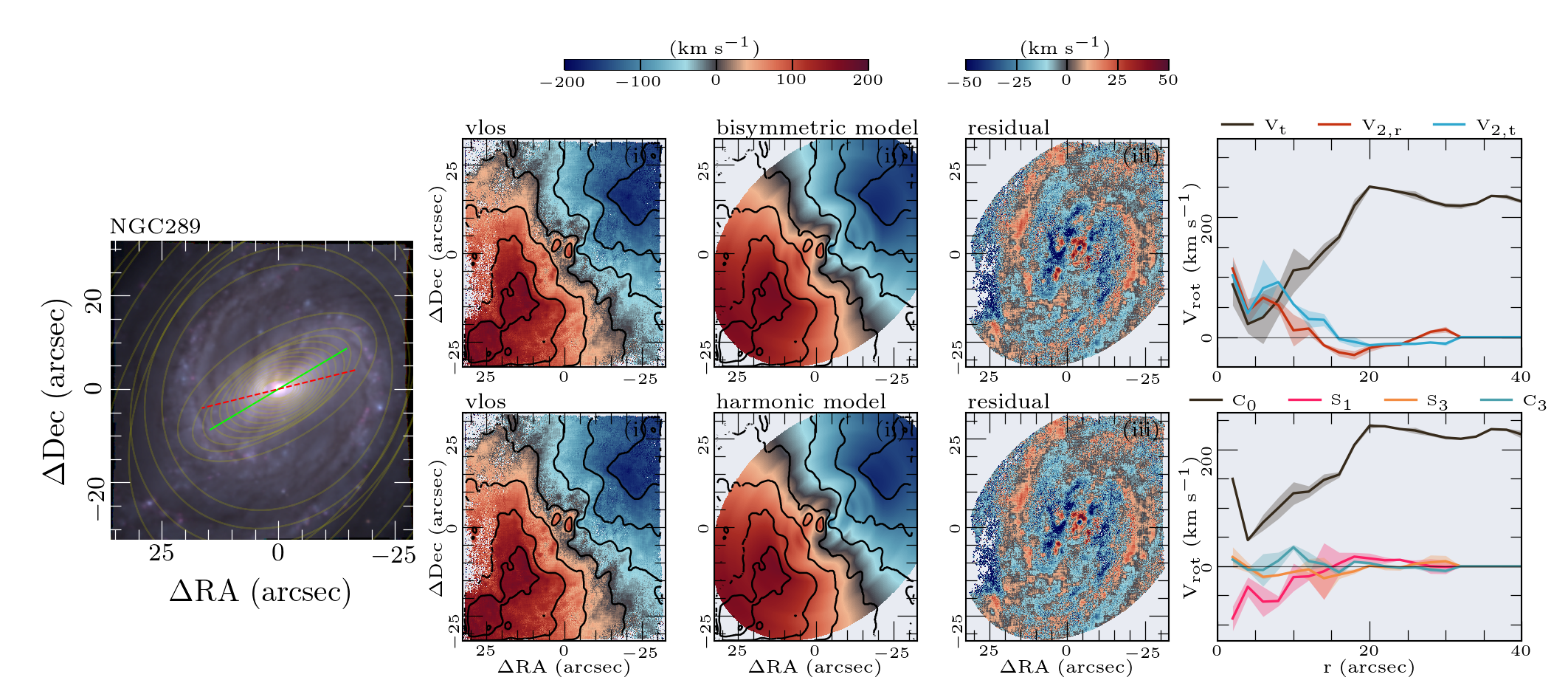}
\figsetgrpnote{Same  as Fig.~\ref{fig:appendix1} but this time for the \ha~velocity maps. We show results only for those galaxies with \ha~emission detected along the bars.}
\figsetgrpend

\figsetend

\newpage
\bibliography{main.bbl}
\bibliographystyle{aasjournal}



\end{document}